\newcommand\aap{{A\&A}}
\newcommand\aaps{{A\&AS}}
\newcommand\aj{{AJ}}
\newcommand\apj{{ApJ}}
\newcommand\apjl{{ApJ}}
\newcommand\apjs{{ApJS}}
\newcommand\araa{{ARA\&A}}

\newcommand\mnras{{MNRAS}}
\newcommand\nat{{Nat}}

\newcommand\pasp{{PASP}}

\documentclass[usenatbib]{mn2e}

\usepackage{epsfig}
\usepackage{rotating}
\usepackage{amssymb}
\usepackage{multirow}

\title[Distant radio galaxies in the southern hemisphere]{A new search for distant radio galaxies in the southern hemisphere  -- I. Sample definition and radio properties}
\author[J. W. Broderick et al.]{
\parbox[t]{\textwidth}{
J.~W.\ Broderick\thanks{E-mail: jess@physics.usyd.edu.au}, J.~J.\ Bryant, R.~W.\
Hunstead, E.~M. Sadler and T.\ Murphy}
\vspace*{6pt} \\
School of Physics, University of Sydney, NSW 2006, Australia \\}
\pubyear{2007}
\begin{document}
\maketitle

\begin{abstract}

This paper introduces a new program to find high-redshift radio galaxies in the southern hemisphere through ultra-steep spectrum (USS) selection. We define a sample of 234 USS radio sources with spectral indices $\alpha^{843}_{408} \leq -1.0$ ($S_{\nu} \propto \nu^{\alpha}$) and flux densities $S_{408} \geq 200$ mJy in a region of 0.35 sr, chosen by cross-correlating the revised 408 MHz Molonglo Reference Catalogue, the 843 MHz Sydney University Molonglo Sky Survey and the 1400 MHz NRAO VLA Sky Survey in the overlap region $-40\degr < \delta < -30\degr$. We present Australia Telescope Compact Array (ATCA) high-resolution 1384 and 2368 MHz radio data for each source, which we use to analyse the morphological, spectral index and polarization properties of our sample. We find that 85 per cent of the sources have observed-frame spectral energy distributions that are straight over the frequency range 408--2368 MHz, and that, on average, sources with smaller angular sizes have slightly steeper spectral indices and lower fractional linear polarization. Fractional polarization is anti-correlated with flux density at both 1400 and 2368 MHz. We also use the ATCA data to determine observed-frame Faraday rotation measures for half of the sample.

\end{abstract}


\section{Introduction}\label{408_paper1_introduction}

High-redshift radio galaxies (HzRGs; $z > 2$) play a fundamental role in our understanding of the formation and evolution of the most massive galaxies in the early Universe. As inferred from the Hubble $K$--$z$ diagram \cite[e.g.][]{eales97,vanbreugel98,jarvis01,debreuck02a,inskip02,willott03,rocca04}, galaxies with powerful radio emission are among the most massive systems at each cosmic epoch, and are 2--3 mag brighter than optically-selected galaxies at $z > 1$. Explaining how HzRGs formed is a crucial test for hierarchical models of galaxy formation \citep[e.g.][]{cole00,kauffmann00,baugh03,somerville04,bower06}; the discovery of quasars at $z > 6$ \citep[e.g.][]{fan04,fan06,mcgreer06} implies that the central supermassive ($\sim$$10^{9}$ M$_{\sun}$) black hole must be in place $<$ 1 Gyr after the Big Bang. Moreover, given the correlation between black hole and galaxy bulge mass \citep[e.g.][]{gebhardt00,mclure02}, we can trace a direct evolutionary path from the most massive galaxies at high redshift to powerful radio galaxies in the local Universe, which are well known to be giant ellipticals \citep*{matthews64}.

As HzRGs are rare objects, it is necessary to filter out low-redshift objects from large radio catalogues in order to maximize the efficiency of follow-up observations. The most successful selection method has proven to be ultra-steep spectrum (USS) selection. This technique is based on the results of \citet*[][]{tielens79} and \citet[][]{blumenthal79}, who found that radio sources with the steepest spectral indices\footnote{We define the radio spectral index $\alpha$ by the relation $S_{\nu} \propto \nu^{\alpha}$, where $S_{\nu}$ is the flux density at frequency $\nu$.} were not detected on Palomar Observatory Sky Survey (POSS) plates. The implied correlation between redshift and spectral index (the $z$--$\alpha$ correlation) is regularly used to optimize HzRG searches; USS samples with $\alpha \lesssim -1$ \citep[e.g.][]{roettgering94,chambers96,blundell98,debreuck00} have yielded over 30 $z > 3$ radio galaxies, out to $z=5.19$ \citep{vanbreugel99}.

Given that Malmquist bias is an inherent feature of any flux-limited sample, and an upper limit exists in radio luminosity, we must use radio surveys with faint flux density limits to find the most distant massive galaxies. Indeed, HzRG searches have been primarily limited to the northern hemisphere due to a paucity of sensitive radio surveys in the south. However, with the completion of the 843 MHz Sydney University Molonglo Sky Survey \citep*[SUMSS;][]{bock99,mauch03}, and a deeper re-analysis of its predecessor, the 408 MHz Molonglo Reference Catalogue \citep[MRC;][]{large81}, this is no longer the case. Using these two surveys, and the 1400 MHz NRAO VLA Sky Survey \citep[NVSS;][]{condon98}, we are carrying out extensive searches for southern HzRGs, which will be used to better constrain galaxy formation and evolution theories. We select samples of USS sources from the revised MRC (henceforth referred to as MRCR), SUMSS and NVSS in the overlap region $-40\degr < \delta < -30\degr$.  Initially, we \citep{debreuck04} defined a pilot sample of 76 USS sources with $\alpha^{1400}_{843} \leq -1.3$ and $S_{1400} \geq 15$ mJy by cross-correlating the SUMSS and NVSS catalogues. Follow-up spectroscopy of this sample (henceforth referred to as the SUMSS--NVSS sample) resulted in the discovery of five radio galaxies at $z > 3$, the highest redshift being $z=3.976$ \citep{debreuck06}. A further seven galaxies may be at $z \gtrsim 6$ or heavily obscured by dust \citep{debreuck06}. In this paper, we define the follow-up MRCR--SUMSS sample, which consists of 234 USS sources with $\alpha^{843}_{408} \leq -1.0$ and $S_{408} \geq 200$ mJy selected by cross-matching the MRCR, SUMSS and NVSS catalogues.

Our observing strategy is the same as that used previously for the SUMSS--NVSS sample. After defining our sample through USS selection, we use high-resolution radio observations from the Australia Telescope Compact Array \citep*[ATCA;][]{frater92} in conjunction with near-infrared (NIR) $K$-band imaging to unambiguously identify the host galaxy of each radio source. Redshifts are then determined by optical or NIR spectroscopy. In this paper, we focus on the radio properties of the MRCR--SUMSS sample. $K$-band and spectroscopic data will be presented in a subsequent paper (Bryant et al., in preparation), henceforth referred to as Paper II.

This paper is set out as follows. In Section~\ref{408_paper1_sample_definition1}, we define the MRCR--SUMSS sample. In Section~\ref{408_paper1_ATCA_observations}, we describe the high-resolution ATCA observations of the sample. We analyse the radio properties of the sample in Section~\ref{408_paper1_analysis_radio_properties}, before discussing in Section~\ref{408_paper1_discussion} both the efficiency of our search and the implications of the radio properties. Finally, we present our conclusions in Section~\ref{408_paper1_conclusions}. Throughout we assume a flat $\Lambda$ cold dark matter cosmology with  $H_0=71$ km s$^{-1}$ Mpc$^{-1}$, $\Omega_{\rm M}=0.27$ and $\Omega_{\Lambda}=0.73$ \citep{spergel03}.

\section{Sample definition}\label{408_paper1_sample_definition1}

\subsection{Input catalogues}

\subsubsection{MRCR}\label{408_paper1_MRC_summary}

The MRC \citep{large81} is the product of a 408 MHz continuum survey undertaken with the Molonglo Cross Telescope \citep{mills63}. The survey was conducted in the region $-85 \degr < \delta < 18 \fdg 5$ for Galactic latitudes $|b| > 3 \degr$.  As a result of a re-analysis of the digitised data from the Molonglo Cross, a much deeper version of the MRC is now available, with the flux density limit lowered from 1 Jy to $\sim$200 mJy ($\sim$5$\sigma$). The MRCR comprises $\sim$77\,000 sources with $S_{408} \geq 200$ mJy. The angular resolution is $2.86 \sec (\delta + 35 \fdg 5) \times 2.62$ arcmin$^{2}$, the typical positional uncertainty near the flux density limit is $\sim$20 arcsec, and the source density is $\sim$2.5 deg$^{-2}$. We stress that the MRCR is incomplete: only data from the best-quality scans have been included. Details of this catalogue will be published elsewhere (Crawford, in preparation).

\subsubsection{SUMSS}\label{408_paper1_SUMSS_summary}

SUMSS \citep{bock99,mauch03} is an 843 MHz continuum survey of the entire sky south of declination $\delta = -30\degr$ with the Molonglo Observatory Synthesis Telescope \citep[MOST;][]{mills81,robertson91}. The angular resolution is $45 \: \rm cosec |\delta| \times 45$ arcsec$^{2}$. For $-40\degr < \delta < -30\degr$, the median rms noise level is 1.9 mJy beam$^{-1}$, and the completeness limit is 18 mJy. Positional uncertainties are 1--2 arcsec for sources with peak brightnesses greater than 20 mJy beam$^{-1}$. In this paper, we use version 2.0 of the SUMSS catalogue\footnote{From version 2.0 of the SUMSS catalogue onwards, an error in the fitting routine that resulted in erroneous flux densities for some sources north of $\delta=-40\degr$ has been rectified.}, which consists of $\sim$210\,000 sources, with surface density $\sim$20 deg$^{-2}$ above 10 mJy beam$^{-1}$.

\subsubsection{NVSS}\label{408_paper1_NVSS_summary}

NVSS \citep{condon98} is a 1400 MHz continuum survey of the entire sky north of $\delta = -40\degr$ with the Very Large Array \citep*[VLA;][]{thompson80}. The angular resolution is $45 \times 45$ arcsec$^{2}$, the rms noise level is 0.45 mJy, and the completeness threshold is 2.5 mJy. The rms positional uncertainties in right ascension and declination are $\lesssim 1$ arcsec for sources stronger than 15 mJy. In this paper, we use version 41 of the NVSS catalogue, which consists of $\sim$$2 \times 10^6$ sources with surface density $\sim$$50$ deg$^{-2}$.

\subsection{MRCR--SUMSS sample}\label{408_paper1_sample_definition}

\begin{table*}
\begin{minipage}{175mm}
\caption{Sources in the SUMSS--NVSS sample \citep{debreuck04} that also have $\alpha^{843}_{408} \leq -1.0$ and $S_{408} \geq 200$ mJy.}\label{408_paper1_table:sources_common_with_SUMSS-NVSS}
\begin{tabular}{|c|r|r|r|r|r|r|}
\hline
\hline
\multicolumn{1}{|c|}{Source} & \multicolumn{1}{|c|}{$S_{408}$} & \multicolumn{1}{|c|}{$S_{843}$} & \multicolumn{1}{|c|}{$S_{1400}$} & \multicolumn{1}{|c|}{$\alpha^{843}_{408}$} & \multicolumn{1}{|c|}{$\alpha^{1400}_{843}$} & \multicolumn{1}{|c|}{$z$} \\
& \multicolumn{1}{|c|}{(mJy)} & \multicolumn{1}{|c|}{(mJy) } & \multicolumn{1}{|c|}{(mJy) }  & & & \\
\hline
NVSS J011606$-$331241 & $220 \pm 38$ & $44.1 \pm 1.8$ & $22.1 \pm 0.8$ & $-2.21 \pm 0.24$ & $-1.36 \pm 0.11$ & $0.352 \pm 0.001$ \\
NVSS J202856$-$353709 & $480 \pm 47$ & $195.5 \pm 6.3$ & $98.1 \pm 3.7$ & $-1.24 \pm 0.14$ & $-1.36 \pm 0.10$ & Obscured by Galactic star \\
NVSS J202945$-$344812 & $244 \pm 29$ & $104.0 \pm 3.6$ & $52.5 \pm 2.0$ & $-1.18 \pm 0.17$ & $-1.35 \pm 0.10$ & $1.497 \pm 0.002$ \\
NVSS J225719$-$343954 & $258 \pm 27$ & $85.8 \pm 2.9$ & $37.5 \pm 1.2$ & $-1.52 \pm 0.15$ & $-1.63 \pm 0.09$ & $0.726 \pm 0.001$ \\
NVSS J230527$-$360534 & $205 \pm 20$ & $63.5 \pm 2.4$ & $31.3 \pm 1.0$ & $-1.61 \pm 0.14$ & $-1.39 \pm 0.10$ & Not detected; $z \gtrsim 6$ candidate \\
NVSS J230846$-$334810 & $276 \pm 22$ & $119.8 \pm 4.0$ & $63.1 \pm 2.3$ & $-1.15 \pm 0.12$ & $-1.26 \pm 0.10$ & Not observed \\
NVSS J231727$-$352606 & $265 \pm 32$ & $114.8 \pm 3.8$ & $59.2 \pm 1.8$ & $-1.15 \pm 0.17$ & $-1.31 \pm 0.09$ & $3.874 \pm 0.002$ \\
NVSS J232058$-$365157 & $257 \pm 36$ & $104.8 \pm 3.8$ & $51.4 \pm 1.6$ & $-1.24 \pm 0.20$ & $-1.40 \pm 0.09$ & Not observed \\
NVSS J233729$-$355529 & $685 \pm 60$ & $267.0 \pm 8.1$ & $123.0 \pm 4.4$ & $-1.30 \pm 0.13$ & $-1.53 \pm 0.09$ & Not observed \\
NVSS J234145$-$350624 & $8330 \pm 500$ & $3401 \pm 102$ & $1823.2 \pm 54.7$ & $-1.23 \pm 0.09$ & $-1.23 \pm 0.08$ & $0.644 \pm 0.001$ \\
\hline
\multicolumn{7}{p{175mm}}{{\em Notes.} Flux densities are from a preliminary version of the MRCR, version 2.0 of the SUMSS catalogue and version 41 of the NVSS catalogue. Revisions to the SUMSS and NVSS flux densities have resulted in two sources falling outside of the $\alpha^{1400}_{843} \leq -1.3$ cutoff used to define the SUMSS--NVSS sample. Redshifts are from \citet{debreuck06}.}  \\
\end{tabular}
\end{minipage}
\end{table*}

We cross-matched the MRCR, SUMSS and NVSS catalogues in the overlap region $-40\degr < \delta < -30\degr$ in order to define a sample of USS sources with $\alpha^{843}_{408} \leq -1.0$ and $S_{408} \geq 200$ mJy. We chose to use three low-frequency catalogues in order to investigate the spectral energy distributions (SEDs) of our targets. While the spectral index cutoff is flatter than the $\alpha^{1400}_{843} \leq -1.3$ condition used in our pilot study, we note that a number of $z > 2$ radio galaxies in the SUMSS--NVSS sample were found to have spectral indices as flat as $-0.90$ after SEDs were constructed over a wider frequency range \citep{klamer06}. The Galactic plane was avoided by restricting the cross-matching to the right ascension regions 20--4$^{\rm h}$ and 10--16$^{\rm h}$ ($|b| > 10\degr$). Note that the former right ascension zone could in principle be extended from 4$^{\rm h}$ to 7$^{\rm h}$ but the SUMSS catalogue did not cover this region when we defined our sample.

We adopted a 60 arcsec search radius when searching for SUMSS matches around the position of each MRCR source. As the SUMSS source density is $\sim$20 deg$^{-2}$, the probability of an MRCR--SUMSS match occurring by chance within 60 arcsec is roughly 2 per cent. We then used a 10 arcsec radius to search for SUMSS--NVSS matches; the probability of a SUMSS--NVSS match occurring by chance within 10 arcsec is $\sim$0.1 per cent. Thus, the probability of a chance match among all three catalogues is $\sim$0.002 per cent, which implies that the expected number of spurious MRCR--SUMSS--NVSS matches in a sample of 234 sources is much less than one.

Due to the difference in resolution of the MRCR, SUMSS and NVSS catalogues, source confusion will affect the reliability of the spectral indices. Therefore, we visually inspected each individual SUMSS field and removed any sources that were obviously blended at 408 MHz. In addition, as the resolution of SUMSS varies from $70 \times 45$ arcsec$^{2}$ to $90 \times 45$ arcsec$^{2}$ over the declination range $-40\degr$ to $-30\degr$, we removed any SUMSS--NVSS matches with an additional NVSS match within 100 arcsec, except for five cases where high-resolution ATCA imaging confirmed that multiple NVSS components were indeed matched to a single SUMSS source (see Table 3). While the removal of such SUMSS--NVSS matches imposes an implicit angular size cutoff in the sample, we do not expect a significant number of HzRGs to be excluded, if any. Studies of powerful radio sources have found that linear size decreases with increasing redshift \citep*[e.g.][]{barthel88,neeser95,blundell99}, and no HzRGs with angular sizes greater than 1 arcmin have been reported in the literature.

To optimize the optical/NIR spectroscopy of our USS sample, it was essential to remove any low-redshift interlopers. We used SuperCOSMOS Sky Survey \citep[][]{hambly01} scans of UK Schmidt IIIaJ (blue, UKJ) and IIIaF (red, UKR) plates (limiting magnitudes $B_{\rm J} \approx 22.5$ and $R \approx 21.5$, respectively) to identify any sources with bright optical identifications, which are most likely at $z \ll 1$. However, for 32 sources, candidate optical counterparts could only be unambiguously identified after high-resolution ATCA images were obtained, leading to the inclusion of some low-redshift sources in the sample (see Section~\ref{408_paper1_morphologies}).

The MRCR--SUMSS sample consists of 234 sources for which we have obtained follow-up ATCA imaging at both 1384 and 2368 MHz. The sky area covered by our sample is 0.35 sr. As the sample was assembled over a three year period, revisions to the 408, 843 and 1400 MHz catalogues have meant that a number of sources no longer meet our original selection criteria. Using a preliminary version of the MRCR, version 2.0 of the SUMSS catalogue and version 41 of the NVSS catalogue, we find that 205 sources still have $\alpha^{843}_{408} \leq -1.0$. We decided to retain the remaining 29 sources ($-0.53 \leq \alpha^{843}_{408} < -1.0$) because of the broadness of the $z$--$\alpha$ correlation, and also to increase our spectral index baseline for seeking trends with other radio properties.

The MRCR--SUMSS and SUMSS--NVSS samples have been defined over the same declination range, and therefore some USS sources satisfy the selection criteria of both samples. Note that both samples do not fully overlap because the sources in the SUMSS--NVSS sample are fainter (see Section~\ref{408_paper1_fluxes}). In Table~\ref{408_paper1_table:sources_common_with_SUMSS-NVSS}, we list the flux densities, spectral indices and redshifts for an additional ten sources from the SUMSS--NVSS sample that also have $\alpha^{843}_{408} \leq -1.0$ and $S_{408} \geq 200$ mJy. In this paper, we restrict our analysis to the 234 new targets for which high-resolution radio data have been obtained. A full description of the sources in Table~\ref{408_paper1_table:sources_common_with_SUMSS-NVSS} is given in \citet{debreuck04}, \citet{debreuck06} and \citet{klamer06}.

\section{ATCA observations}\label{408_paper1_ATCA_observations}

Table~\ref{408_paper1_table:ATCA_log} contains a summary of the high-resolution radio observations of the MRCR--SUMSS sample obtained with the ATCA.\footnote{Proposal number C1000.} For each session, we list the date of observation, the number of sources observed, the array configuration (including the baselines spanned by the array), and the secondary calibrators observed. In each session, we performed dual-frequency observations at 1384 and 2368 MHz (wavelengths 20 and 13 cm) using a $2\times 128$ MHz bandwidth correlator configuration. We obtained 5--10 cuts of 3--5 min duration of each source using the snapshot imaging technique. Typically, the cuts were spread out over a $\sim$12 h period to optimize the $uv$ coverage. The median integration time per source was $\sim$25 min. PKS B1934$-$638 was used as the primary calibrator in each session.

Data reduction followed standard procedures in {\scriptsize MIRIAD} \citep*{sault95}. The sensitivity was optimized by using natural weighting to construct total intensity (Stokes $I$) dirty images. To ensure that the sidelobes of any nearby bright sources were properly deconvolved, the dirty images were made at least twice the size of the primary beam full-width at half-maximum (FWHM), which is 33 arcmin at 1384 MHz and 22 arcmin at 2368 MHz. It is well known that running the {\scriptsize CLEAN} deconvolution algorithm unconstrained on such images can lead to flux being subtracted from the target source. This effect, known as {\scriptsize CLEAN} bias, is discussed by \citet{condon98} in relation to NVSS data. To resolve the problems encountered with unconstrained {\scriptsize CLEAN}, we have developed a technique using {\scriptsize MIRIAD} that automatically constrains {\scriptsize CLEAN} by selecting {\scriptsize CLEAN} regions around the positions of known sources in the NVSS and/or SUMSS catalogues. This is analogous to the {\scriptsize FACES} task in the {\scriptsize AIPS} data reduction package.

Using our automatically constrained {\scriptsize CLEAN} technique, we defined {\scriptsize CLEAN} regions around all of the NVSS sources in each field with flux density $S_{1400} \geq 5$ mJy. We chose NVSS rather than SUMSS for this task because of its lower flux density limit. After the {\scriptsize CLEAN} regions had been defined, {\scriptsize CLEAN} was run for 2500 iterations or until the peak residuals were no larger than 0.5 mJy. The default loop gain of 0.1 was used. We then performed five iterations of phase-only self-calibration to improve the image dynamic range. The median angular resolution is $15.8 \times 9.4$ arcsec$^{2}$ at 1384 MHz and $9.2 \times 5.6$ arcsec$^{2}$ at 2368 MHz, while the median rms noise levels are 0.32 mJy beam$^{-1}$ at 1384 MHz and 0.27 mJy beam$^{-1}$ at 2368 MHz. We estimate the astrometry to be accurate to $< 1$ arcsec. The resolution and positional accuracy are sufficient to unambiguously identify the host galaxy of each radio source in our $K$-band images (Paper II).

To enable a polarimetric analysis of the sources in the MRCR--SUMSS sample, we produced Stokes $Q$ and $U$ images from the self-calibrated visibilities at 1384 and 2368 MHz. All dirty images were {\scriptsize CLEAN}ed for a maximum of 100 iterations. The {\scriptsize MIRIAD} task {\scriptsize IMPOL} was then used to make total polarized intensity images [flux density = $(S_{Q}^2+S_{U}^2)^{1/2}]$ that were corrected for Ricean bias. The median rms noise levels in the $Q$ and $U$ images, $\sigma_{QU}$, are 0.20 and 0.23 mJy beam$^{-1}$ at 1384 and 2368 MHz, respectively.

\begin{table}
\setlength{\tabcolsep}{3.25pt}
\caption{ATCA observing log.}\label{408_paper1_table:ATCA_log}
\begin{tabular}{|c|c|c|r|}
\hline
\hline
Date of & Number of & Array & \multicolumn{1}{|c|}{Secondary} \\
observations & sources & configuration & \multicolumn{1}{|c|}{calibrators} \\
& observed & (baselines) & \\
\hline
2003 & 26 & 6A & PKS B0153$-$410  \\
December 11--12 &  & (337--5939 m)  & PKS B2339$-$353 \\
\\
2004 & 48 & 6D & PKS B0153$-$410 \\
December 8--9 &  & (77--5878 m)  &  \\
\\
2005 & 72 & 6A & PKS B1232$-$416 \\
April 8--10 &  & (337--5939 m)  &  PMN B1458$-$391 \\
\\
2006 & 88 & 6C & PKS B0008$-$421   \\
April 14--17 &  & (153--6000 m)  & PKS B0153$-$410 \\
&  &  & PKS B1232$-$416 \\
&  &  & PMN B1458$-$391 \\
&  &  & PKS B2211$-$388 \\
\hline
\end{tabular}
\end{table}

\section{Analysis of radio properties}\label{408_paper1_analysis_radio_properties}

\begin{figure*}
\centering
\psfig{file=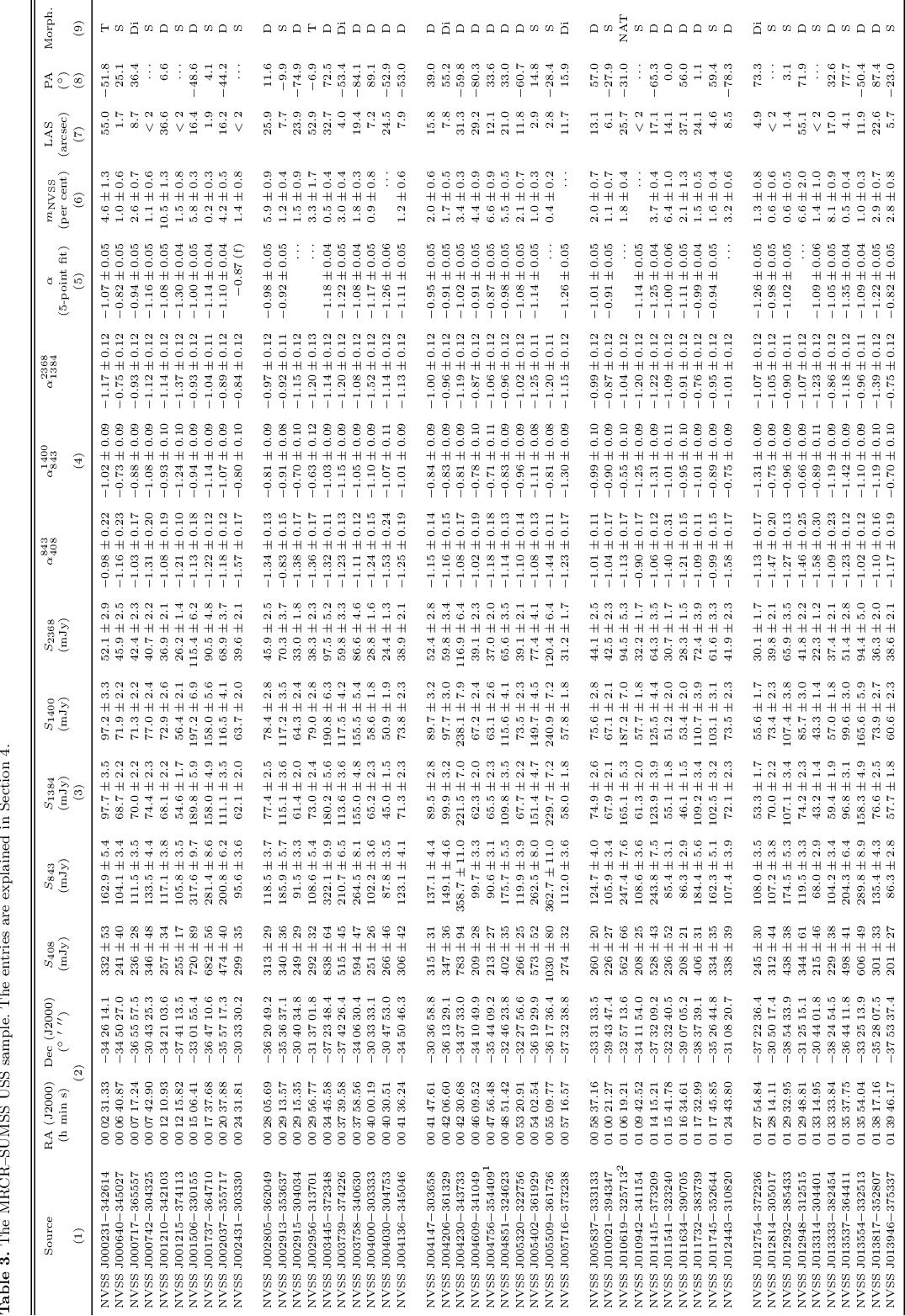,height=23.0cm,angle=180}
\label{fig:408_paper1_table:sample_definition}
\end{figure*}

\begin{figure*}
\centering
\psfig{file=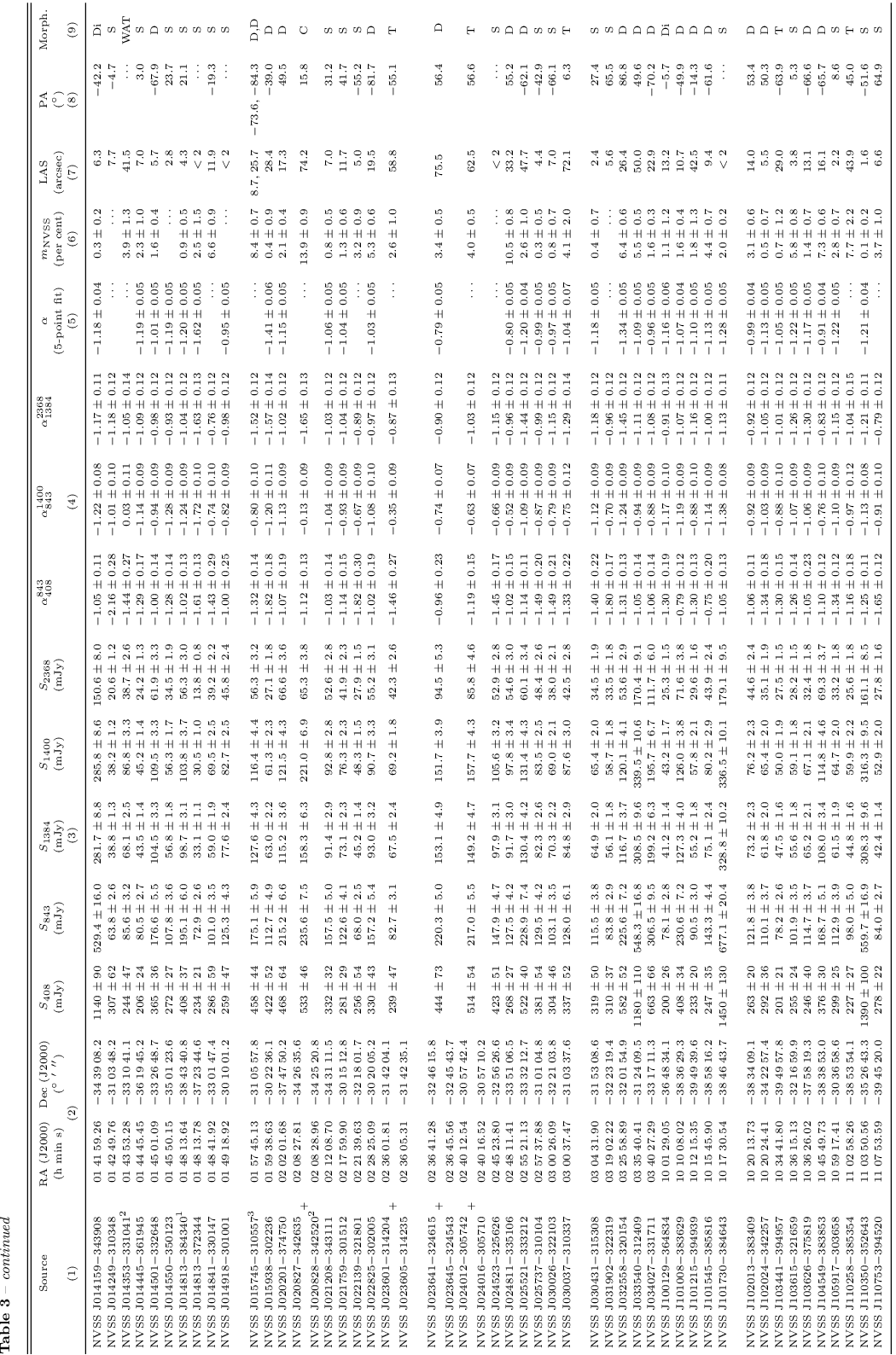,height=23.0cm,angle=180}
\end{figure*}

\begin{figure*}
\centering
\psfig{file=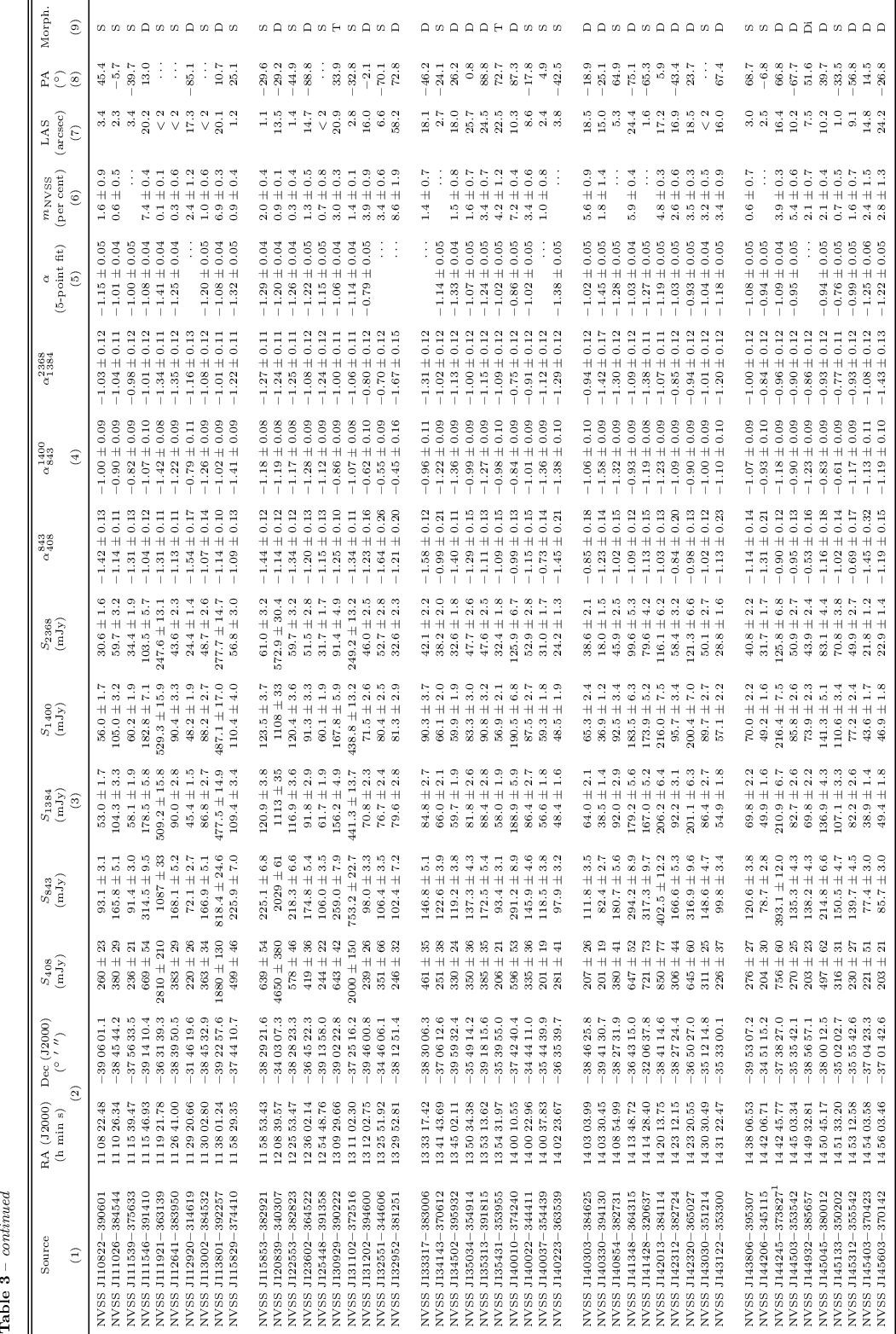,height=23.0cm,angle=180}
\end{figure*}

\begin{figure*}
\centering
\psfig{file=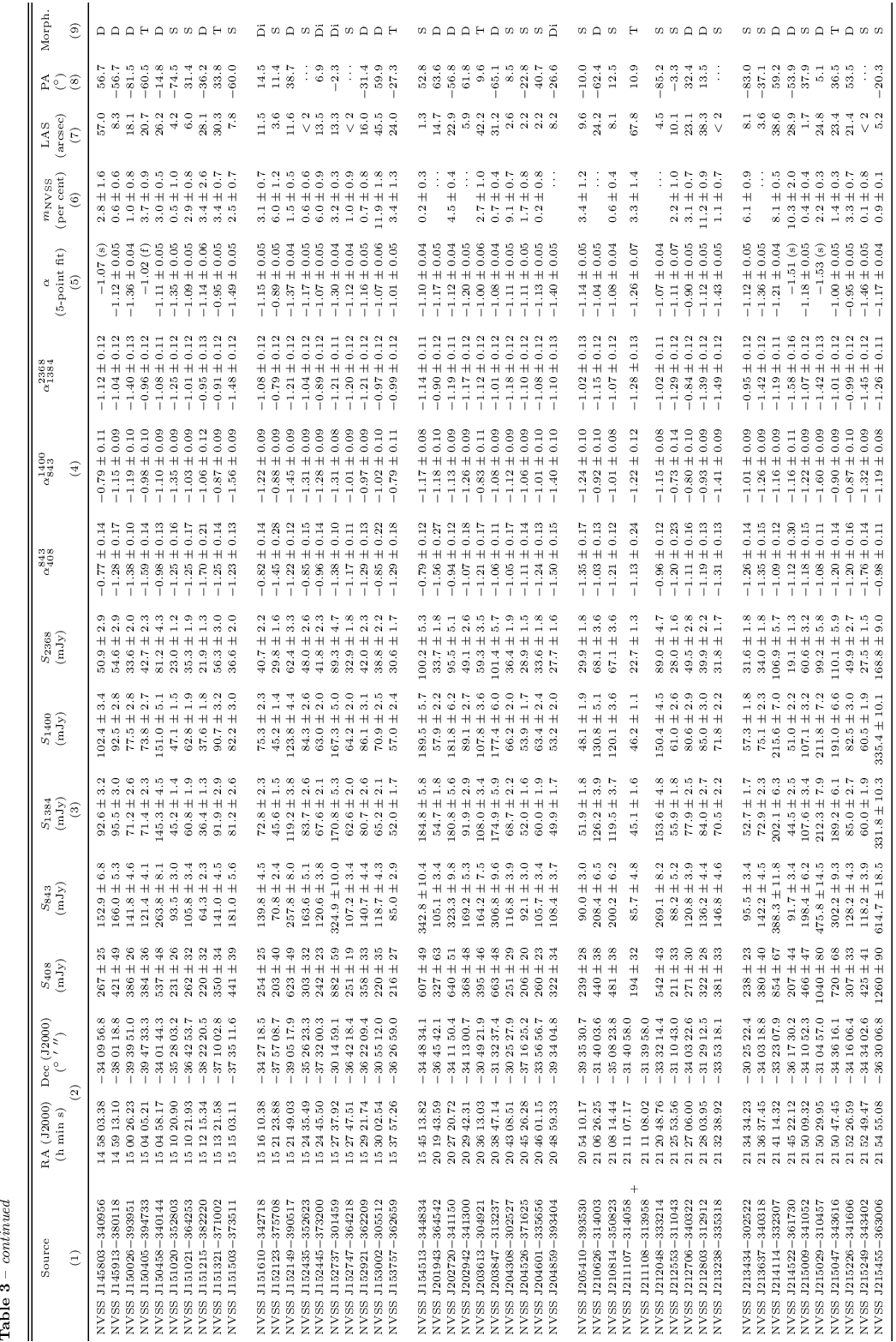,height=23.0cm,angle=180}
\end{figure*}

\begin{figure*}
\centering
\psfig{file=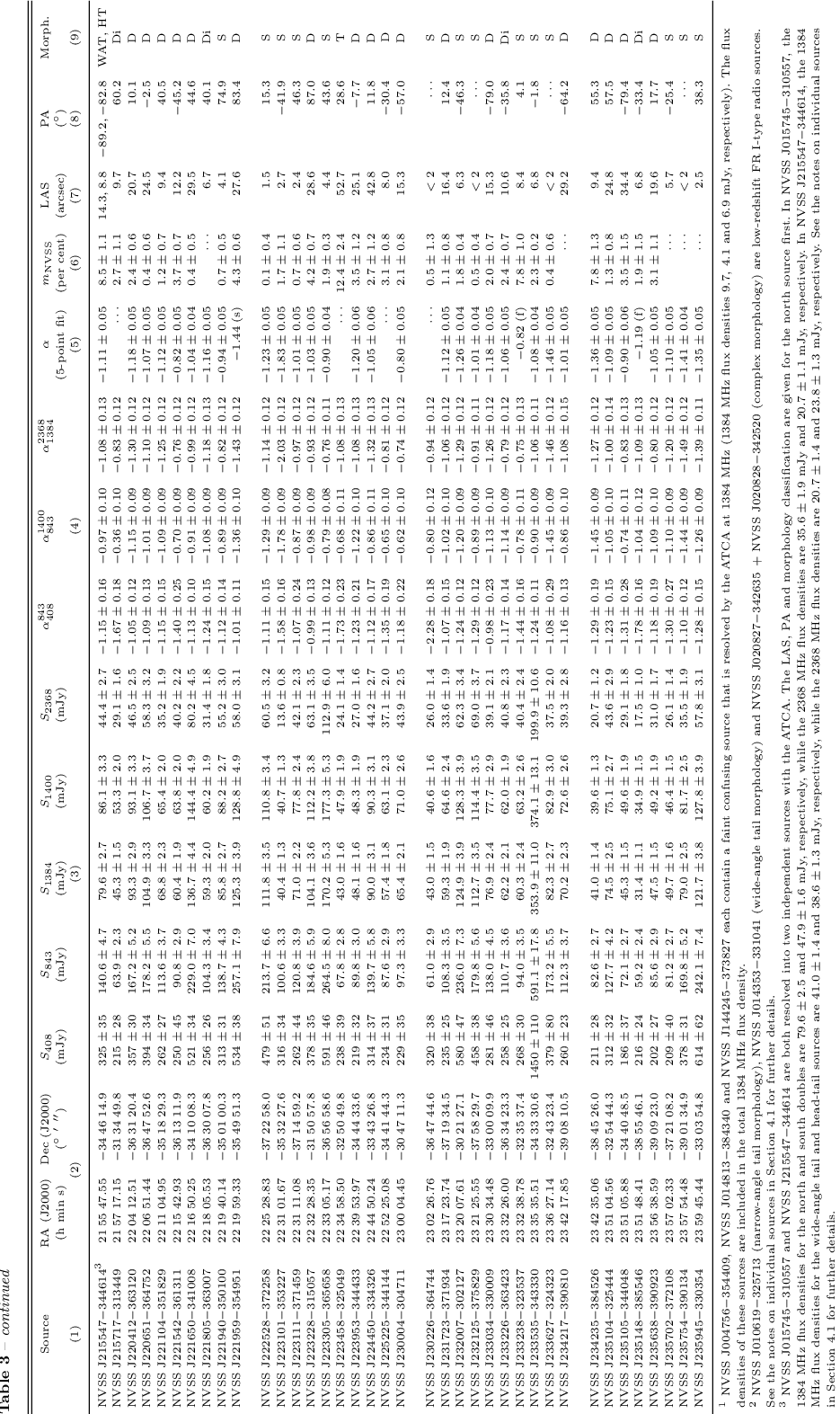,height=23.0cm,angle=180}
\end{figure*}

\begin{figure*}
\centering
\psfig{file=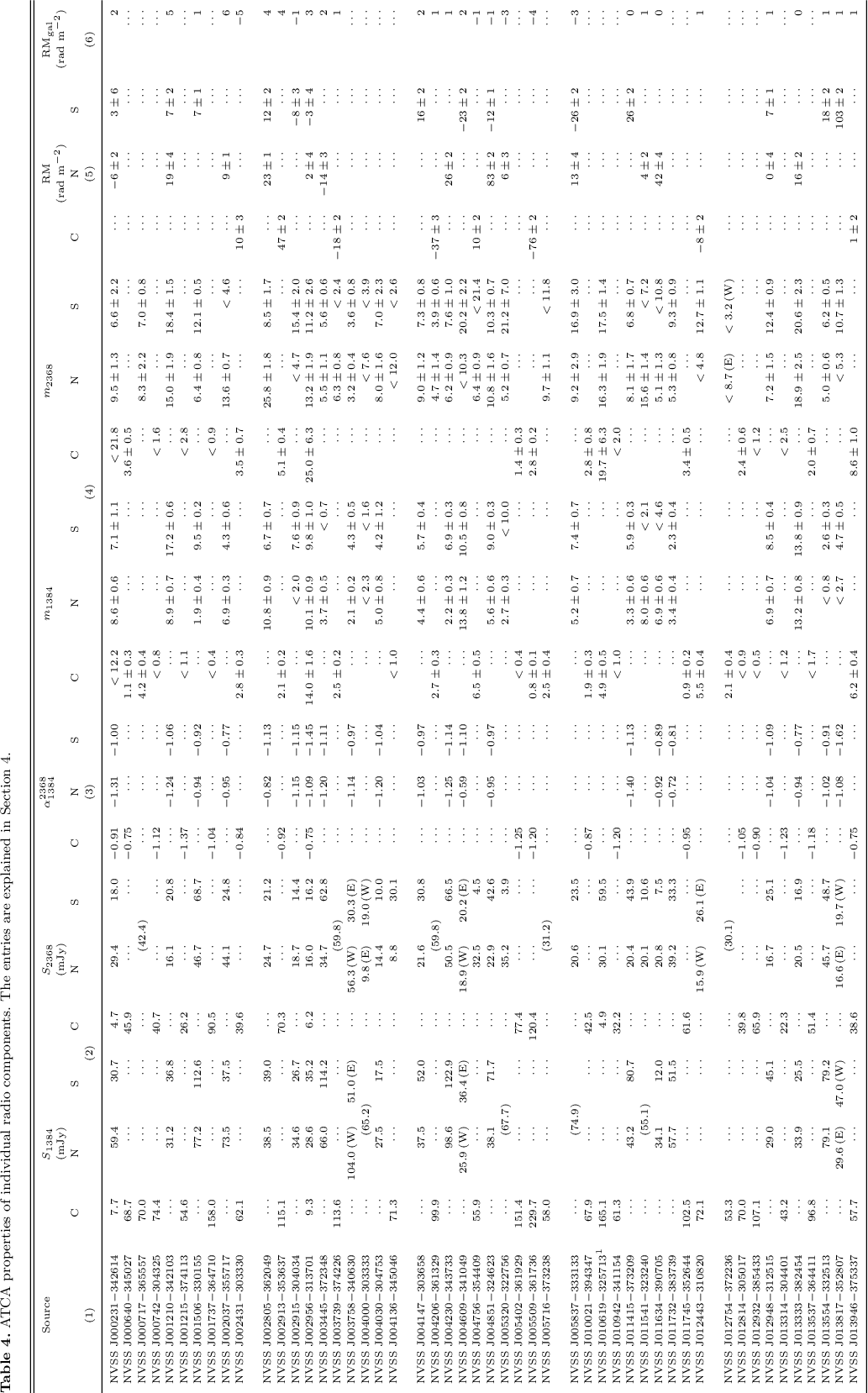,height=24.0cm,angle=180}
\end{figure*}

\begin{figure*}
\centering
\psfig{file=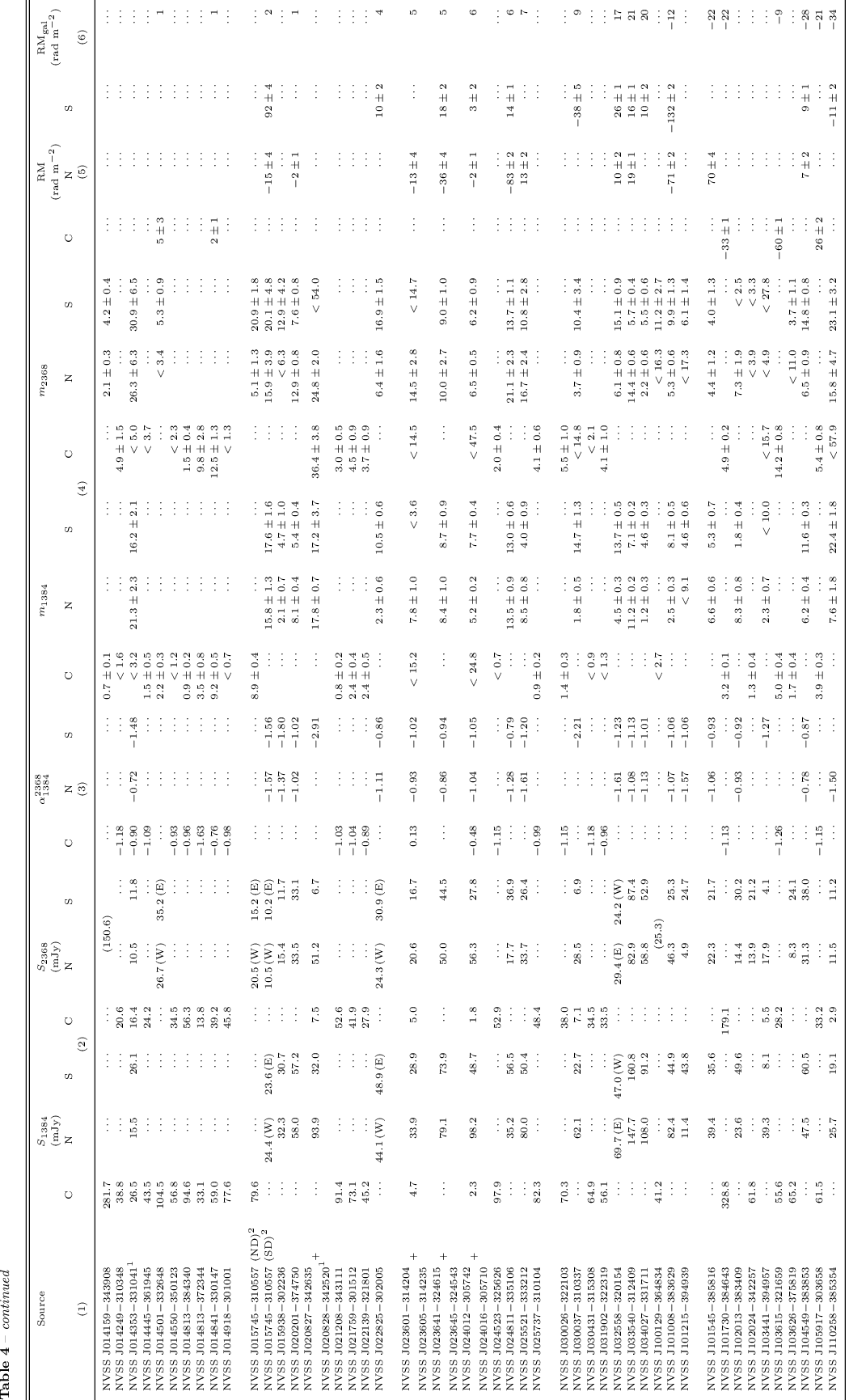,height=24.0cm,angle=180}
\end{figure*}

\begin{figure*}
\centering
\psfig{file=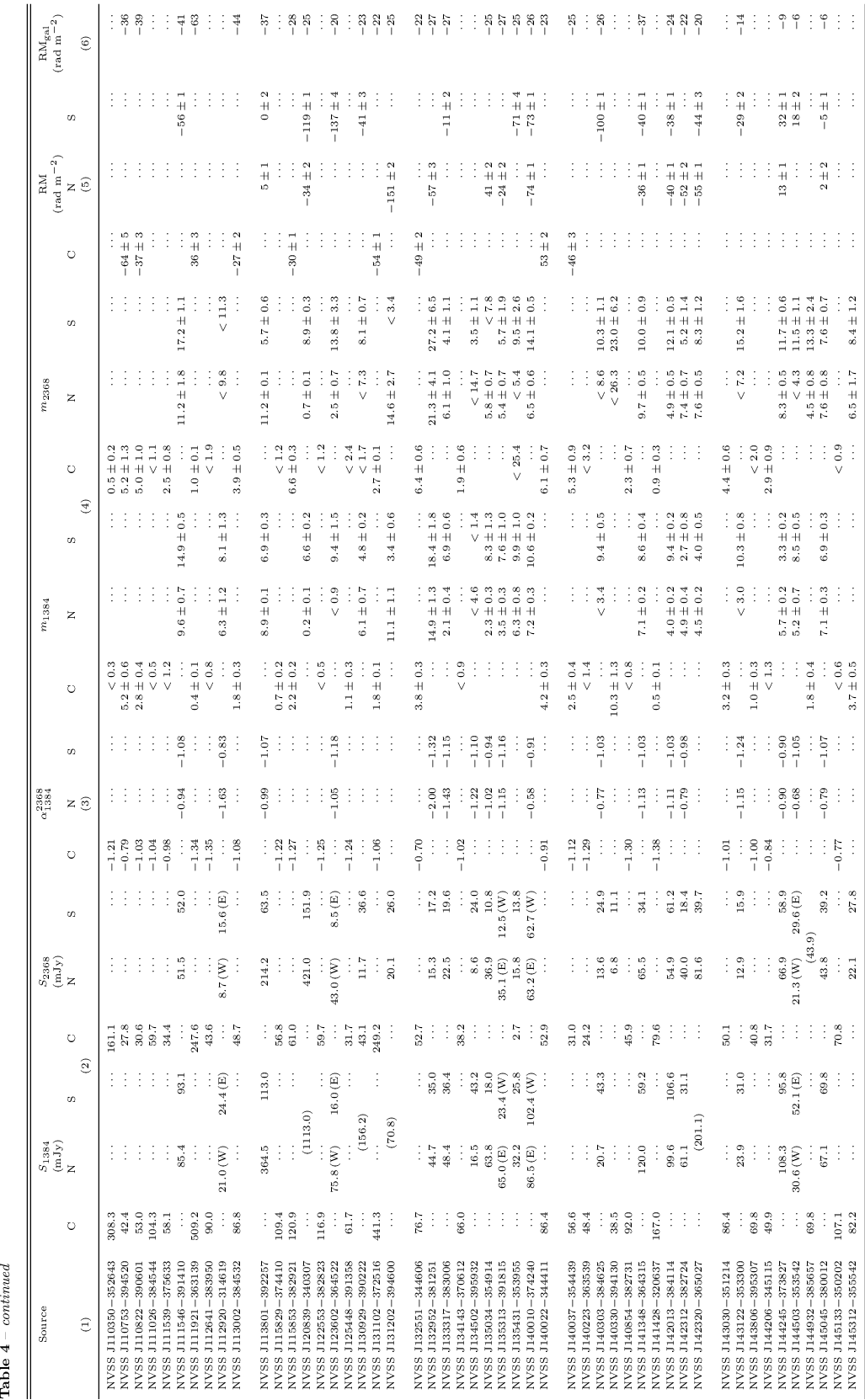,height=24.0cm,angle=180}
\end{figure*}

\begin{figure*}
\centering
\psfig{file=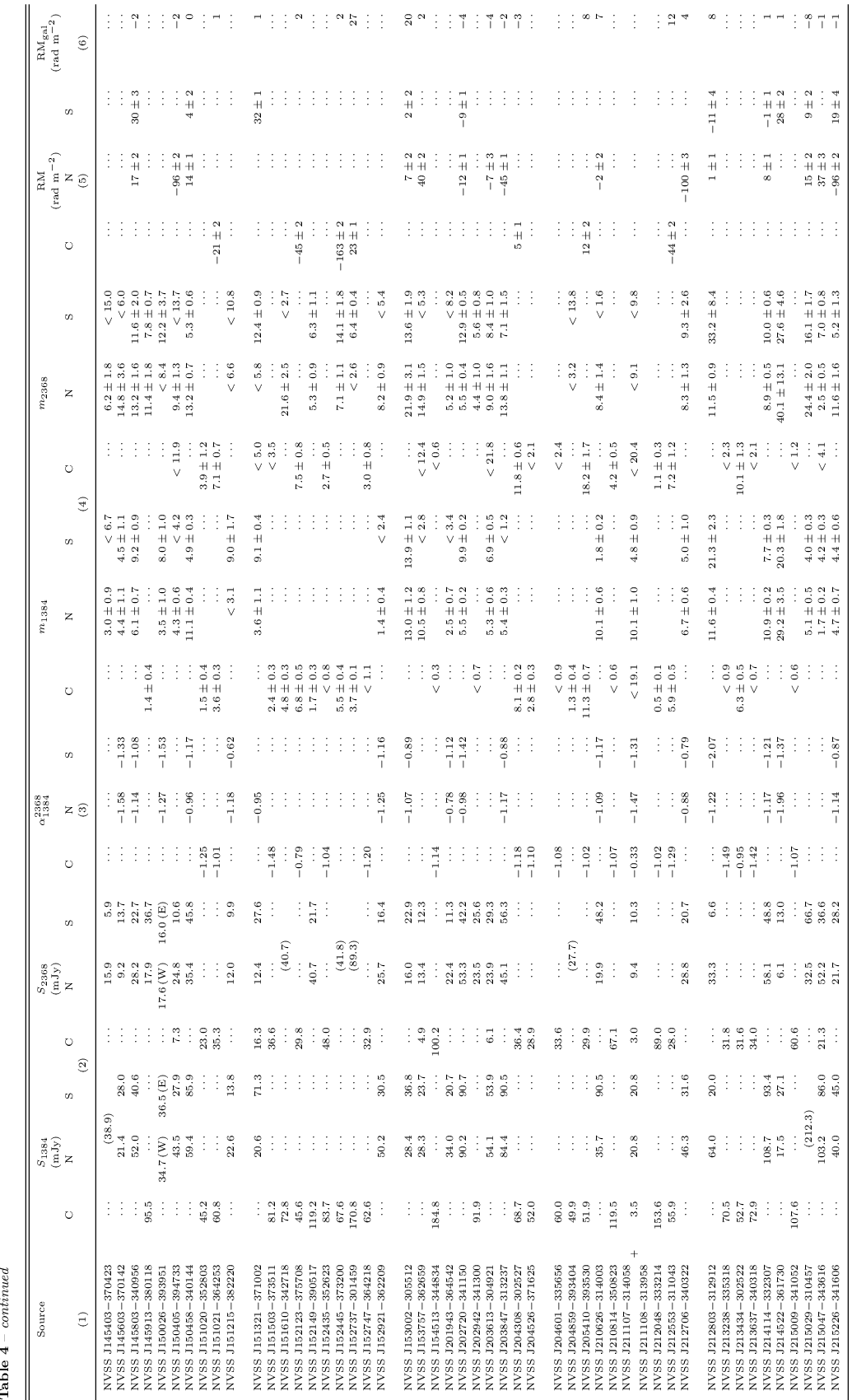,height=24.0cm,angle=180}
\end{figure*}

\begin{figure*}
\centering
\psfig{file=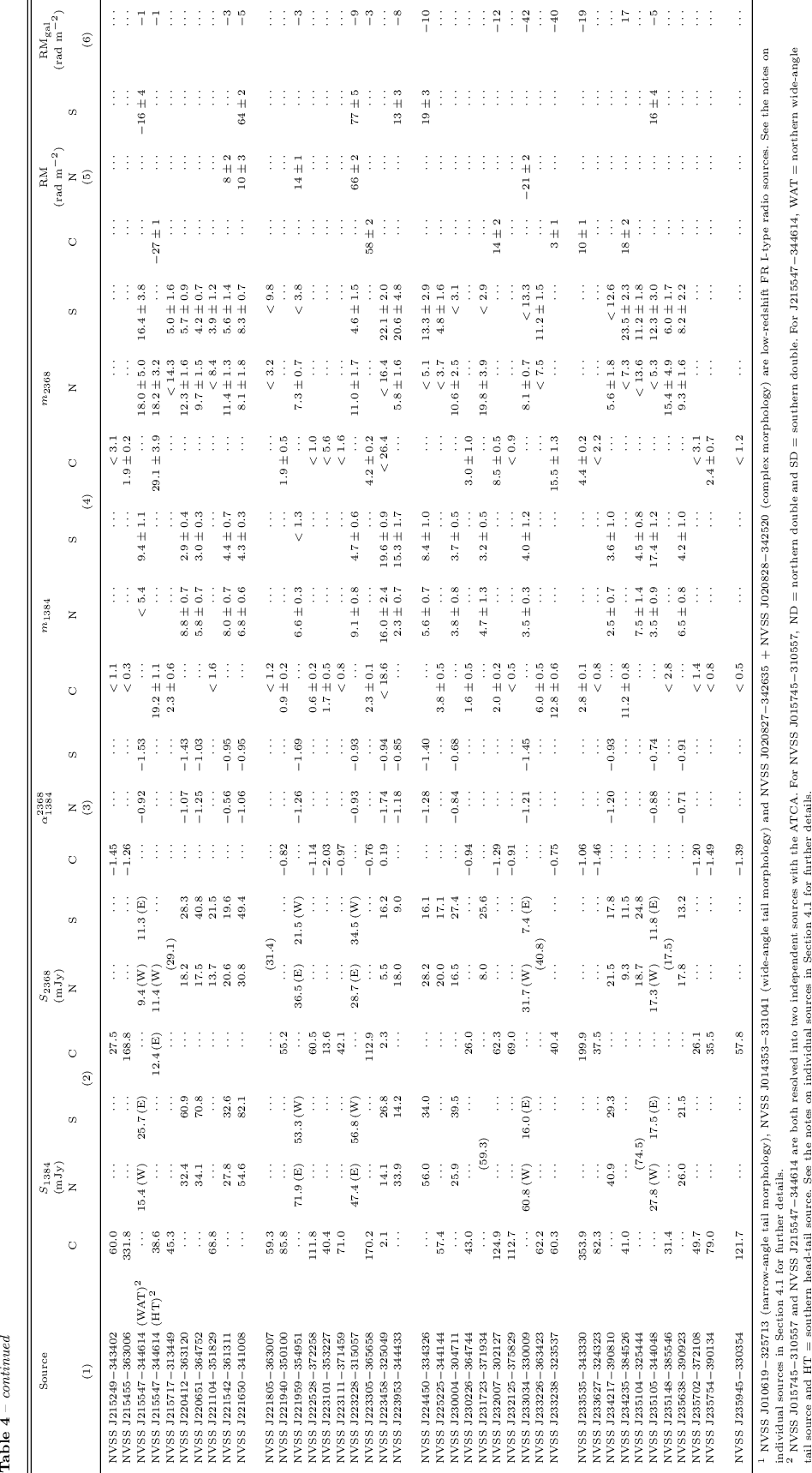,height=24.0cm,angle=180}
\end{figure*}

The radio properties of each source in the MRCR--SUMSS sample are summarised in Tables 3 and 4, and are discussed in the following sections. We define our sample in Table 3, which contains flux density and spectral index properties from 408 to 2368 MHz, as well as morphological and polarization information. In Table 4, we list the ATCA properties of each source in our sample, in particular the flux density, spectral index and polarization properties of the individual radio components.\newline
\newline
In Table 3, the columns are:
\newline
\newline
(1) Name of the source in the NVSS catalogue in IAU J2000 format.\newline
(2) The RA and Dec J2000 coordinates from the NVSS catalogue. \newline
(3) The 408, 843, 1384, 1400 and 2368 MHz integrated flux densities. The 408 MHz flux densities are from a preliminary version of the MRCR, while the 843 and 1400 MHz flux densities are from versions 2.0 and 41 of the SUMSS and NVSS catalogues.  \newline
(4) The two-point observed-frame spectral indices $\alpha^{843}_{408}$, $\alpha^{1400}_{843}$ and $\alpha^{2368}_{1384}$. \newline
(5) The observed-frame spectral index based on a linear fit to all five flux density data points. For those sources that have curved spectra that flatten (f) or steepen (s) at high frequency, we use a quadratic fit to all five flux density data points to derive the observed-frame spectral index at 1400 MHz. \newline
(6) The NVSS fractional linear polarization, determined from the integrated polarized flux densities in version 41 of the NVSS catalogue. We ignore those cases where the polarized flux density has a negative estimated value \citep[see][]{condon98}. \newline
(7) The largest angular size (LAS) at 2368 MHz. For single-component resolved sources, this is the deconvolved major axis of the elliptical Gaussian used to fit the source. For unresolved sources, we have used the distribution of deconvolved sizes to estimate a LAS upper limit of 2 arcsec. For multi-component sources, the LAS is the angular separation of the most widely-separated components.\newline
(8) The deconvolved position angle (PA) of the radio structure at 2368 MHz, measured from the north to the east. For multi-component sources, this is the orientation of the most widely-separated components used to calculate the LAS.\newline
(9) The radio morphology at 2368 MHz: single-component (S), double (D), triple (T), head-tail (HT), narrow-angle tail (NAT), wide-angle tail (WAT) or complex (C). Incipient doubles in which the components are not fully resolved are labelled as Di. \newline
\newline
In Table 4, the columns are:
\newline
\newline
(1) Name of the source in the NVSS catalogue in IAU J2000 format.\newline
(2) 1384 and 2368 MHz integrated flux densities of the radio components in each ATCA image: central component (C), north lobe (N), and south lobe (S). The central component is the only entry for sources with a single-component morphology at either 1384 or 2368 MHz, or, for triple sources, the core. Often there are more components present at 2368 MHz than at 1384 MHz because of the increase in angular resolution. We give the combined flux density of the lobes in incipient doubles at either 1384 or 2368 MHz (enclosed in parentheses). For those sources where the lobes are close to east--west, we also label the eastern lobe (E) and the western lobe (W).
\newline
(3) The two-point spectral index $\alpha^{2368}_{1384}$ for the central component, north lobe and south lobe, assuming that each component is not blended at 1384 MHz.\newline
(4) 1384 and 2368 MHz fractional linear polarization for the central, north lobe and south lobe components, measured at the position of peak intensity. A $3\sigma$ upper limit is given for source components with undetected polarization.\newline
(5) The observed-frame Faraday rotation measure (RM), not corrected for the Galactic Faraday screen, for the central, north lobe and south lobe components.\newline
(6) The Galactic RM determined from the all-sky RM map of \citet*{hollitt04}.\newline
\newline
Note that for the head-tail, narrow-angle tail and wide-angle tail sources, the properties of the core are given under the C label, and the properties of the tail(s) are given under the N and S labels (assuming that these components are resolved).

\subsection{Morphologies}\label{408_paper1_morphologies}

\begin{figure*}
\begin{minipage}{175mm}
\epsfig{file=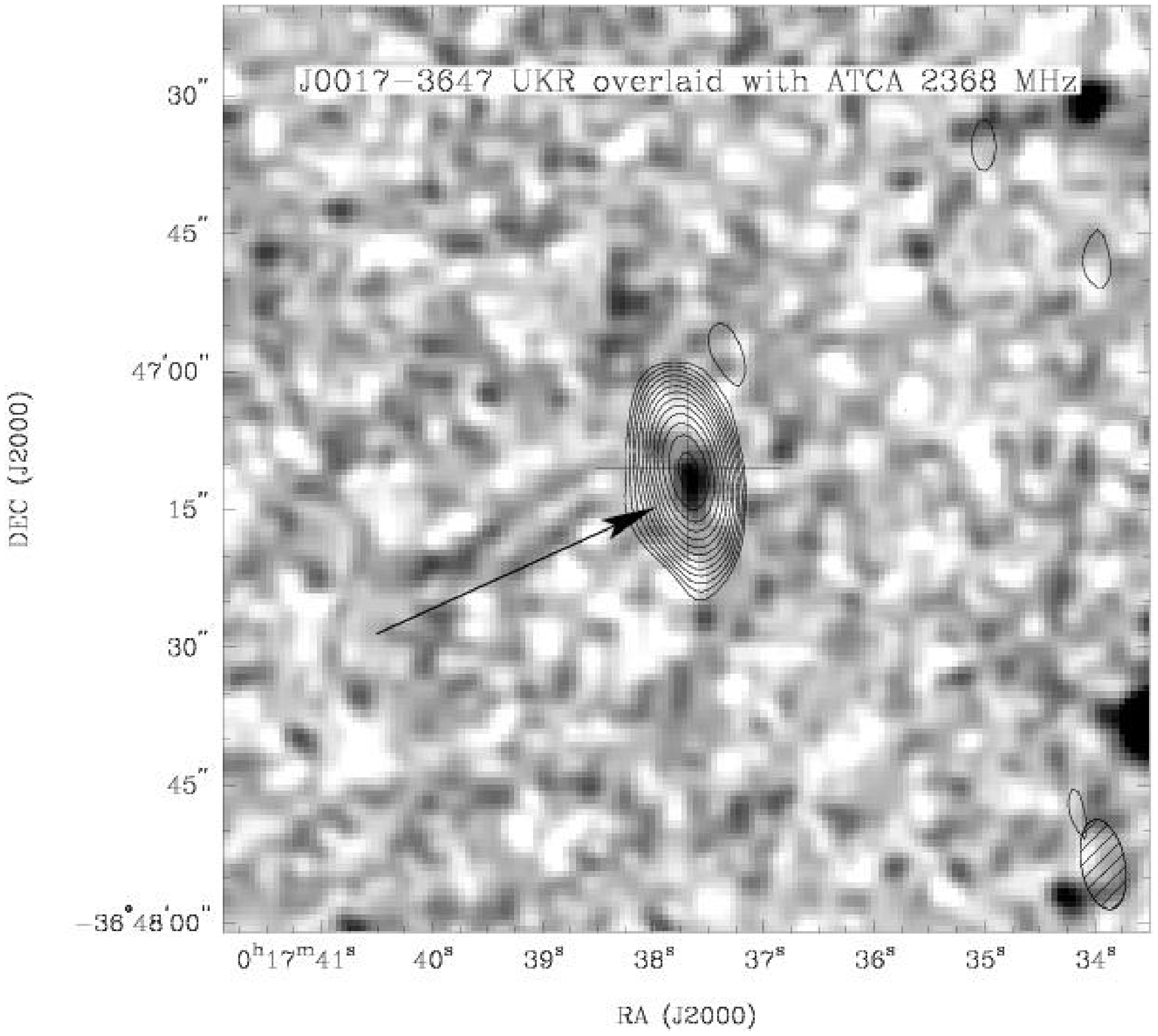,height=5.0cm,width=5.5cm}~\epsfig{file=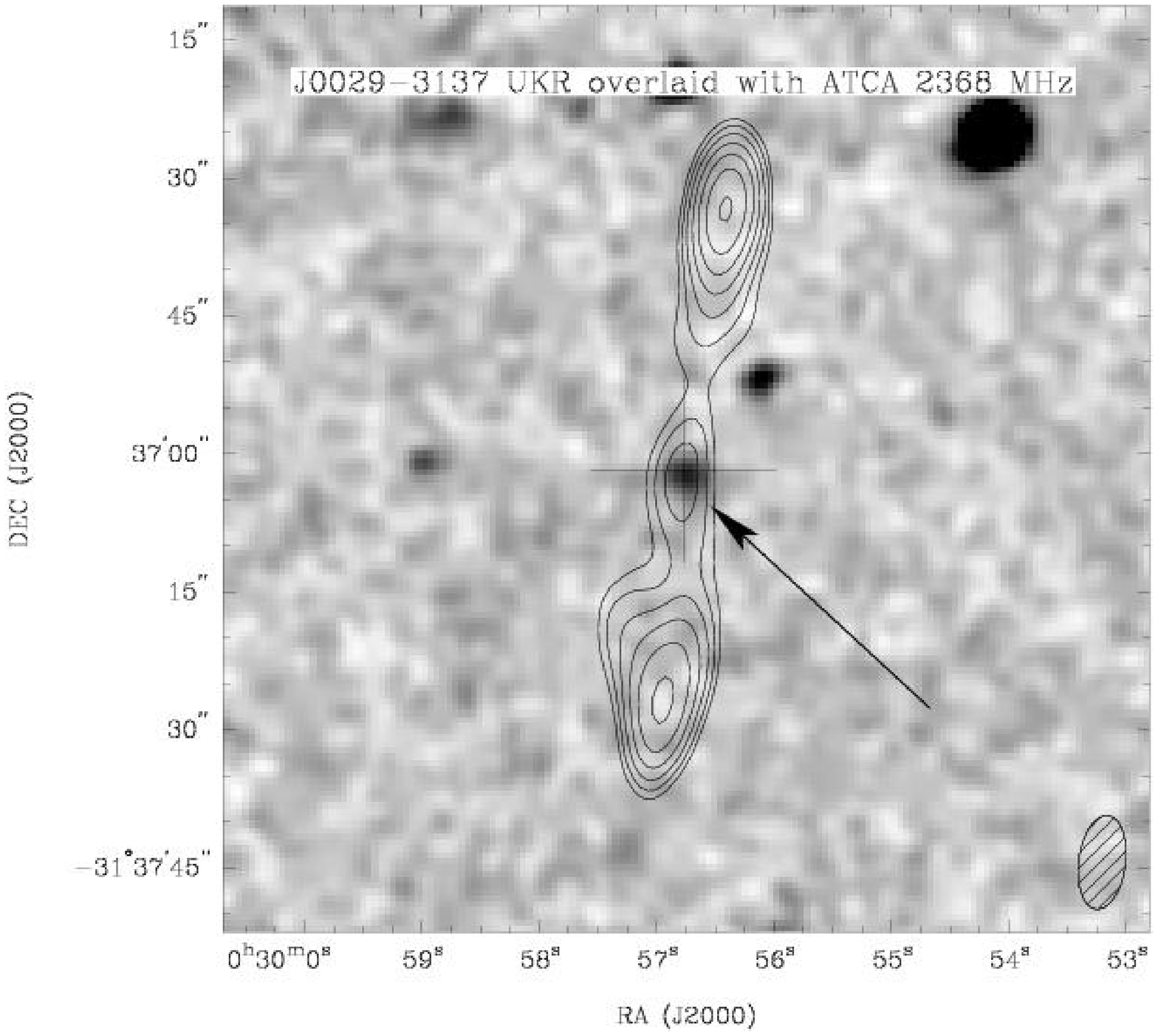,height=5.0cm,width=5.5cm}~\epsfig{file=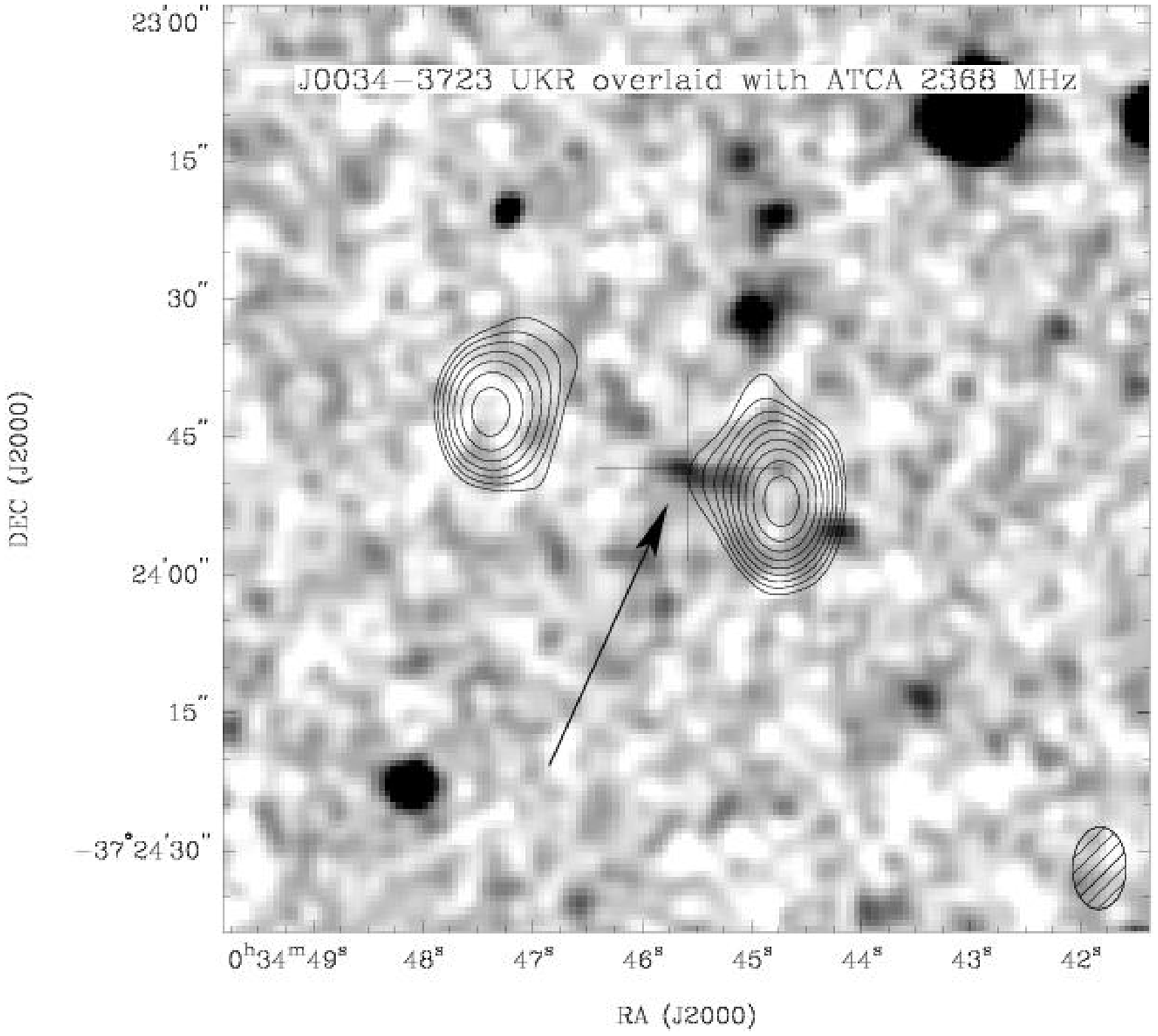,height=5.0cm,width=5.5cm}
\newline
\newline
\epsfig{file=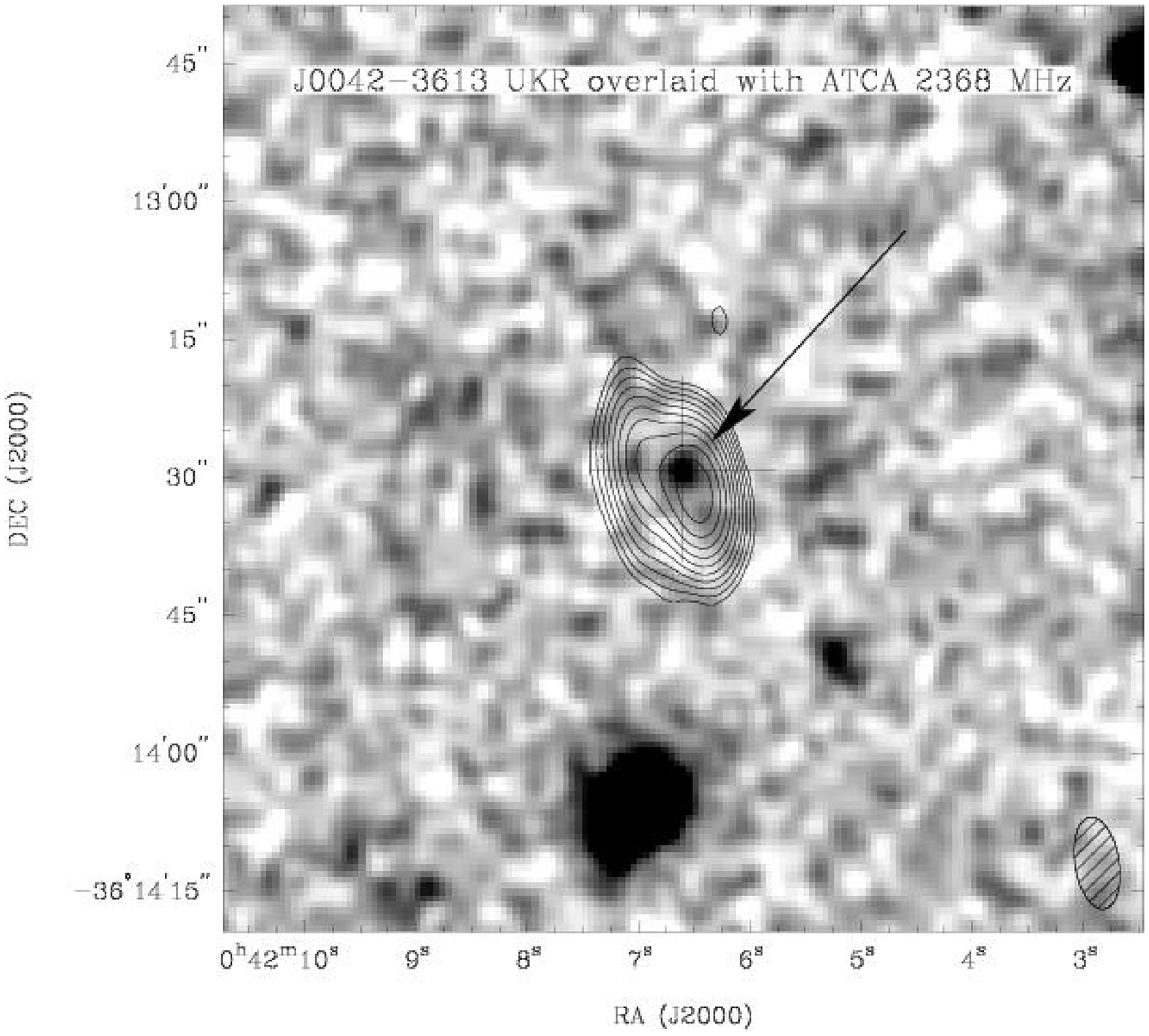,height=5.0cm,width=5.5cm}~\epsfig{file=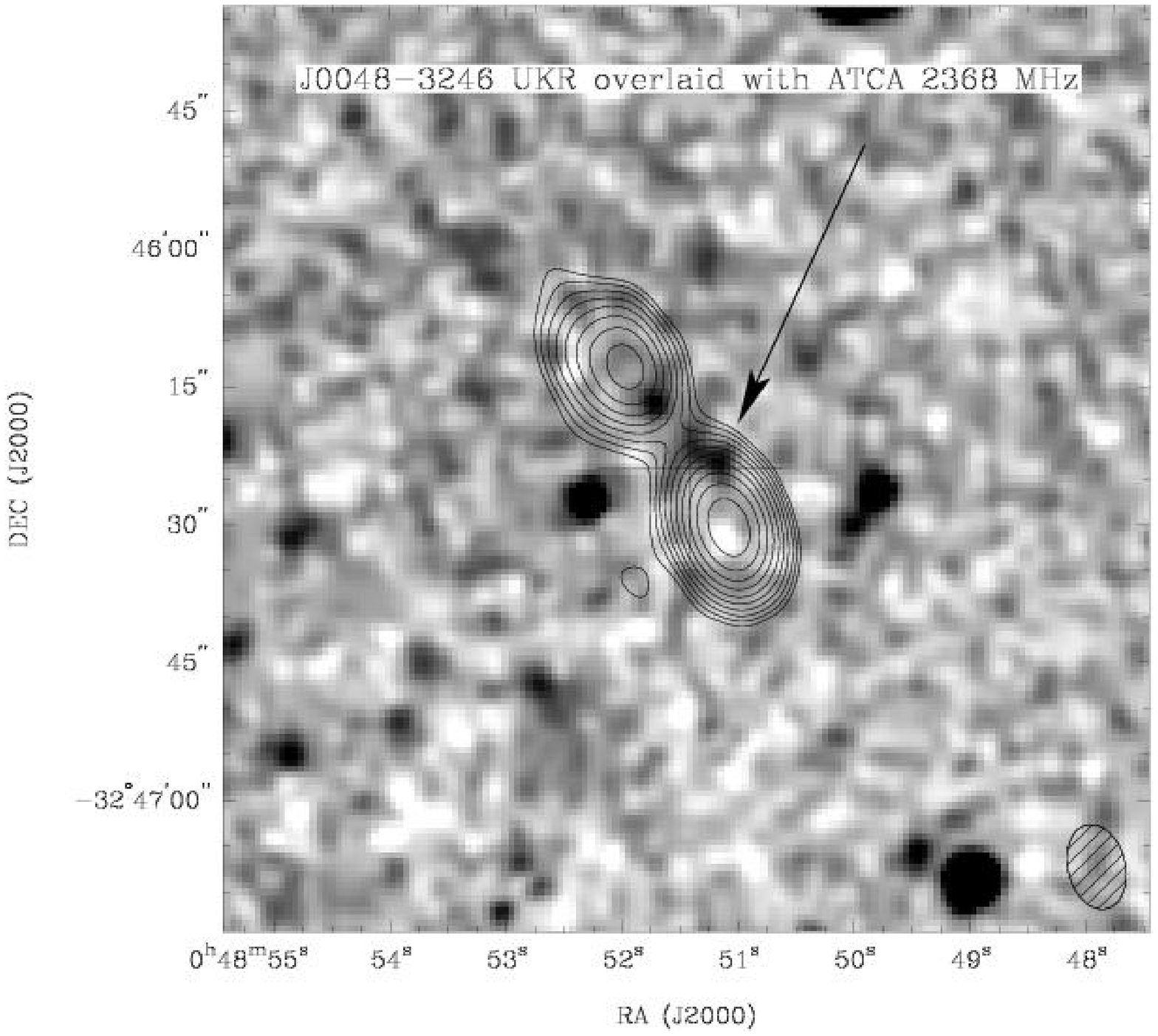,height=5.0cm,width=5.5cm}~\epsfig{file=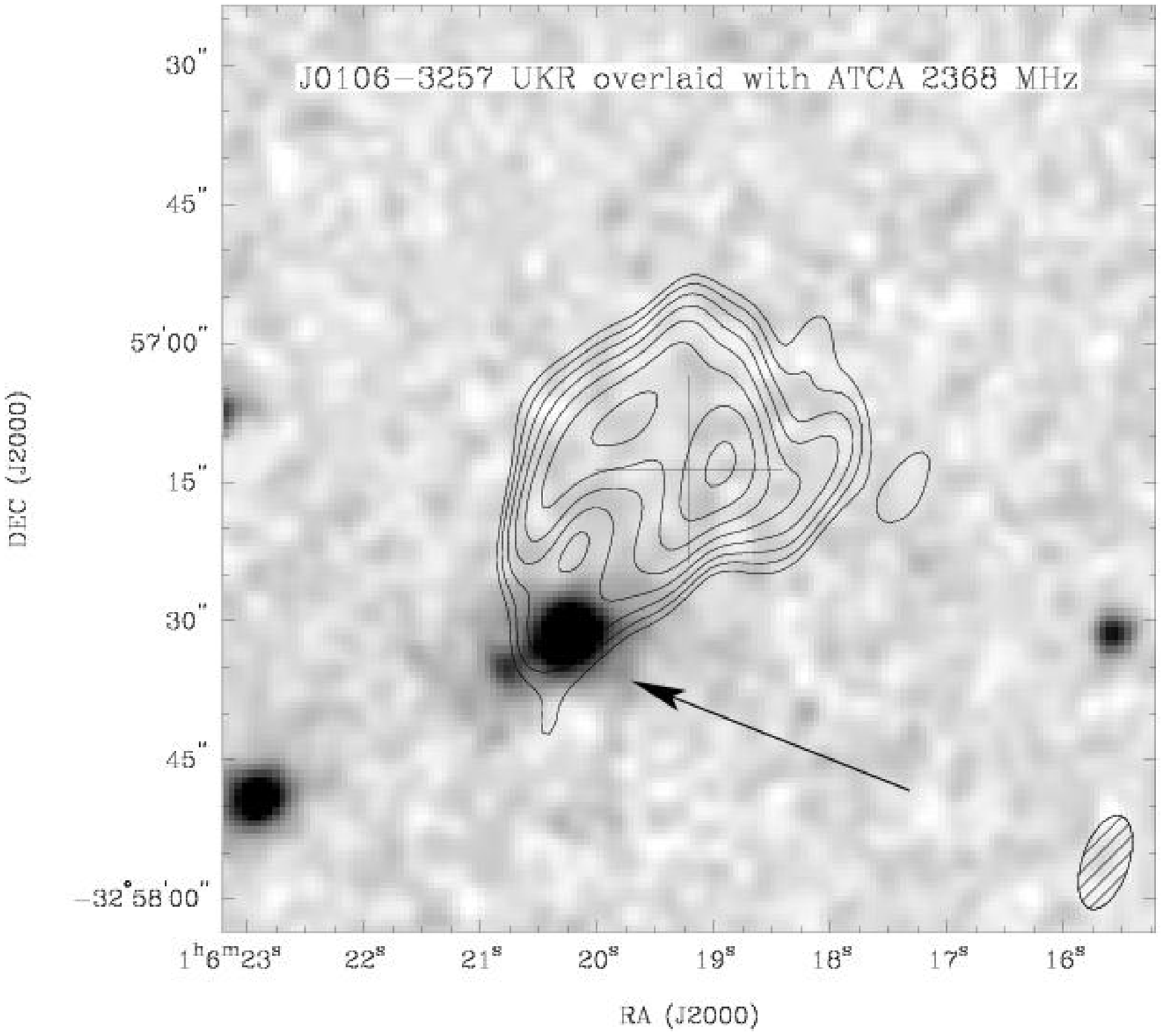,height=5.0cm,width=5.5cm}
\newline
\newline
\epsfig{file=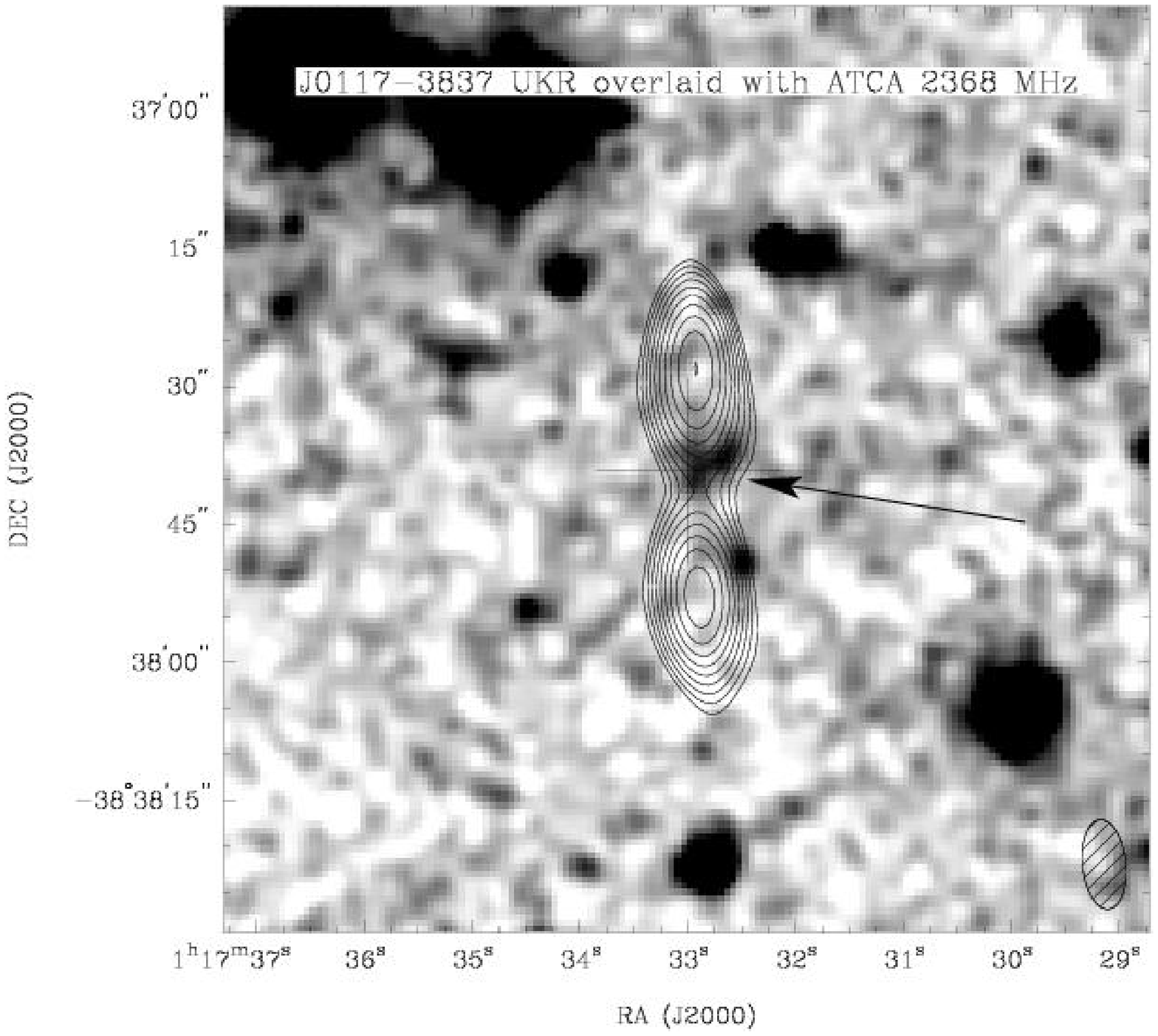,height=5.0cm,width=5.5cm}~\epsfig{file=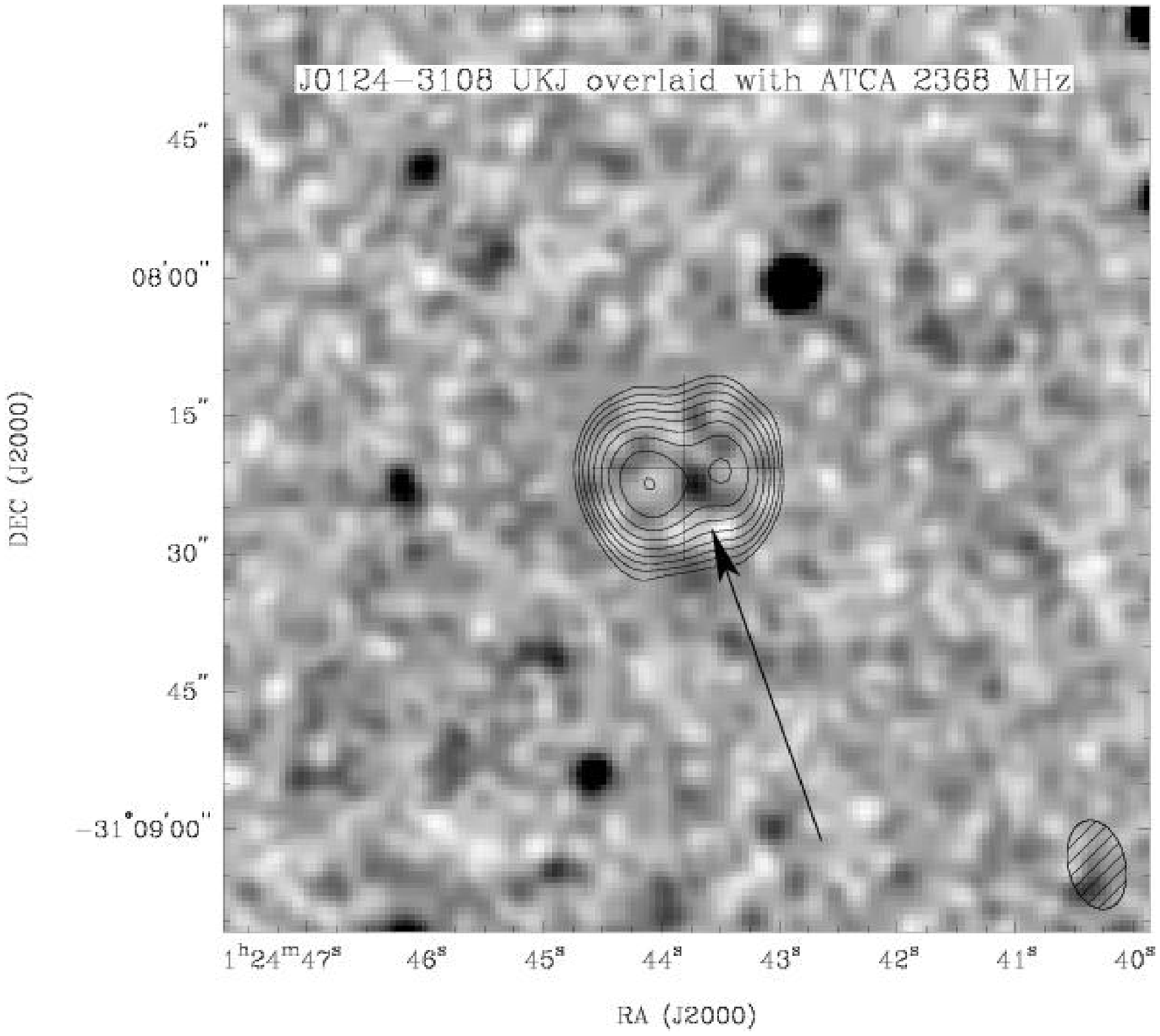,height=5.0cm,width=5.5cm}~\epsfig{file=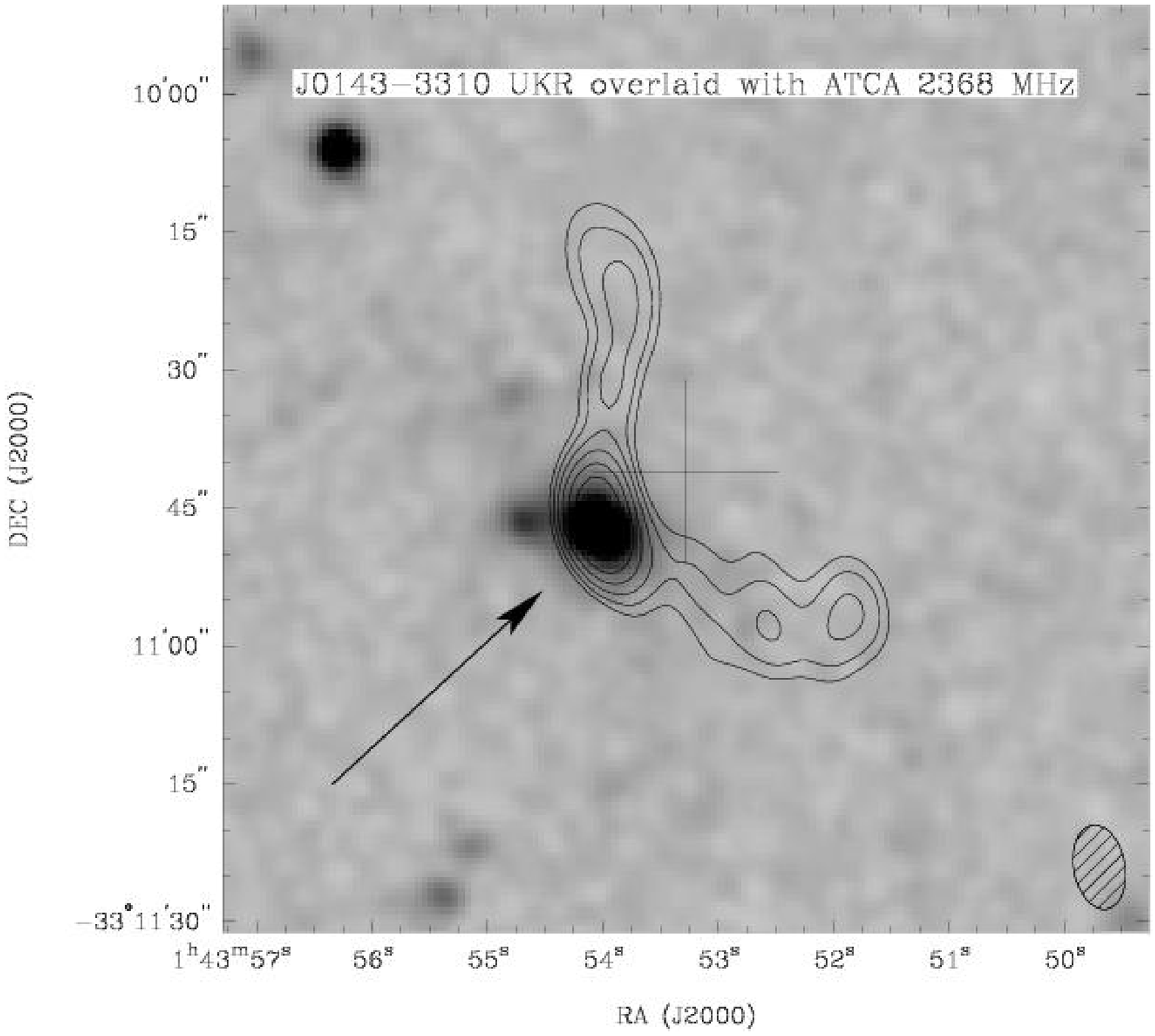,height=5.0cm,width=5.5cm}
\newline
\newline
\epsfig{file=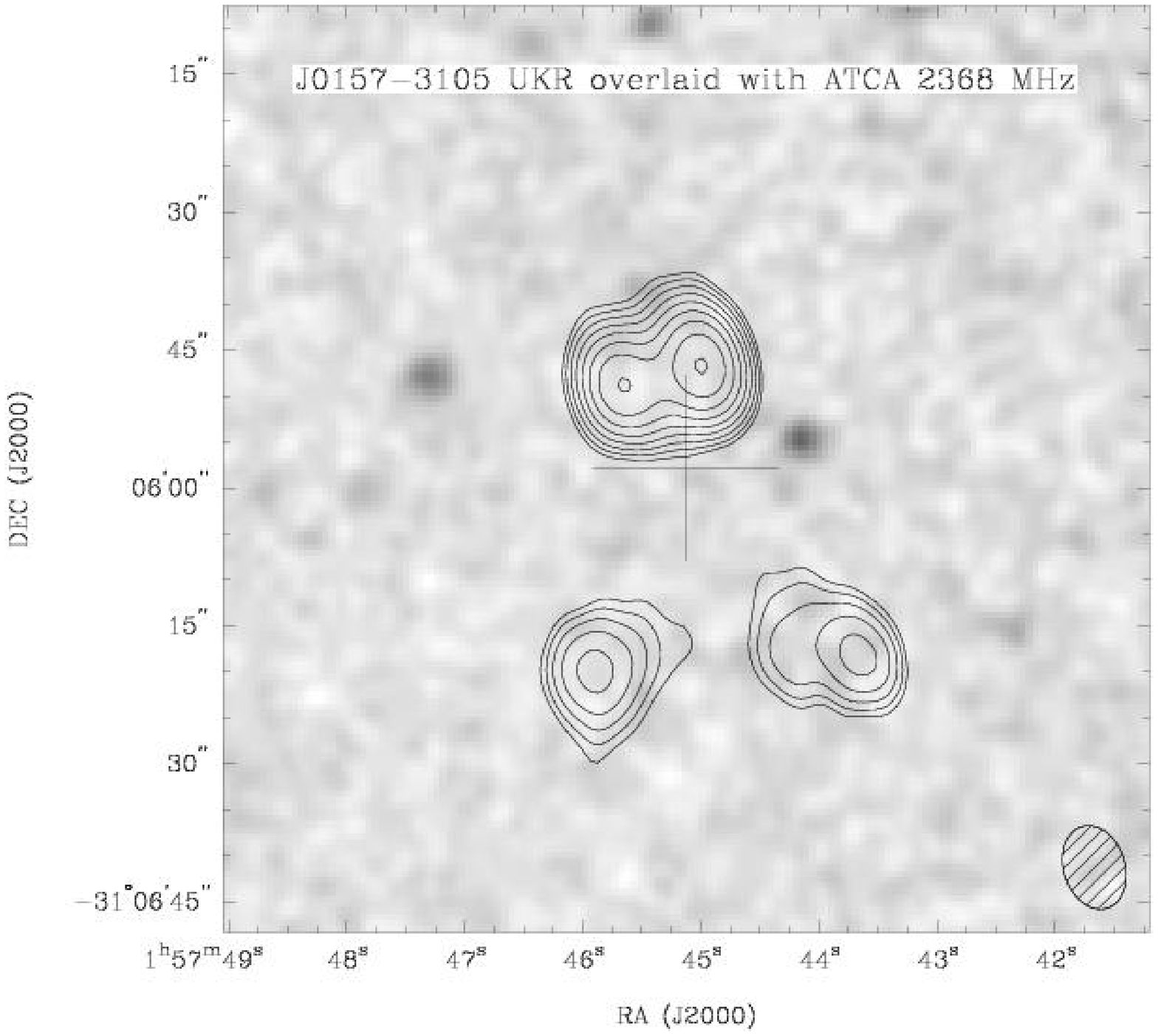,height=5.0cm,width=5.5cm}~\epsfig{file=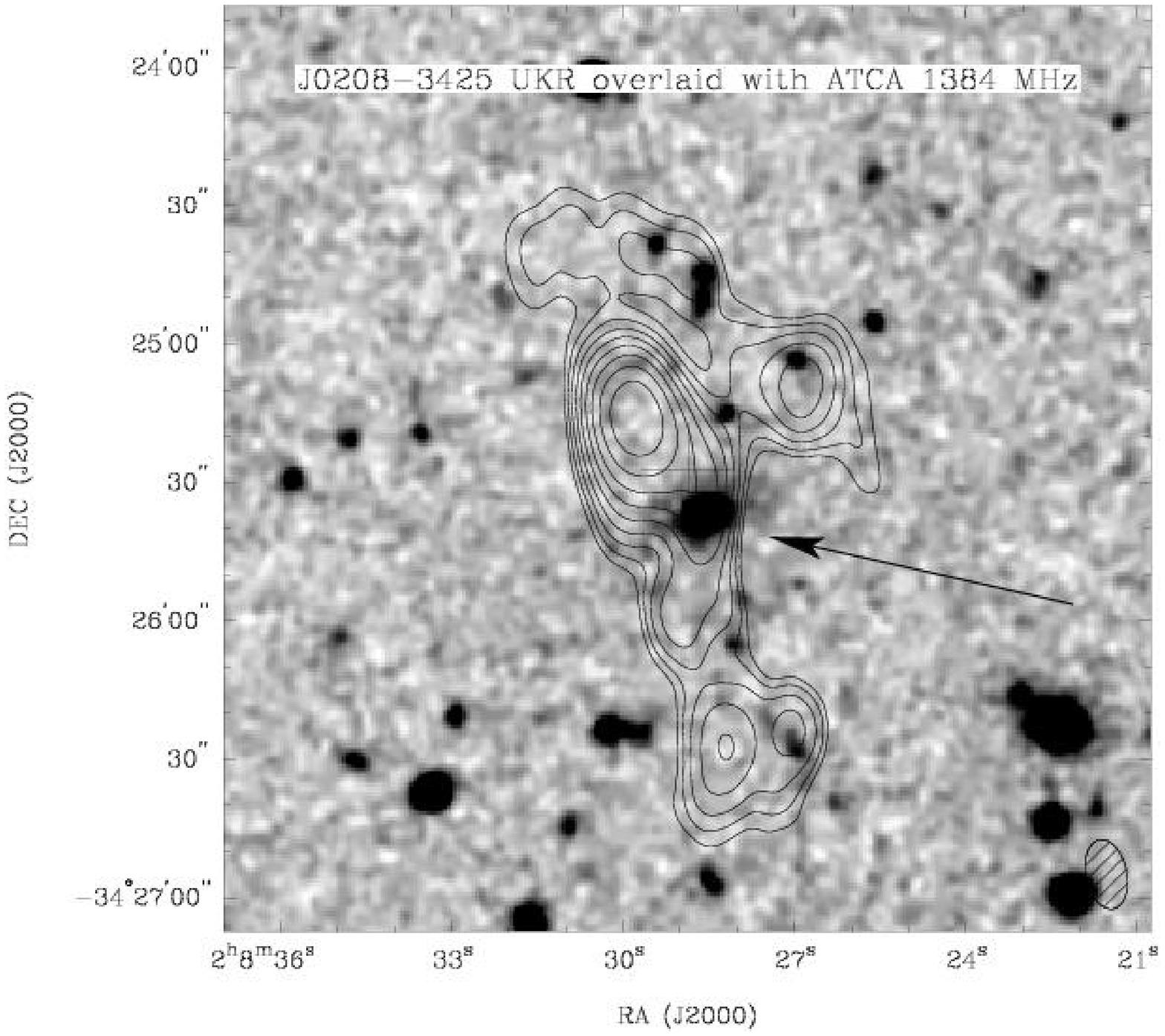,height=5.0cm,width=5.5cm}~\epsfig{file=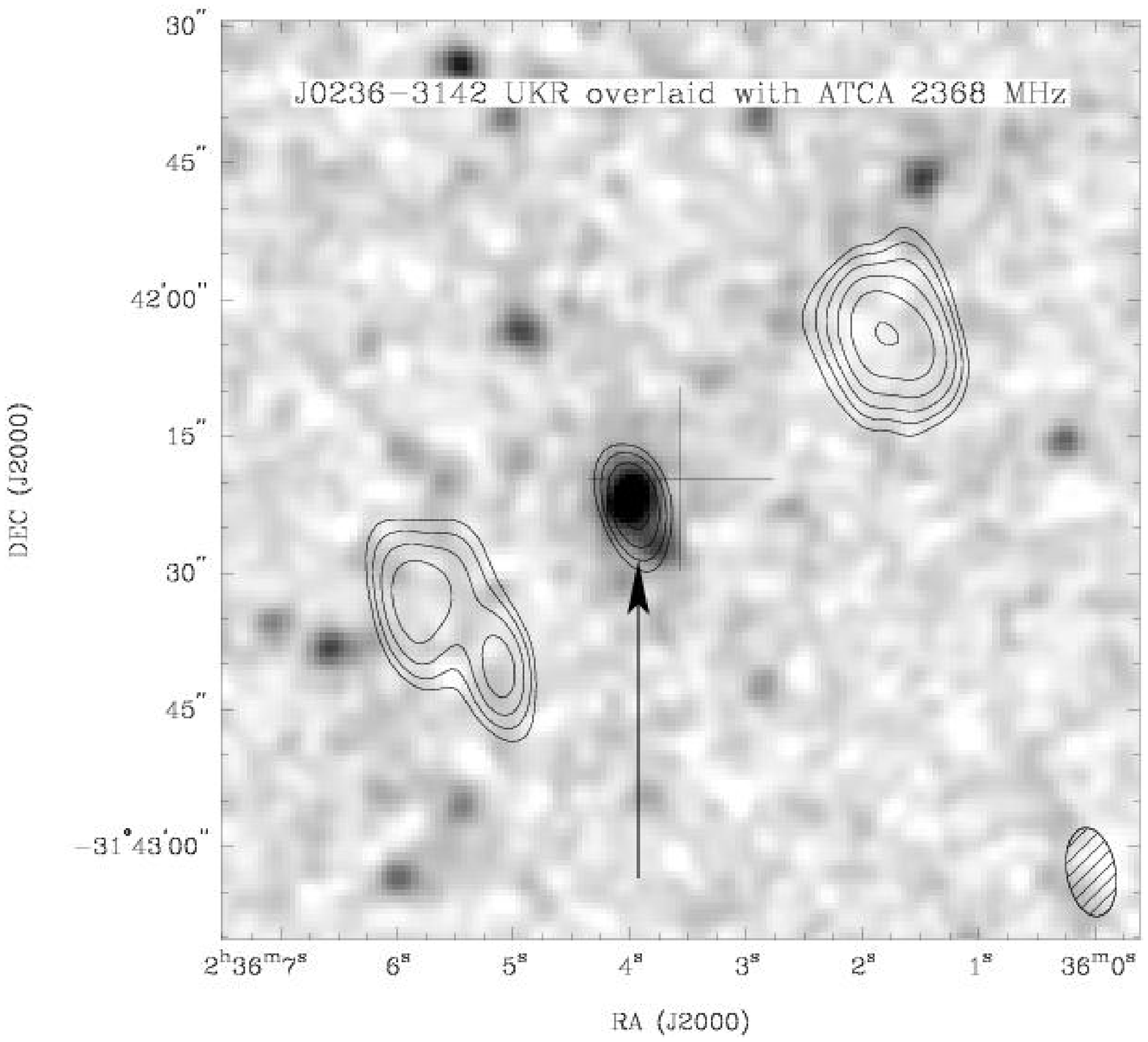,height=5.0cm,width=5.5cm}
\caption{Overlay plots for sources in the MRCR--SUMSS sample with an optical identification, an unusual radio morphology, and/or information available from the literature. ATCA 2368 MHz contours (1384 MHz contours for NVSS J020827$-$342635 $+$ NVSS J020828$-$342520) have been overlaid on smoothed SuperCOSMOS UKJ or UKR images. In each plot, the lowest contour is 1.2 mJy beam$^{-1}$, and each contour represents an increase of $\sqrt{2}$ in surface brightness. The exceptions are NVSS J003445$-$372348, where the lowest contour is 2.4 mJy beam$^{-1}$, and NVSS J215547$-$344614, where we have added 5.0 and 6.0 mJy beam$^{-1}$ contours to make the radio morphology clearer. Crosshairs mark the centroid of each radio source in NVSS, while arrows indicate the host galaxy optical identification. The ATCA synthesized beam is shown in the bottom right-hand corner of each panel.}\label{408_paper1_fig:morph_examples}
\end{minipage}
\end{figure*}

\begin{figure*}
\begin{minipage}{175mm}
\epsfig{file=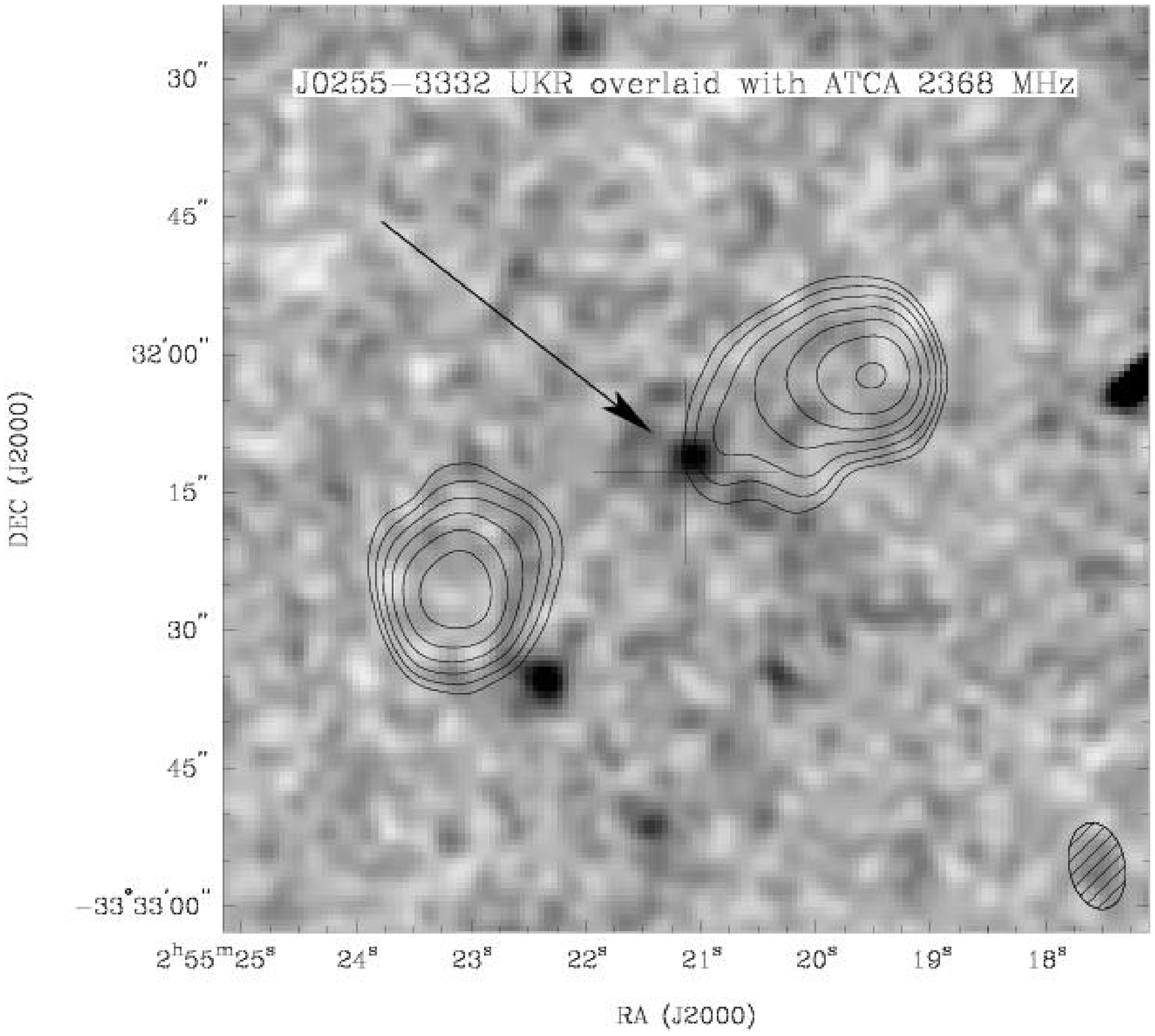,height=5.0cm,width=5.5cm}~\epsfig{file=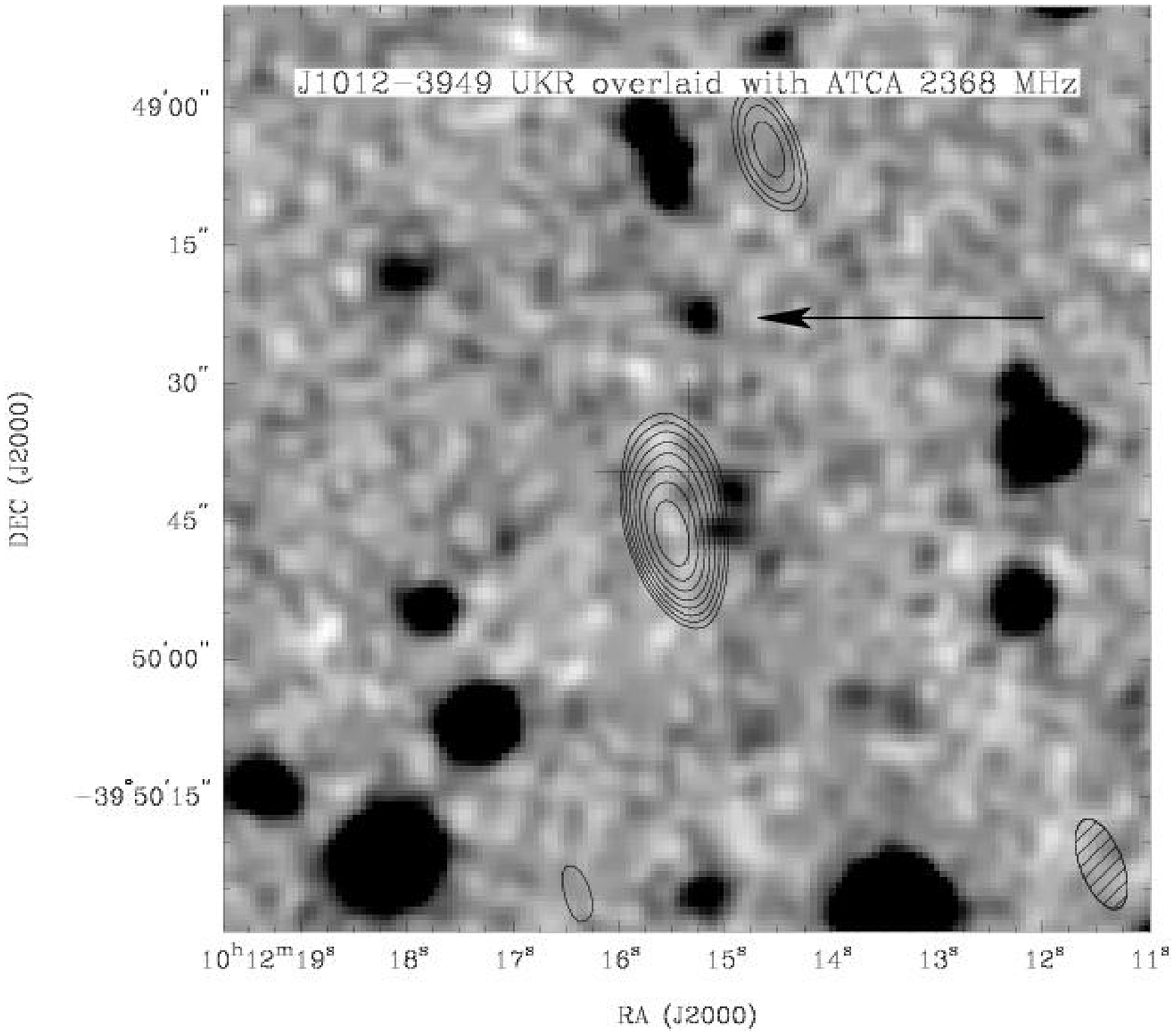,height=5.0cm,width=5.5cm}~\epsfig{file=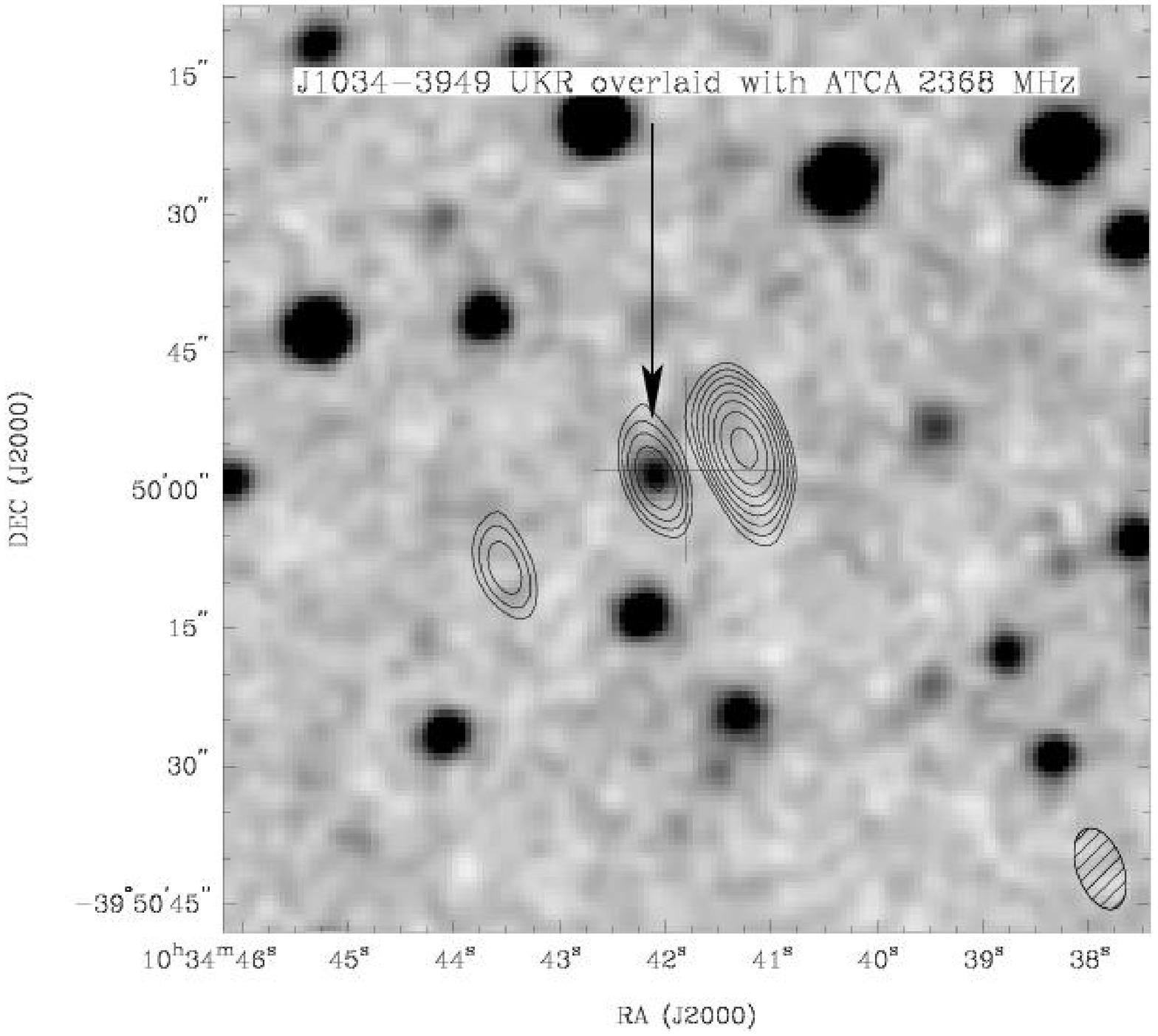,height=5.0cm,width=5.5cm}
\newline
\newline
\newline
\epsfig{file=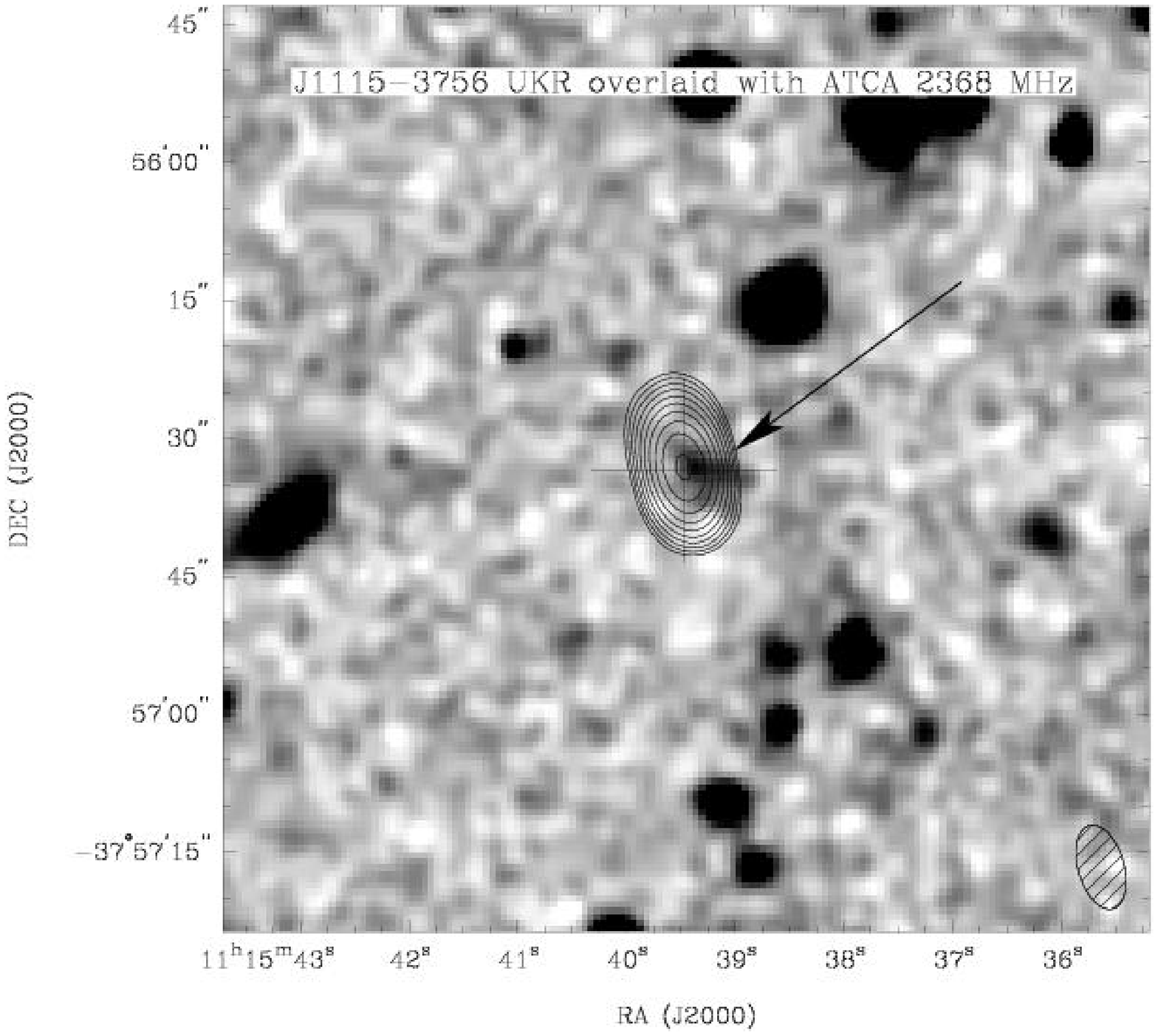,height=5.0cm,width=5.5cm}~\epsfig{file=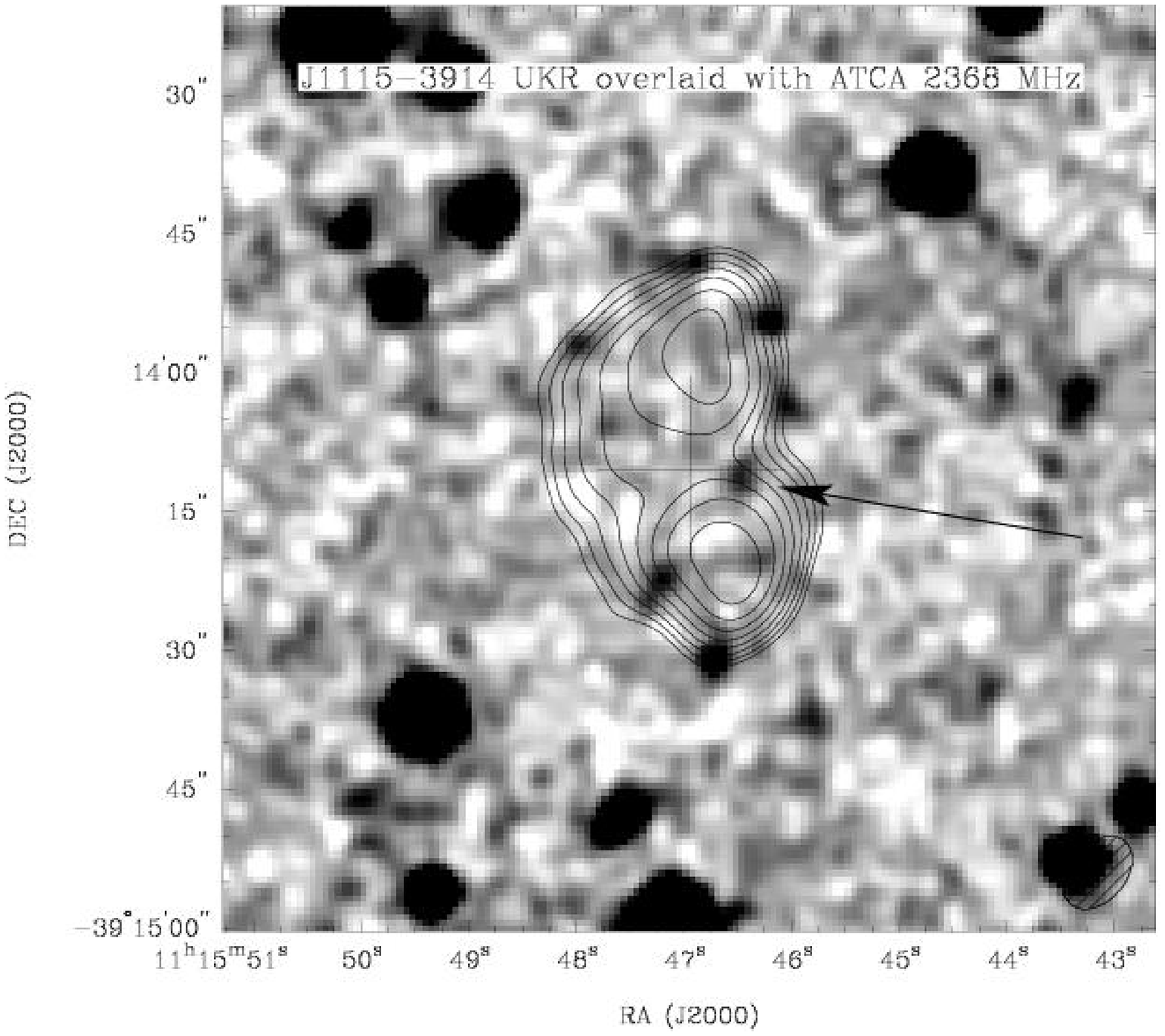,height=5.0cm,width=5.5cm}~\epsfig{file=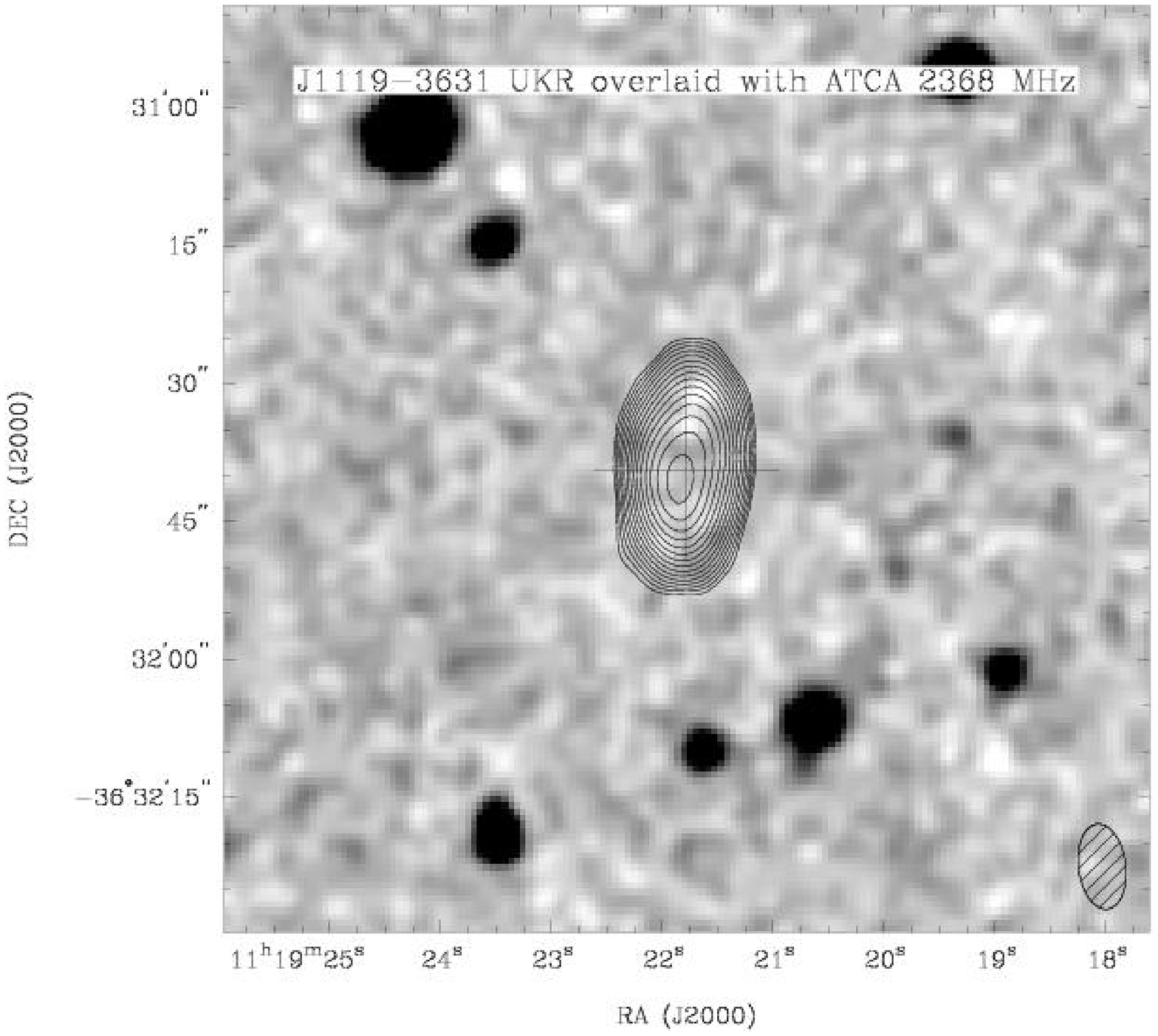,height=5.0cm,width=5.5cm}
\newline
\newline
\epsfig{file=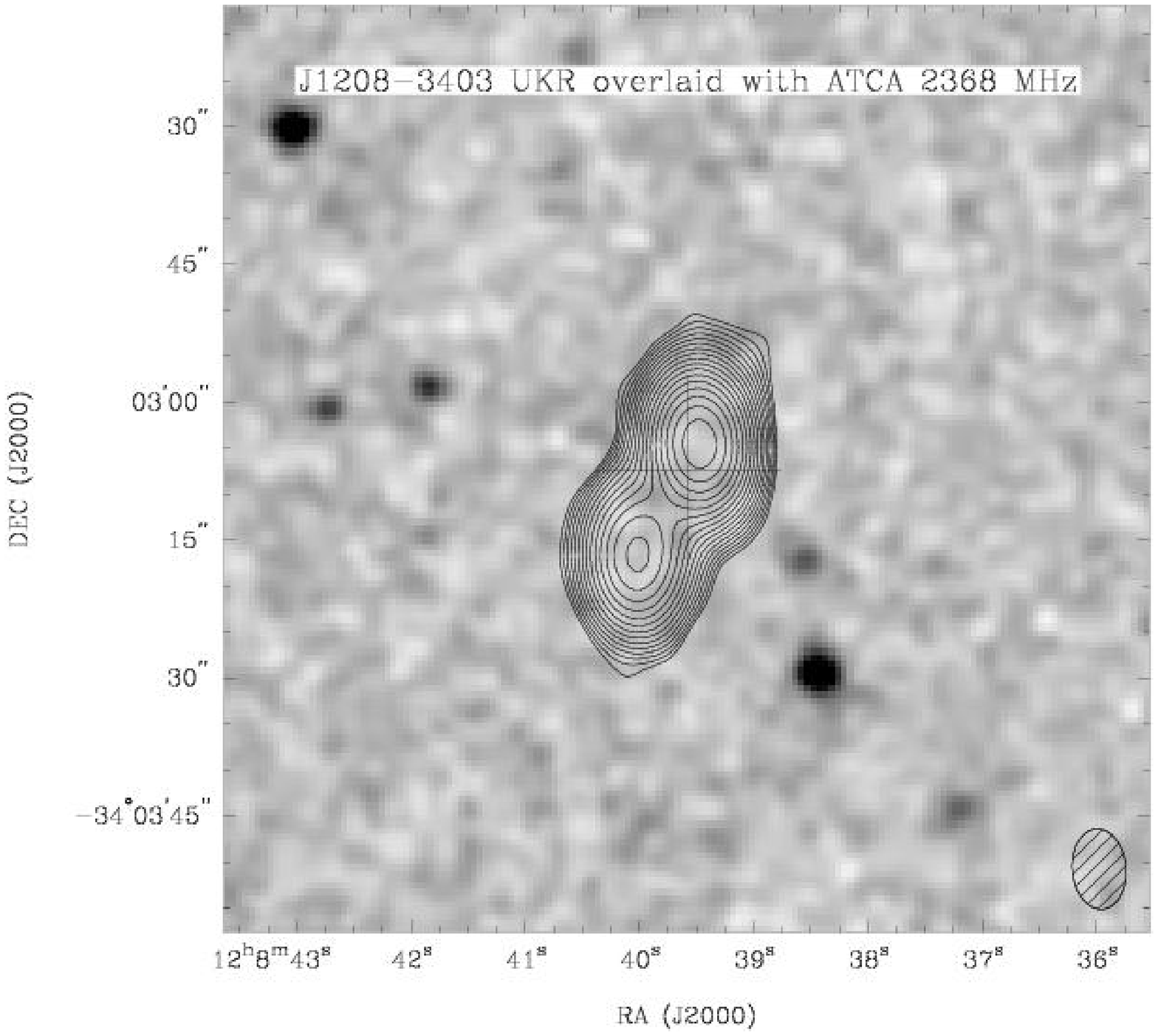,height=5.0cm,width=5.5cm}~\epsfig{file=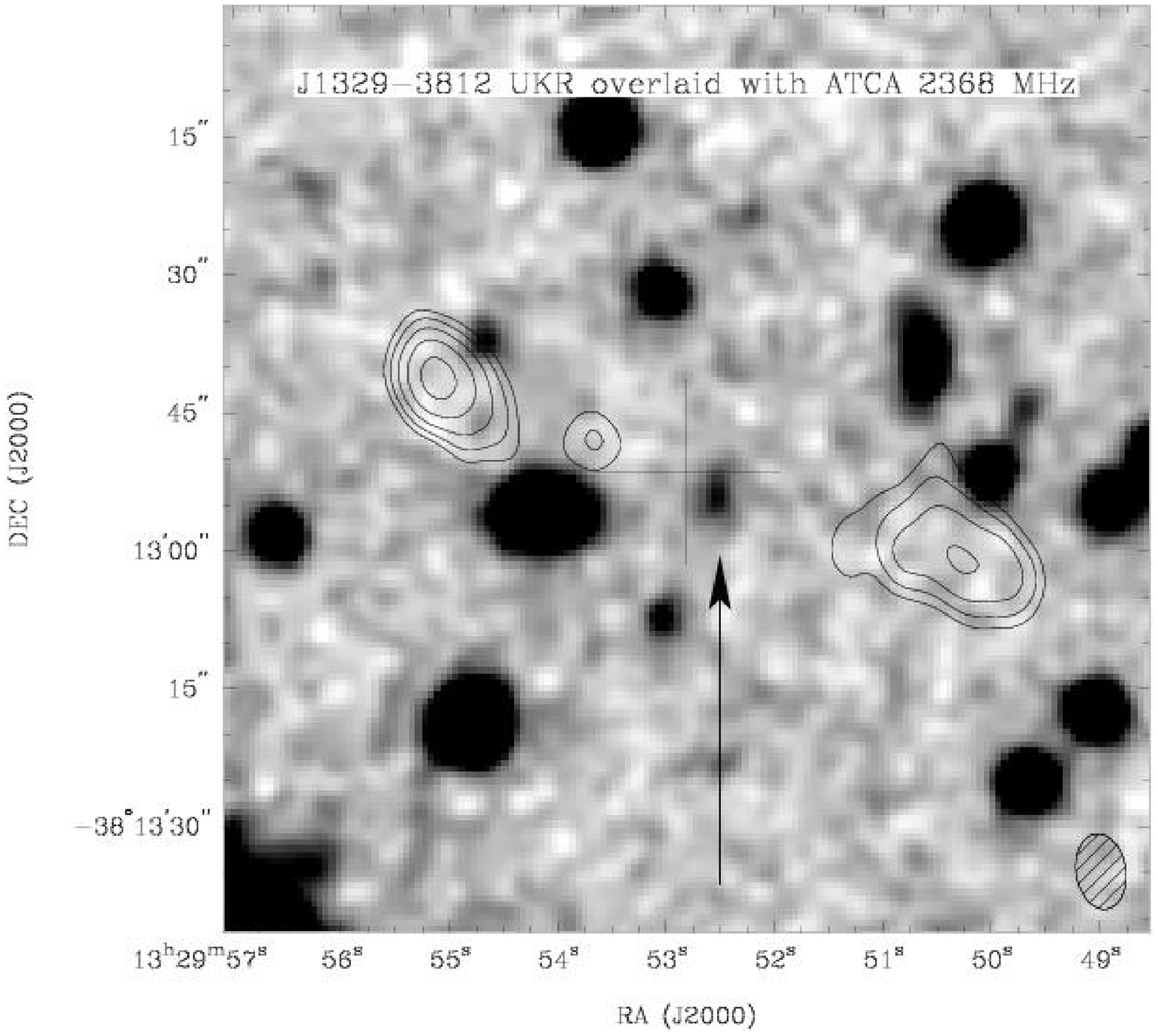,height=5.0cm,width=5.5cm}~\epsfig{file=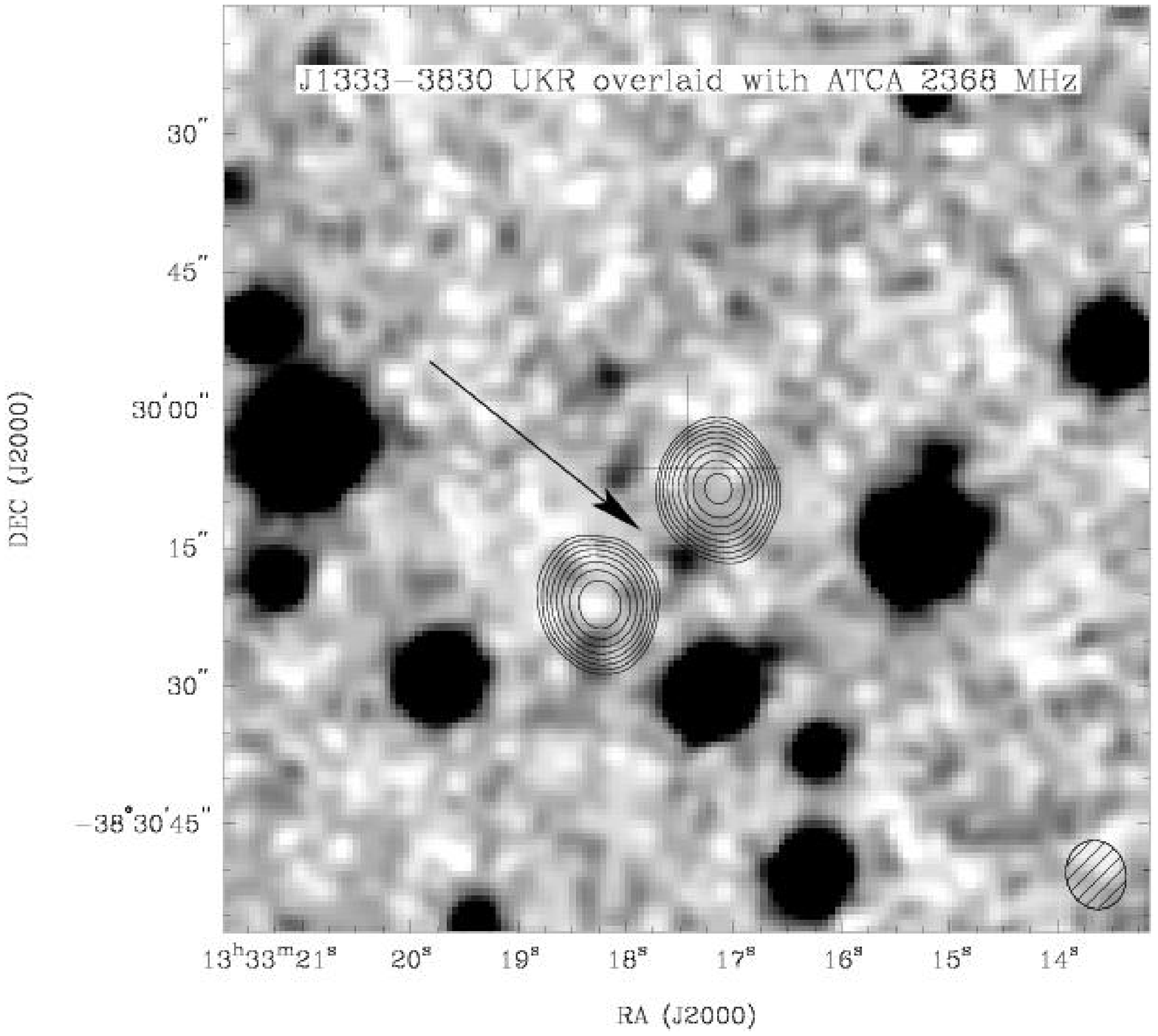,height=5.0cm,width=5.5cm}
\newline
\newline
\epsfig{file=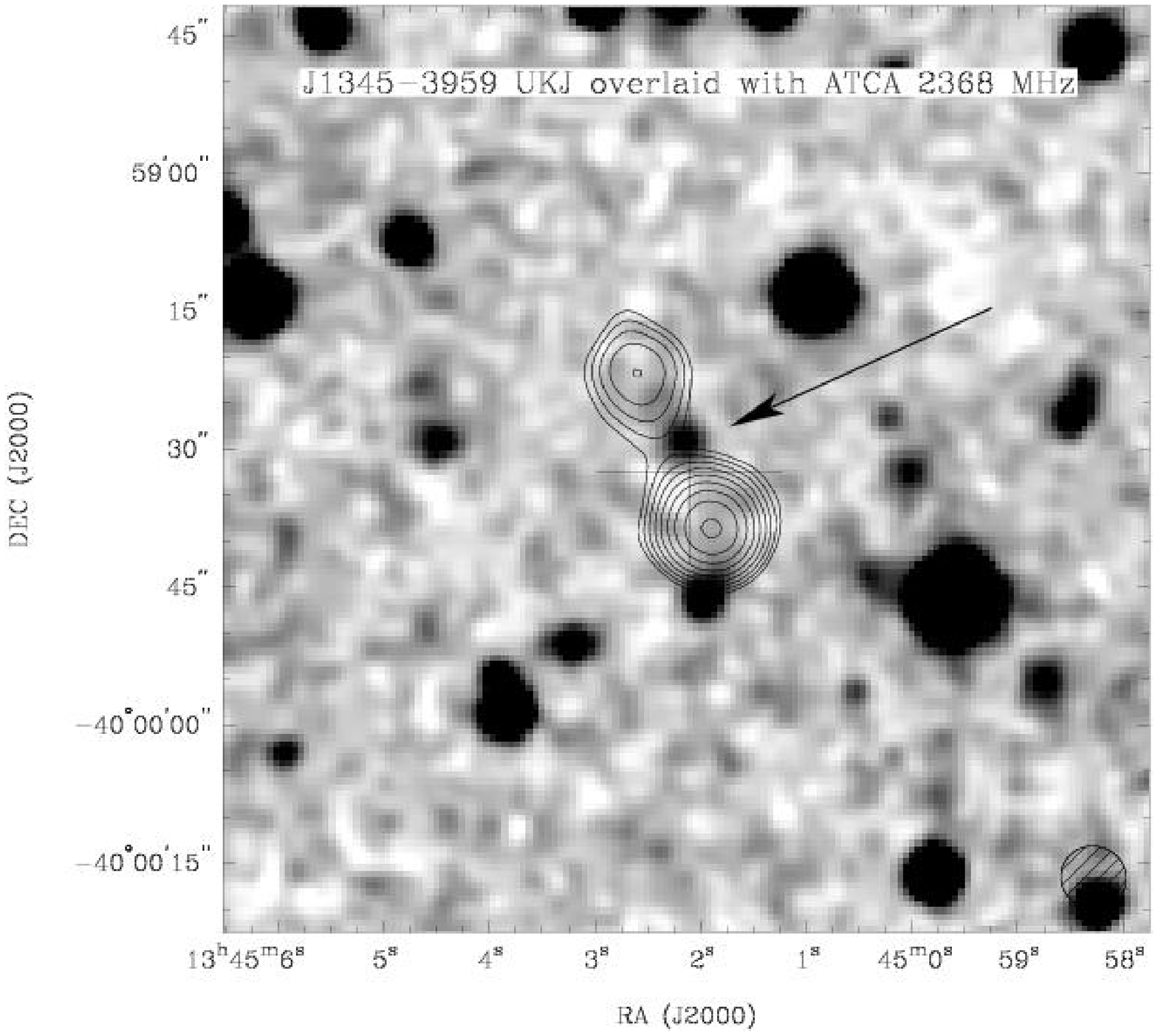,height=5.0cm,width=5.5cm}~\epsfig{file=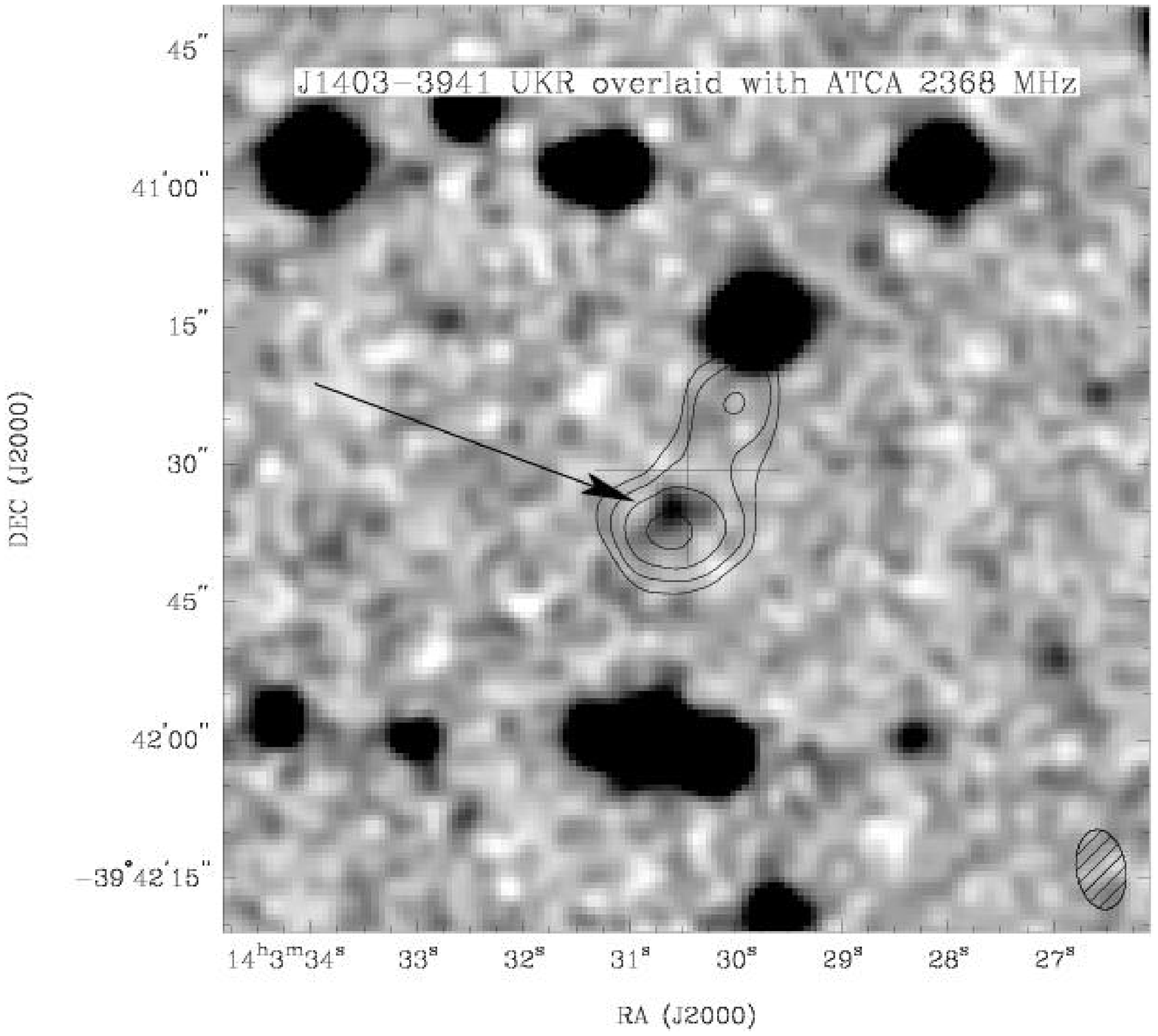,height=5.0cm,width=5.5cm}~\epsfig{file=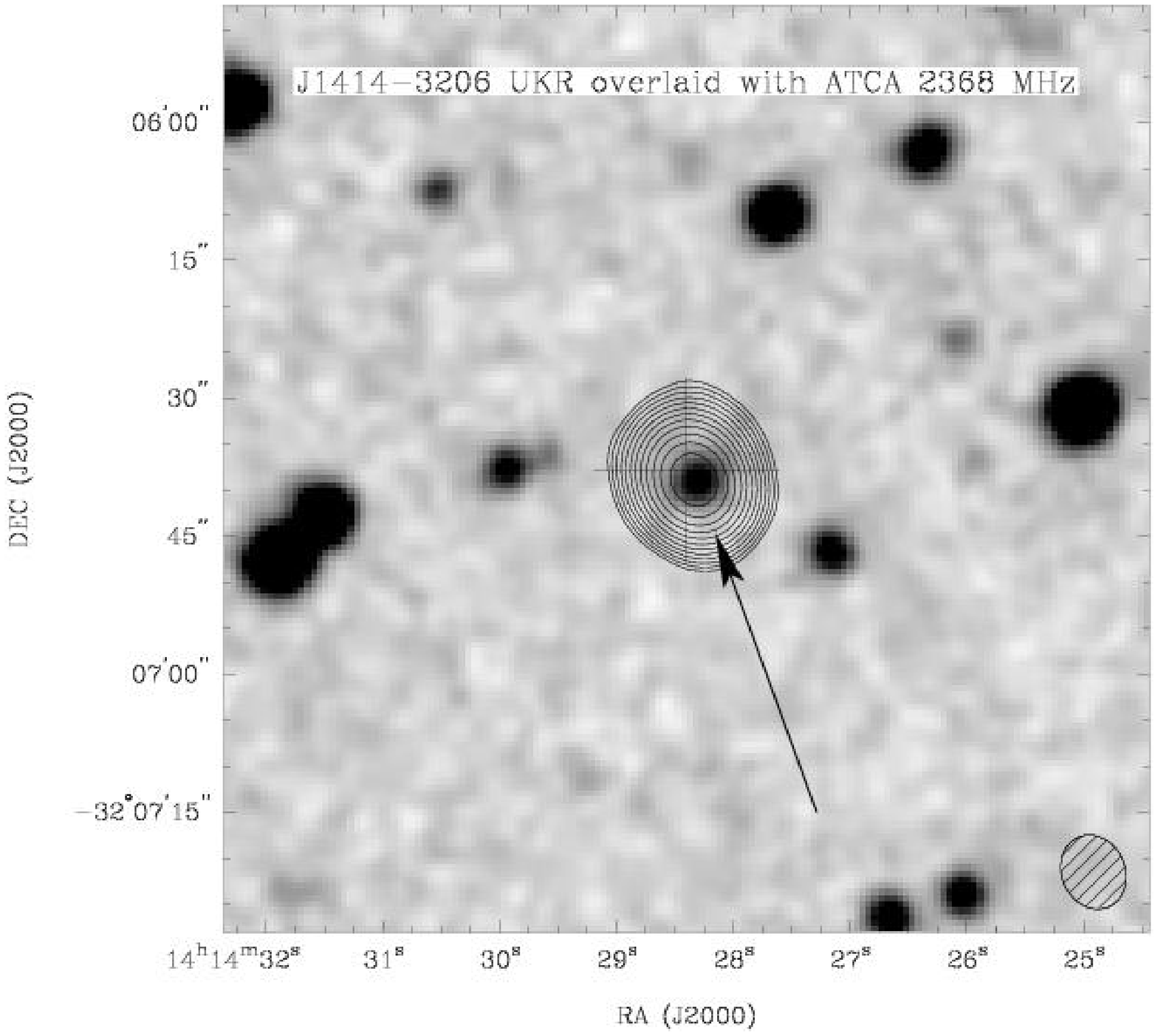,height=5.0cm,width=5.5cm}
\contcaption{}
\end{minipage}
\end{figure*}

\begin{figure*}
\begin{minipage}{175mm}
\epsfig{file=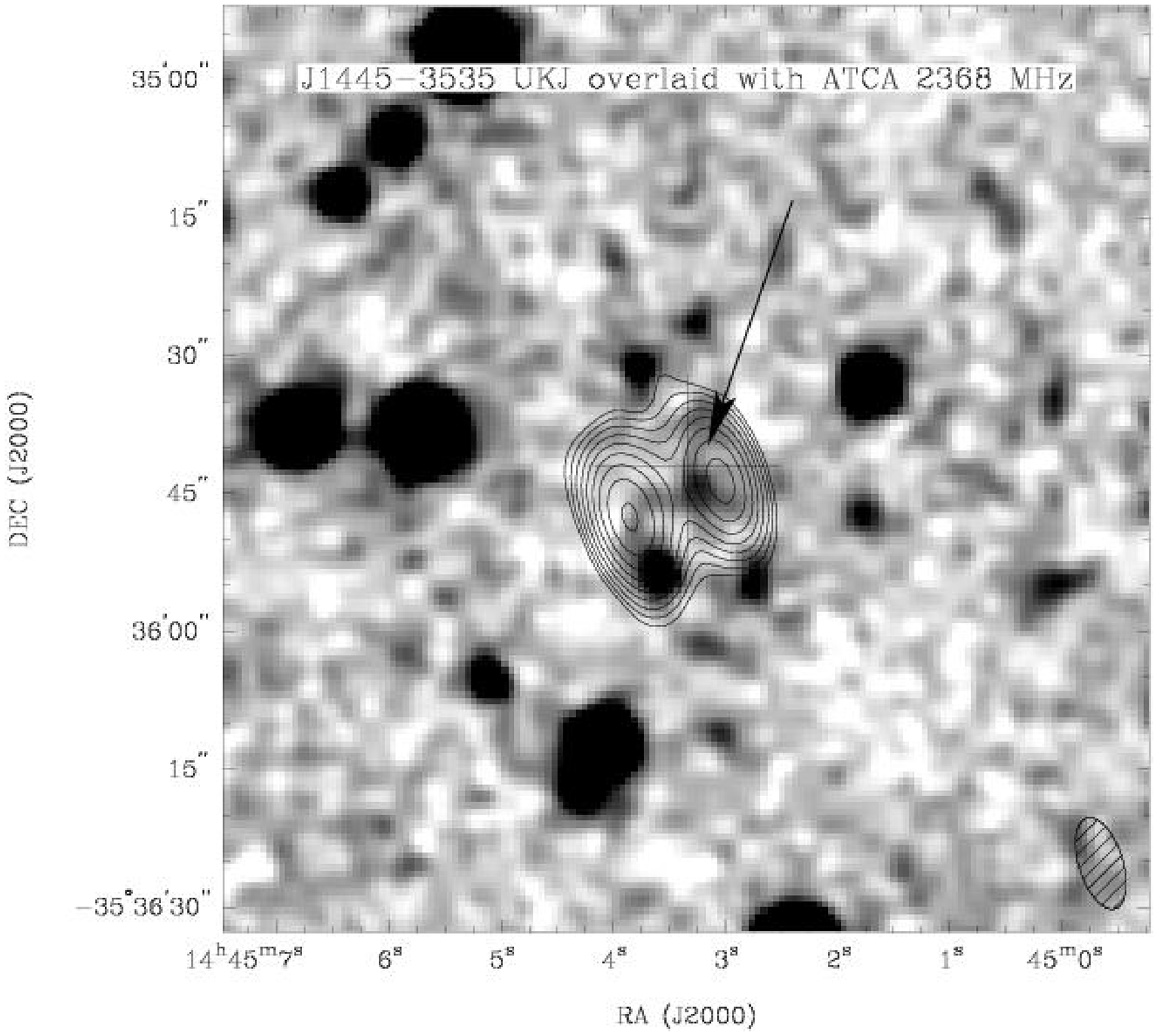,height=5.0cm,width=5.5cm}~\epsfig{file=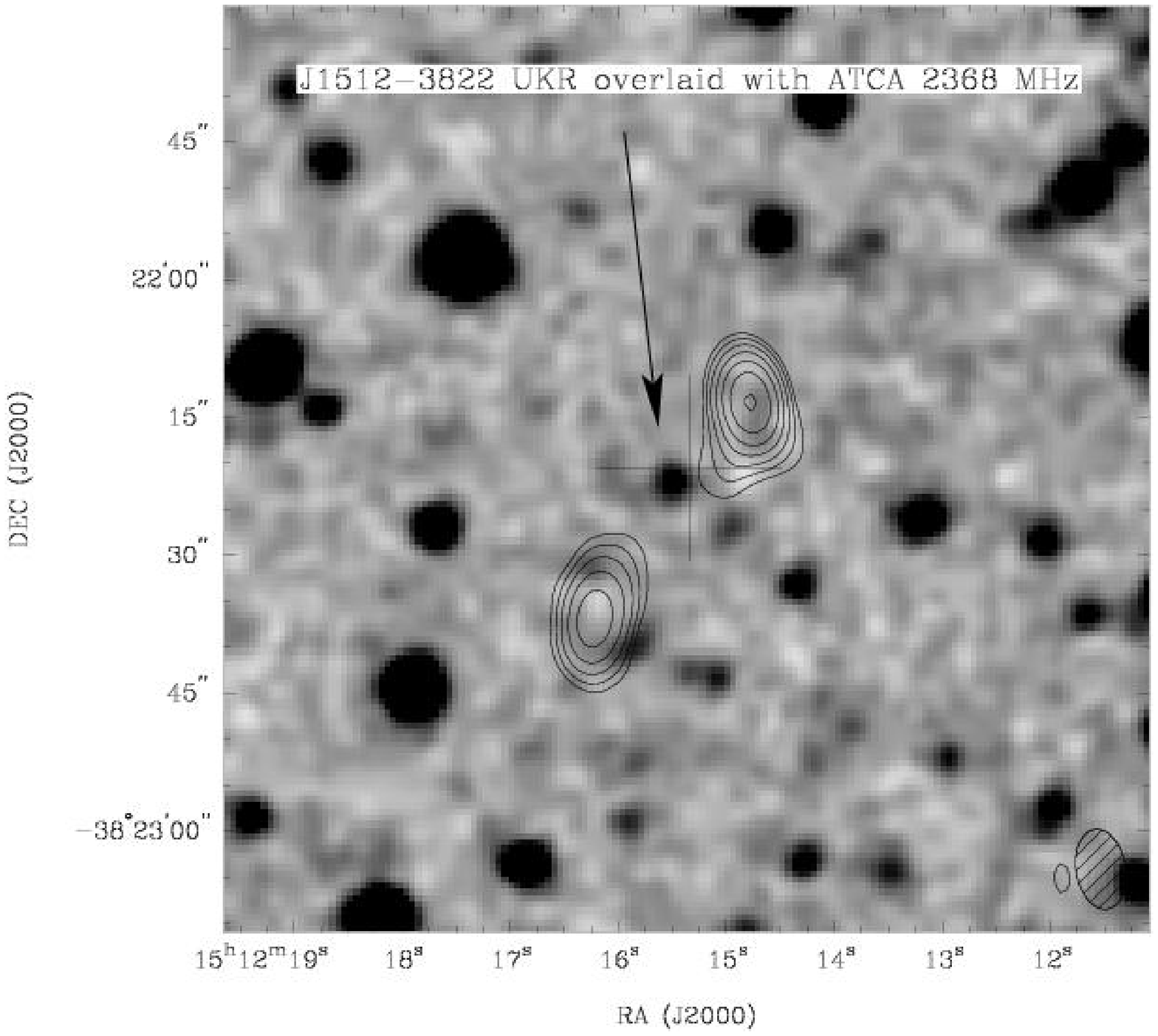,height=5.0cm,width=5.5cm}~\epsfig{file=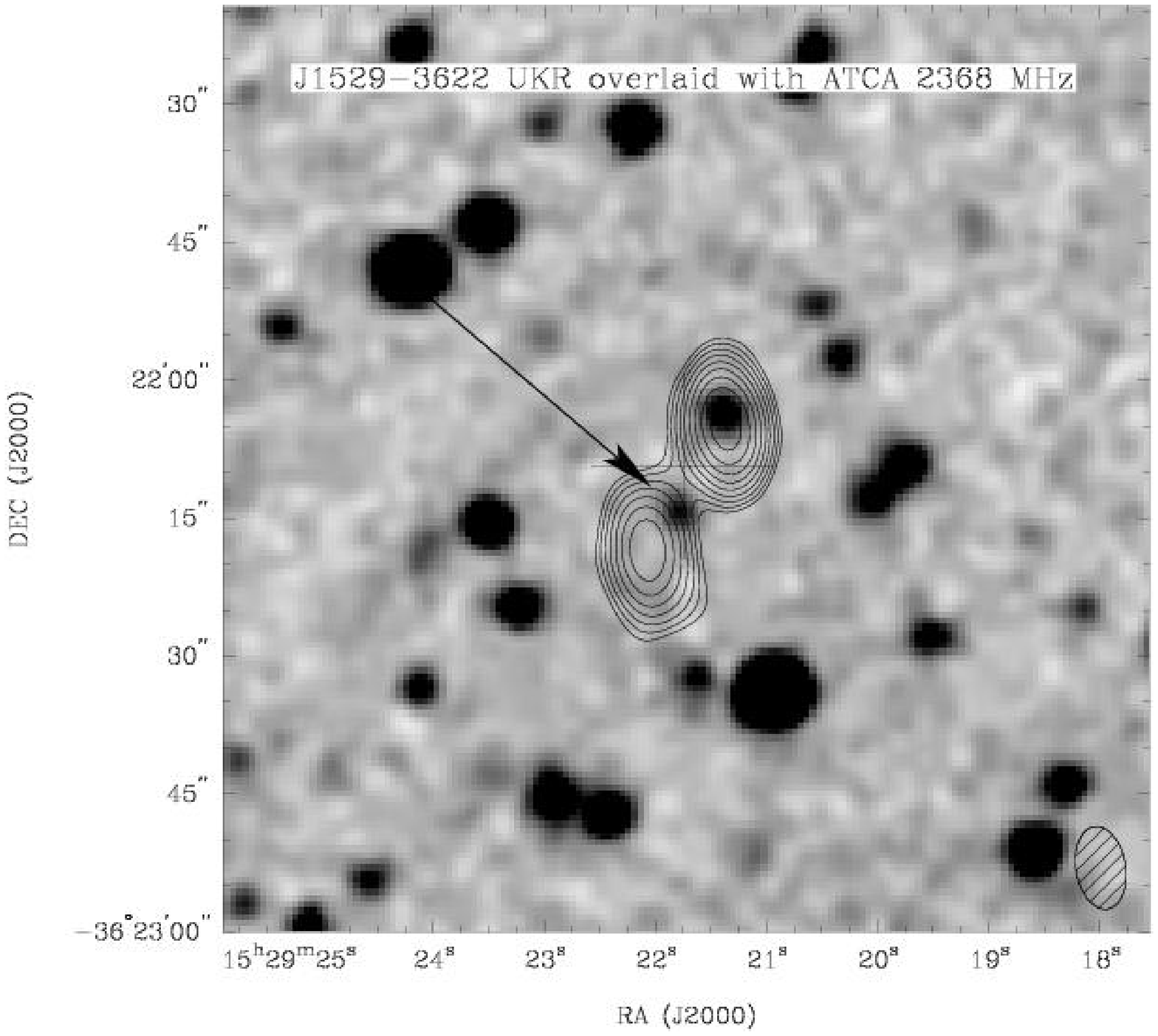,height=5.0cm,width=5.5cm}
\newline
\newline
\epsfig{file=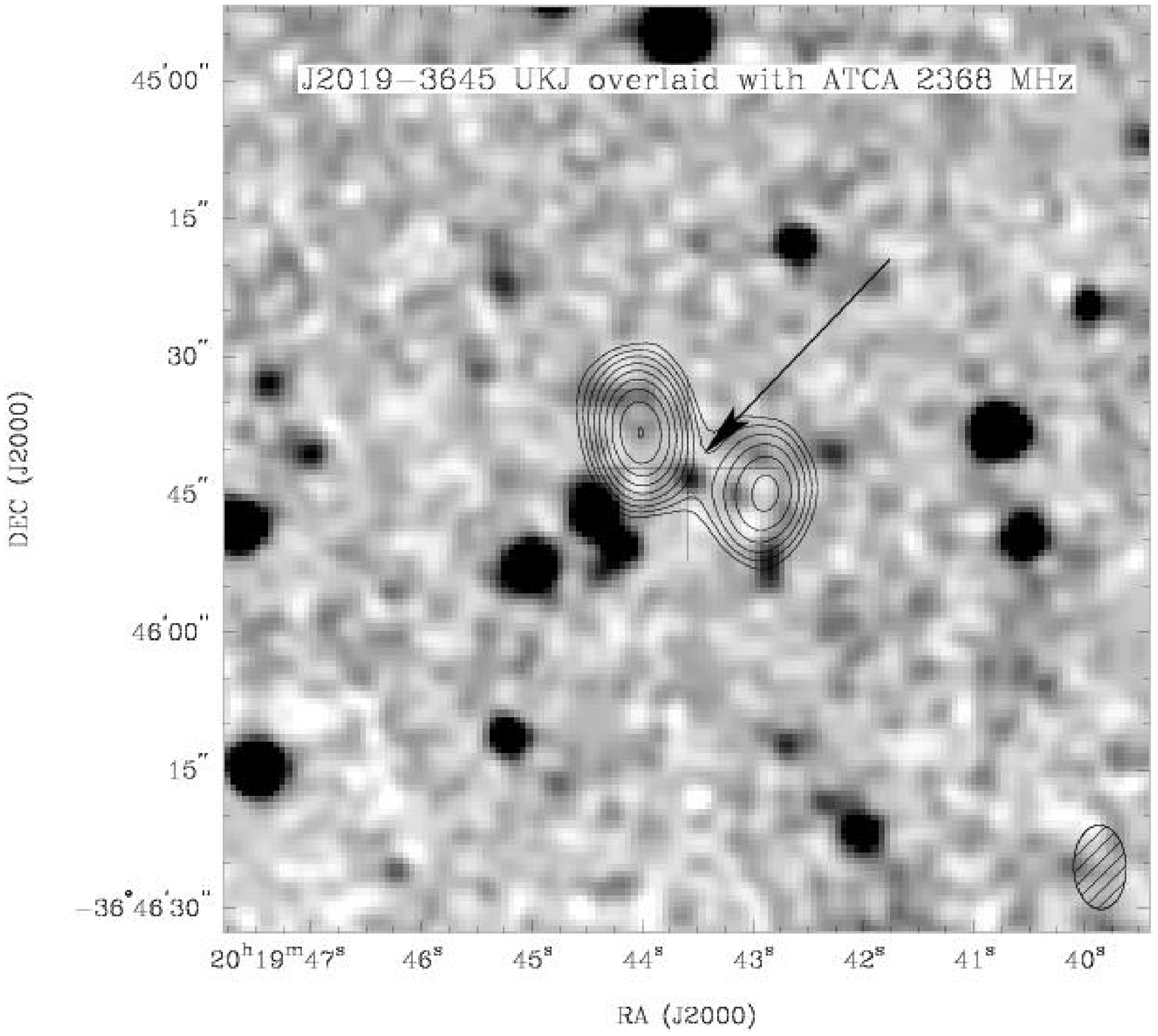,height=5.0cm,width=5.5cm}~\epsfig{file=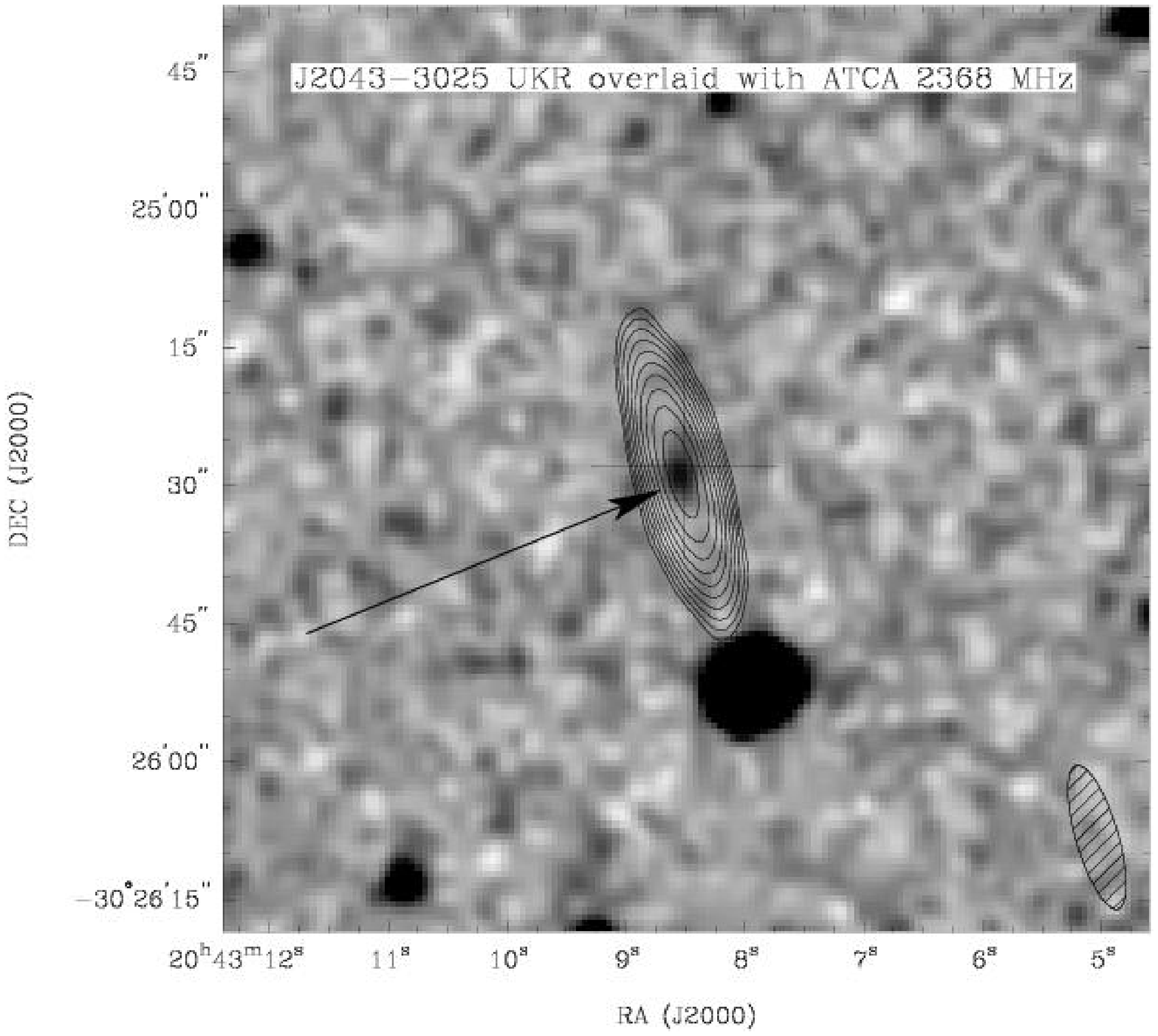,height=5.0cm,width=5.5cm}~\epsfig{file=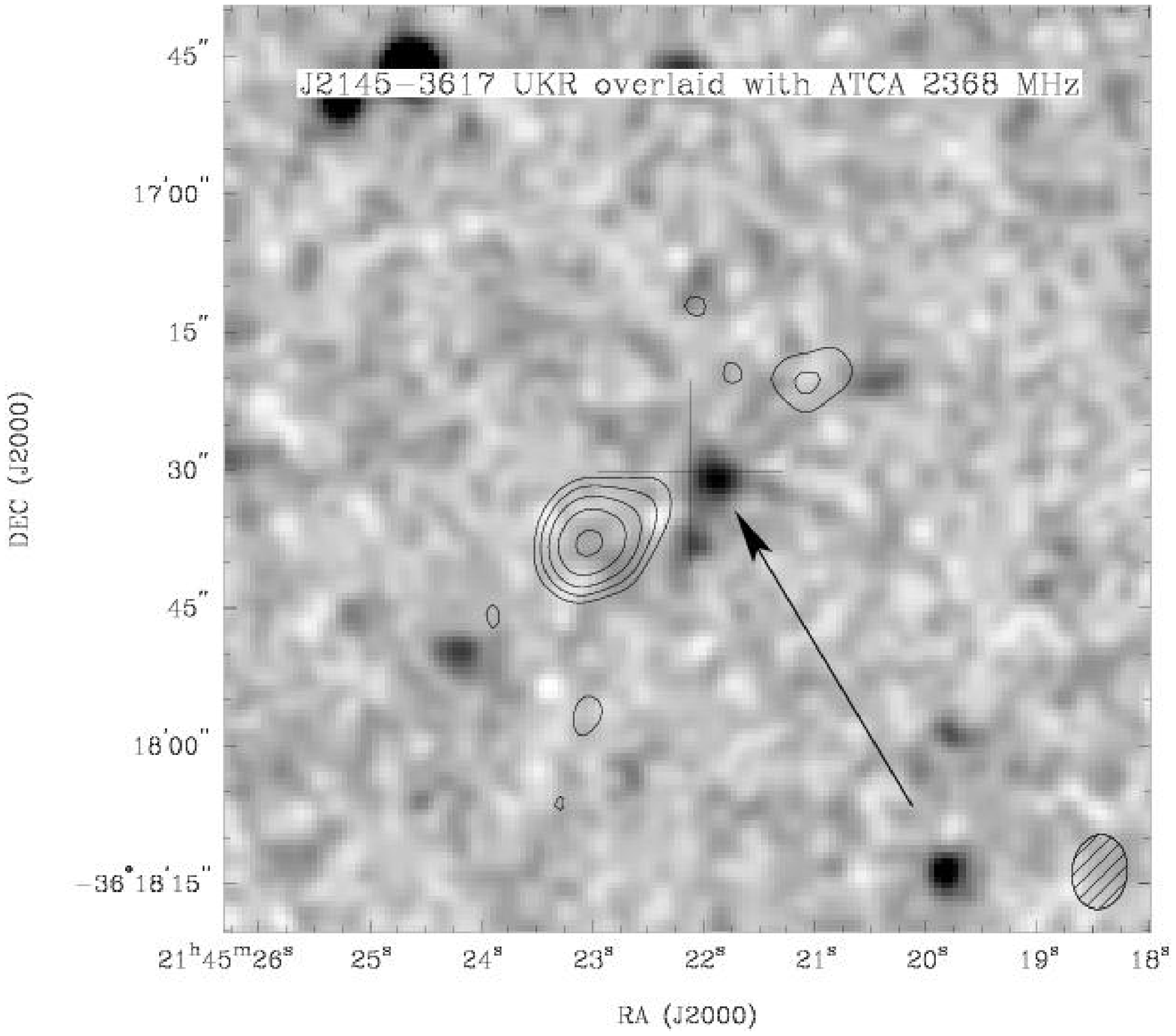,height=5.0cm,width=5.5cm}
\newline
\newline
\epsfig{file=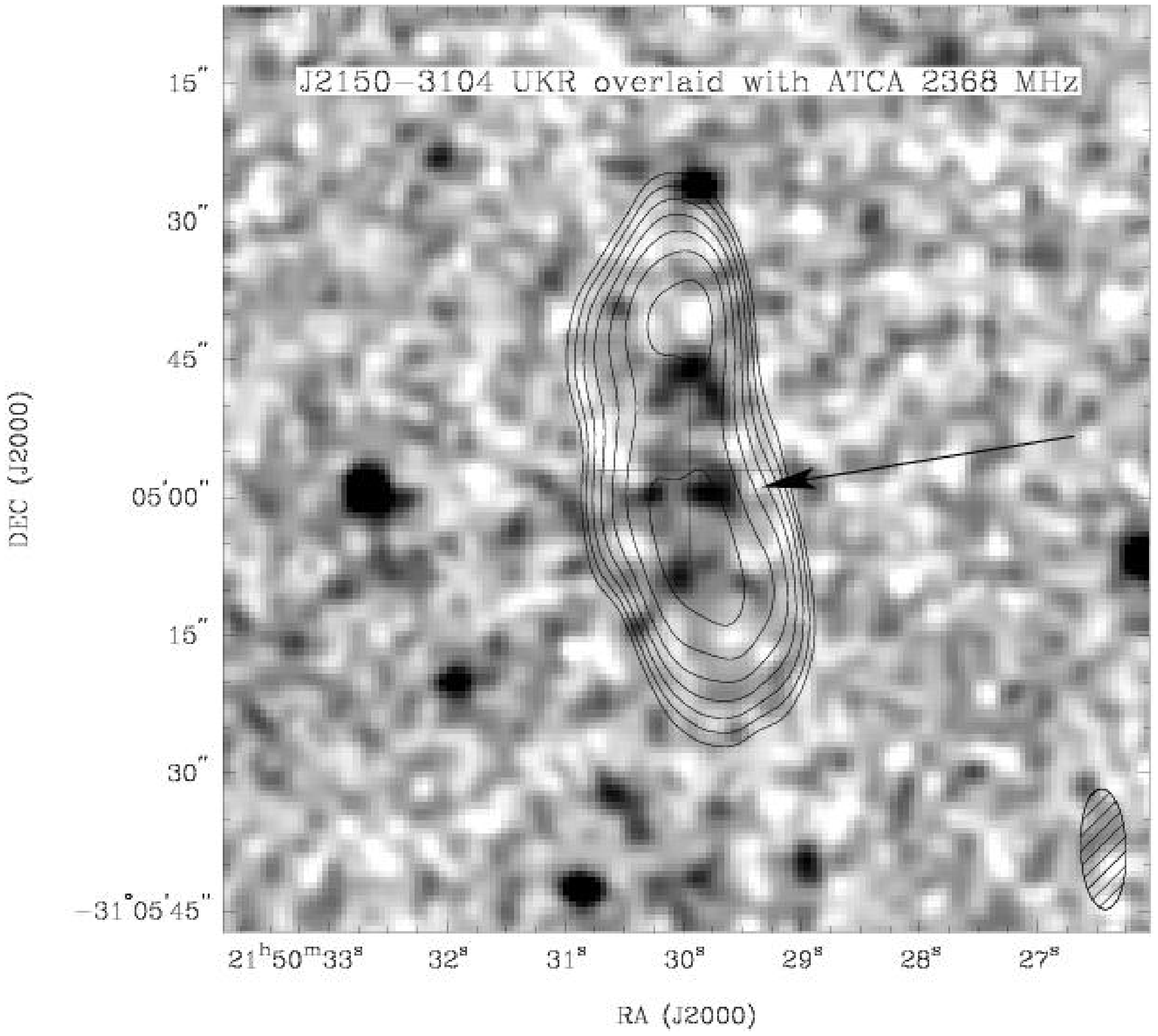,height=5.0cm,width=5.5cm}~\epsfig{file=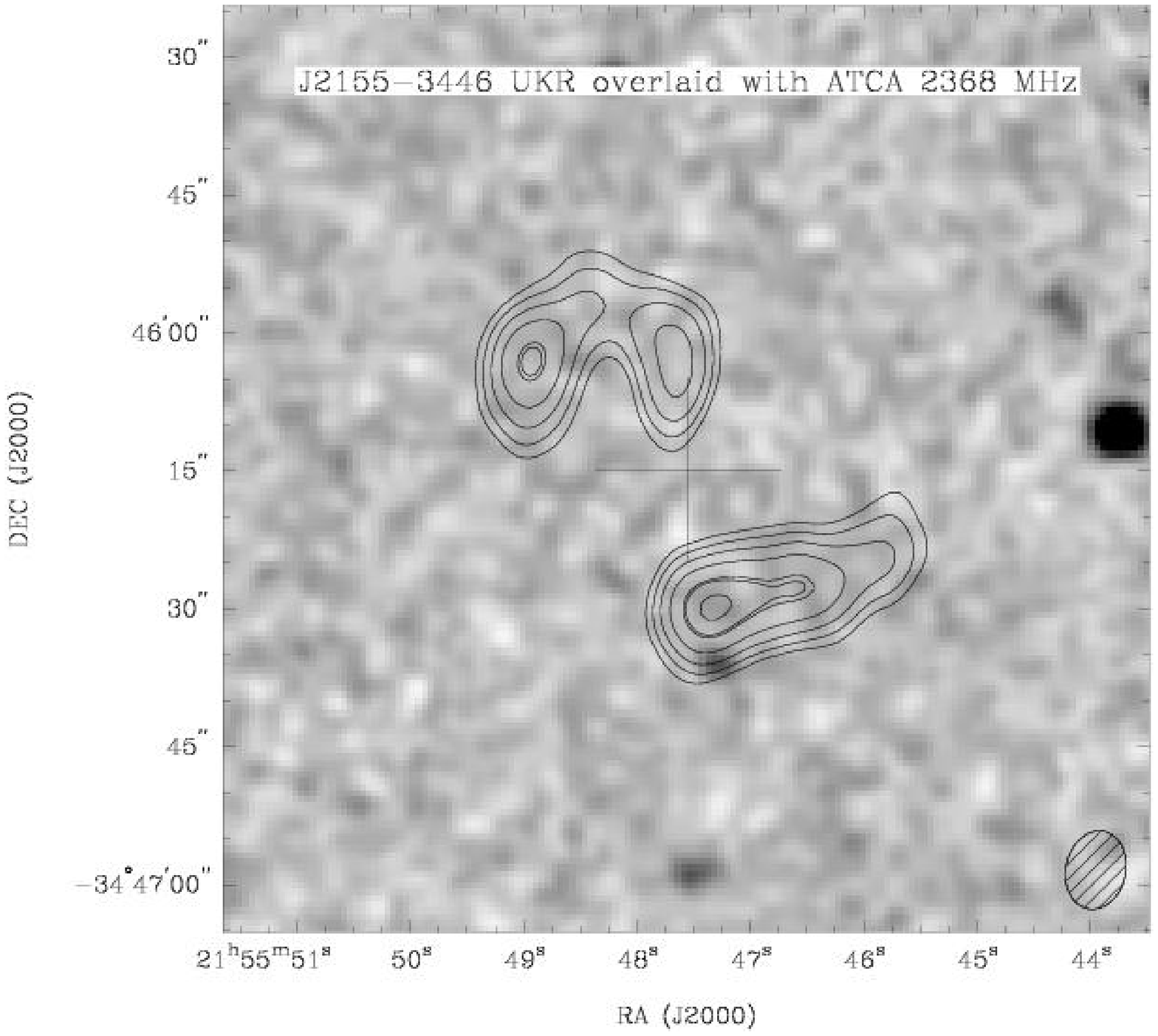,height=5.0cm,width=5.5cm}~\epsfig{file=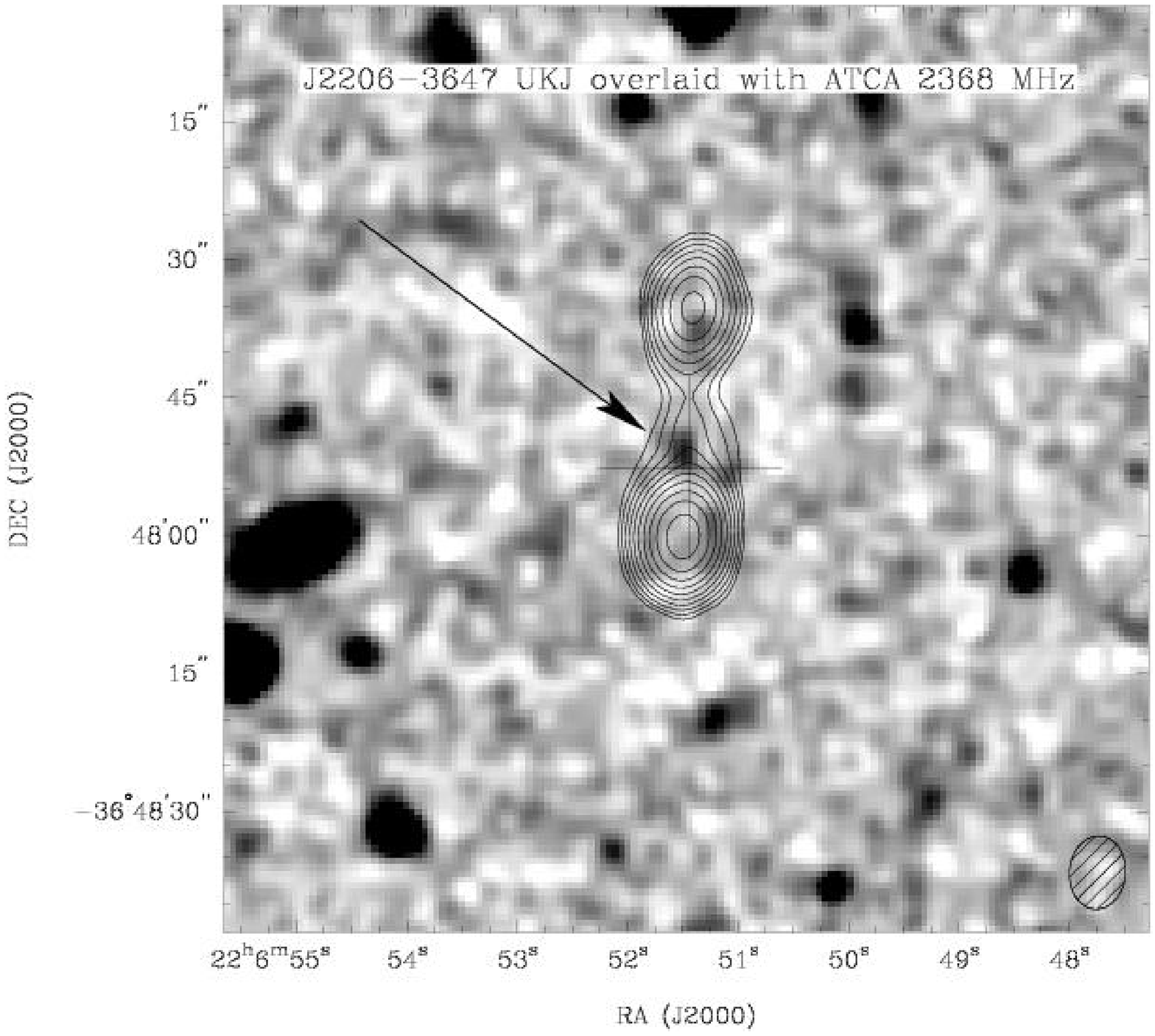,height=5.0cm,width=5.5cm}
\newline
\newline
\epsfig{file=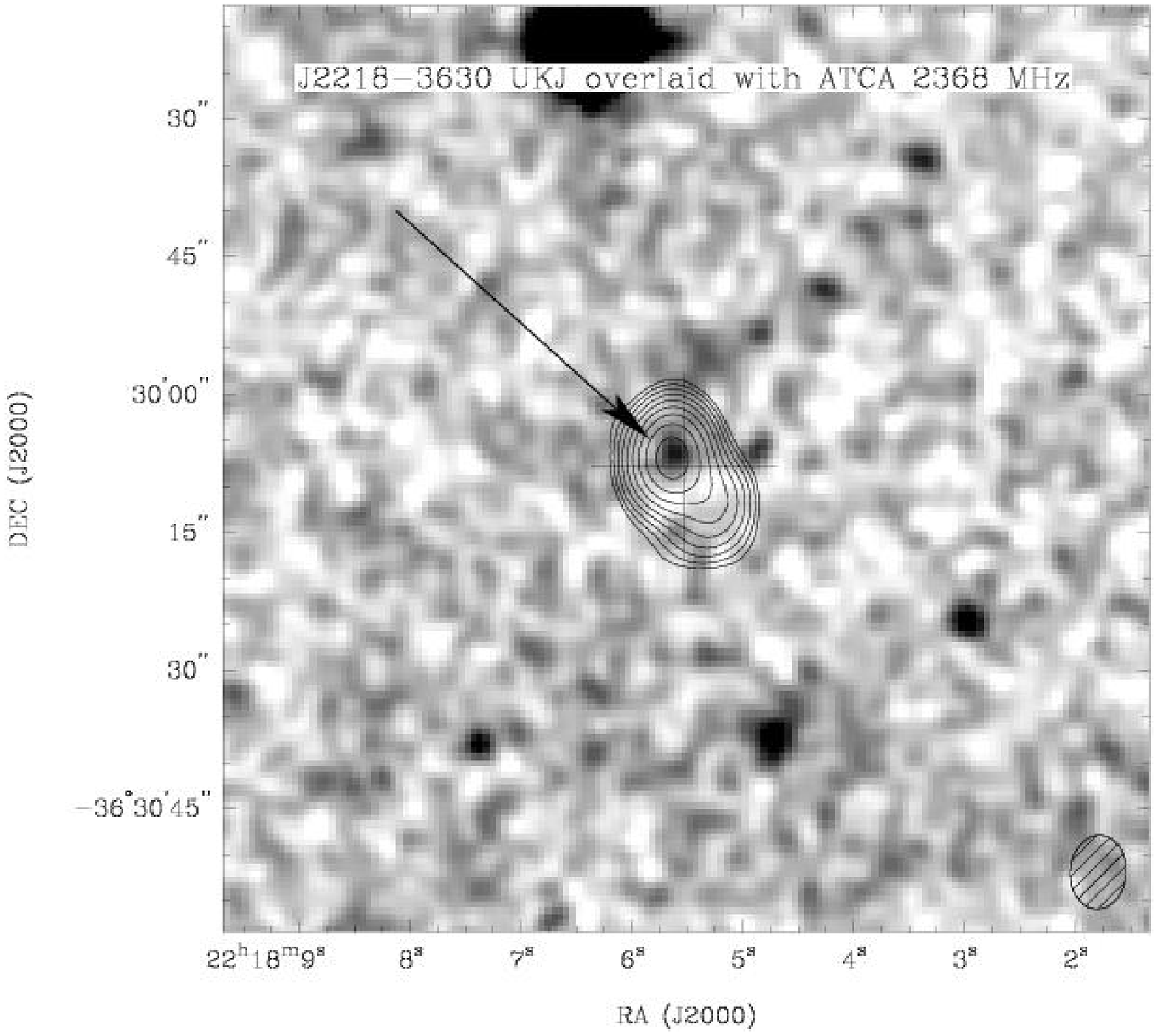,height=5.0cm,width=5.5cm}~\epsfig{file=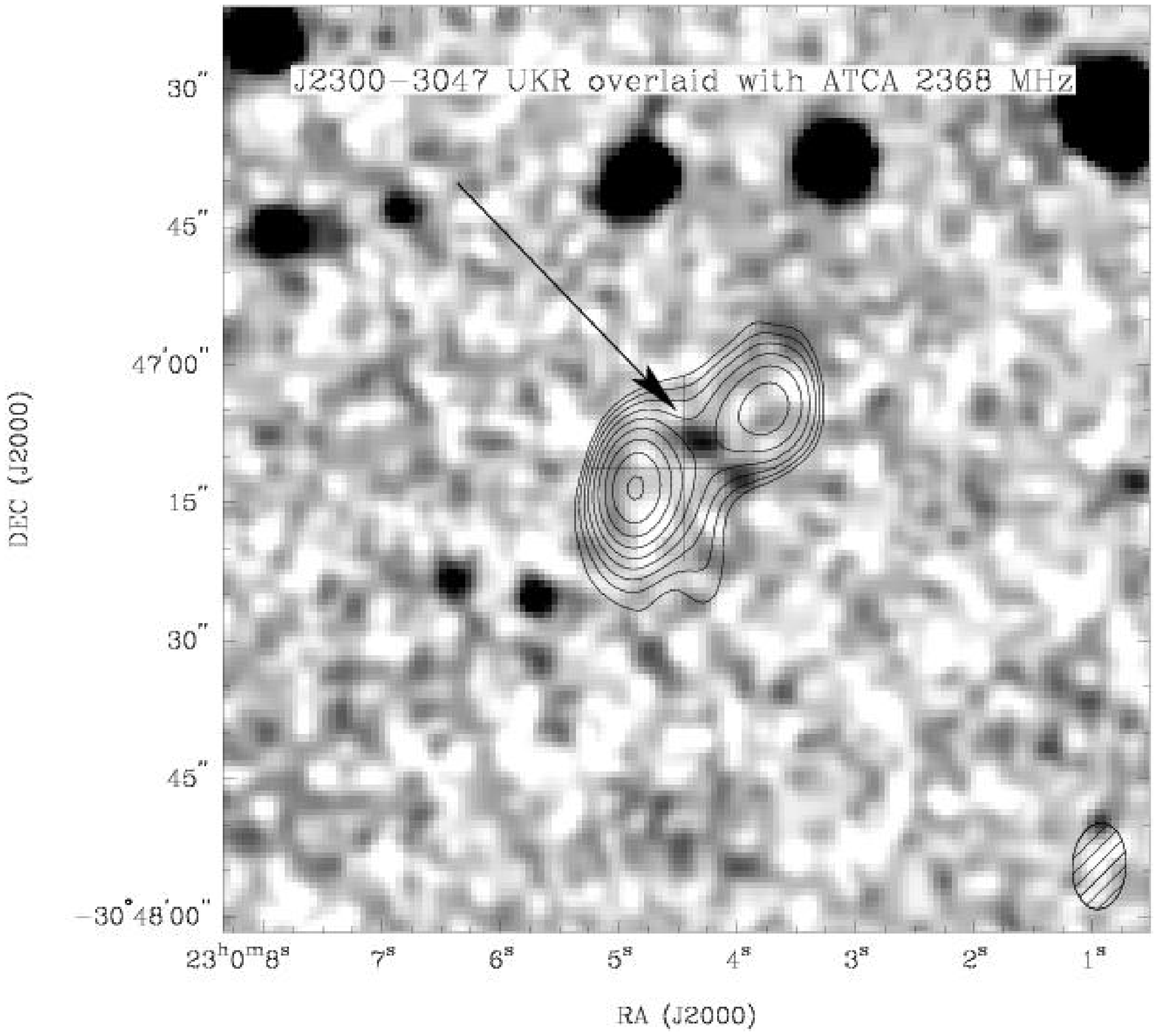,height=5.0cm,width=5.5cm}~\epsfig{file=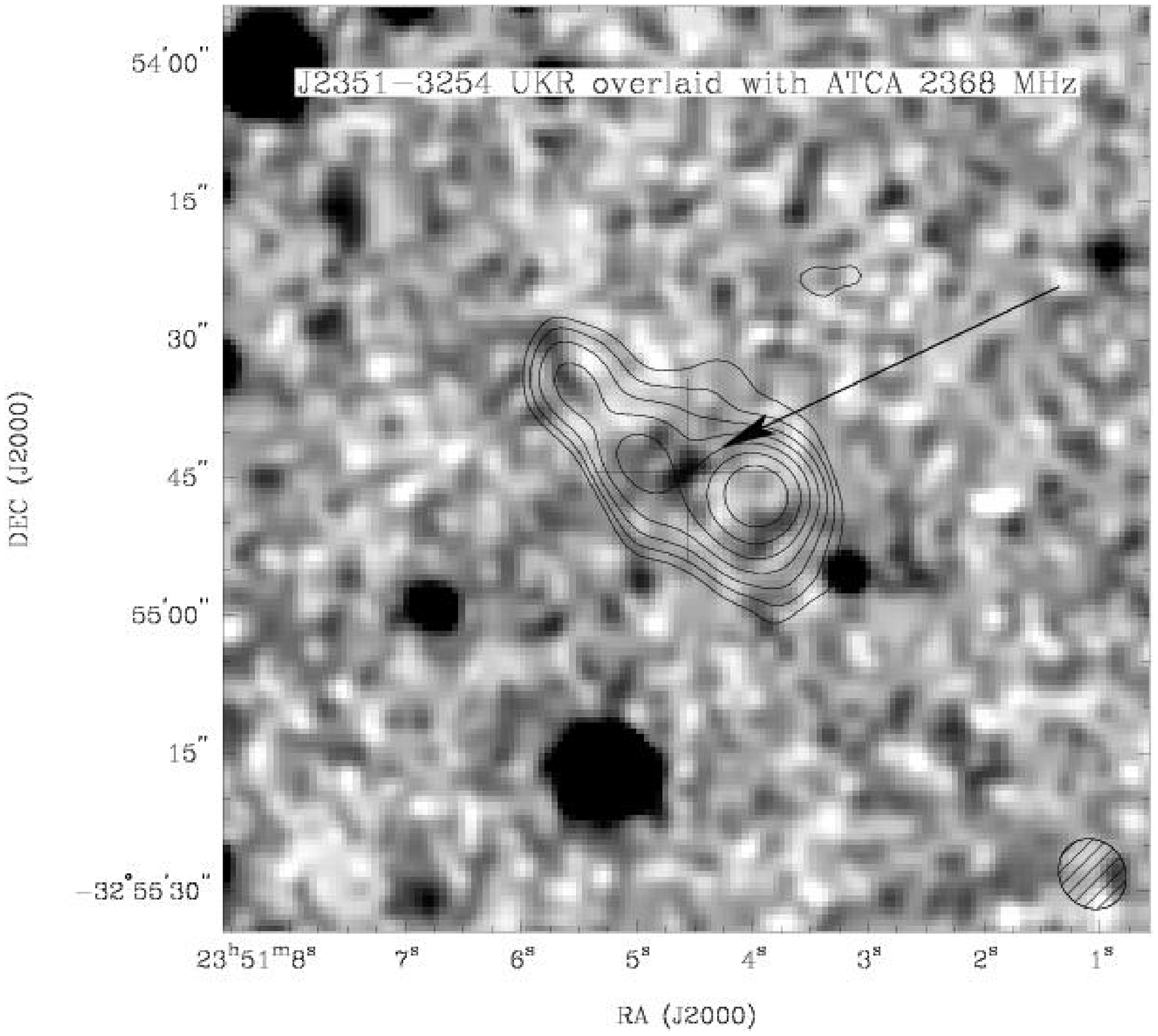,height=5.0cm,width=5.5cm}
\contcaption{}
\end{minipage}
\end{figure*}

At 2368 MHz, 39 per cent of the sources in the MRCR--SUMSS sample have a single-component morphology, 52 per cent are Fanaroff--Riley type II \citep[FR II;][]{fanaroff74} doubles, and 7 per cent are FR II triples. Interestingly, many of the doubles and triples have asymmetric lobe flux densities, which suggests a corresponding asymmetry in the density of the intergalactic medium through which the radio jets propagate. In addition, there are a few FR I sources that have either a head-tail, narrow-angle tail, wide-angle tail or complex morphology; these sources are the expected principal contaminants in a USS-selected sample.

The largest angular sizes range from 1--76 arcsec at 2368 MHz, with a median of 11 arcsec.  Given that the relationship between angular size and projected linear size is relatively flat over the redshift interval $\sim$1--4 ($\sim$7--8.5 kpc arcsec$^{-1}$) in our adopted cosmology, around 20 per cent of the sample would be classified as compact steep-spectrum sources \citep[CSS; see e.g. the review by][]{odea98}, with projected linear sizes $\lesssim$ 20 kpc.

In Section~\ref{408_paper1_sample_definition}, we noted that 32 sources (14 per cent of the sample) were found to have likely optical identifications in SuperCOSMOS after ATCA imaging. In Table~\ref{408_paper1_table:optical IDs}, we give the coordinates and optical magnitudes for each of the potential host galaxies. Typical uncertainties in the optical magnitudes are a few tenths of a magnitude. Most of the optical counterparts are very faint; the median SuperCOSMOS magnitudes are $\widetilde{B_{\rm J}} = 22.2 $ and $\widetilde{R} = 20.1$.

In Fig.~\ref{408_paper1_fig:morph_examples}, ATCA contours have been overlaid on SuperCOSMOS images for sources with a possible optical identification, an unusual radio morphology, and/or information available from the literature. As the optical counterparts are faint, we smoothed the SuperCOSMOS images with a Gaussian kernel with FWHM = 2.1 arcsec (3 pixels). $K$-band-radio overlays for the remaining sources without optical counterparts are presented in Paper II. Notes on individual sources in Fig.~\ref{408_paper1_fig:morph_examples} are given below:
\newline
\newline
{\bf NVSS J010619$-$325713.} Narrow-angle tail source with partially detached tails \citep[e.g. similar morphology to J1324$-$3138 in the Abell cluster A3556;][]{venturi98}. The optical counterpart is located at the centroid of the cluster EDCC 0523 \citep{lumsden92}; using data from the 2dF Galaxy Redshift Survey \citep[2dFGRS;][]{colless01}, \citet{depropris02} determined the cluster centroid to be located at $\alpha=01^{\rm h}\,06^{\rm m}\,20.28^{\rm s}$, $\delta=-32\degr\,57\arcmin\,32.7\,\arcsec$ with a redshift of $z = 0.146 \pm 0.002$. The 2dFGRS redshift of the optical counterpart is $z = 0.1433 \pm 0.0002$.
\newline
{\bf NVSS J014353$-$331041.} Wide-angle tail source with a clear optical identification. The 2dFGRS redshift of the optical counterpart is $z = 0.1756 \pm 0.0003$.
\newline
{\bf NVSS J015745$-$310557.} Two independent, closely-spaced doubles. While the 408, 843 and 1400 MHz flux densities are blended, both doubles are still HzRG candidates: $\alpha^{2368}_{1384} = -1.50 \pm 0.12$ for the north double, and $\alpha^{2368}_{1384} = -1.57 \pm 0.12$  for the south double.
\newline
{\bf NVSS J020827$-$342635 $+$ NVSS J020828$-$342520.} This source has a complex, diffuse, multi-component morphology. There are a number of faint confusing sources at 1384 MHz that are not resolved in NVSS (these are included in the total 1384 MHz flux density in Table 3). Both the southern lobe and most of the confusing sources are either marginally detected or not detected at 2368 MHz, implying that they have very steep spectra. The 2dFGRS redshift of the optical counterpart is $z=0.3044 \pm 0.0003$.
\newline
{\bf NVSS J111921$-$363139.} This single-component source is also in the MRC--PMN USS sample, with spectral index $\alpha^{4850}_{408} = -1.24 \pm 0.05$ \citep{debreuck00}. It was observed with the ATCA at 1420 MHz by \citet{debreuck00}, who obtained a flux density of $S_{1420} = 542 \pm 54$ mJy. This is consistent with our 1384 MHz flux density of $S_{1384} = 509.2 \pm 15.8$ mJy.
\newline
{\bf NVSS J120839$-$340307.} Double source that is also in the Molonglo Southern 4 Jy sample \citep[MS4;][]{burgess06a,burgess06b} and the MRC--PMN USS sample \citep[spectral index $\alpha^{4850}_{408} = -1.28 \pm 0.04$;][] {debreuck00}. \citet{burgess06b} used the Anglo-Australian Telescope (AAT) to detect an optical counterpart between the lobes at $\alpha = 12^{\rm h}\,08^{\rm m}\,39.68^{\rm s}$, $\delta=-34\degr\,03\arcmin\,10.6\,\arcsec$   with $R=22.9 \pm 0.5$ mag. The SuperCOSMOS $B_{\rm J}$ magnitude of the optical counterpart is $> 22.5$ mag.
\newline
{\bf NVSS J215547$-$344614.} Two independent, closely-spaced wide-angle tail (north) and head-tail (south) sources that are potentially members of the same distant cluster \citep[e.g. similar morphology to MRC B1610$-$605 + MRC B1610$-$608 in][]{jones92}; $\alpha^{2368}_{1384} = -1.28 \pm 0.14$ for the wide-angle tail source, and $\alpha^{2368}_{1384}= -0.90 \pm 0.12$ for the head-tail source.
\newline
\newline
As the three low-redshift FR I sources in the MRCR--SUMSS sample are interlopers, we exclude NVSS J010619$-$325713, NVSS J014353$-$331041 and NVSS J020827$-$342635 $+$ NVSS J020828$-$342520 from any subsequent analysis in this paper.

\begin{table}
\scriptsize
\centering
\setcounter{table}{4}
\setlength{\tabcolsep}{3.00pt}
\caption{Sources in the MRCR--SUMSS sample with likely optical identifications in SuperCOSMOS.}\label{408_paper1_table:optical IDs}
\begin{tabular}{|l|r|r|r|r|}
\hline
\hline
& \multicolumn{2}{|c|}{Host galaxy} & \multicolumn{2}{|c|}{Optical} \\
\multicolumn{1}{|c|}{Source} & \multicolumn{2}{|c|}{coordinates$^{1}$} & \multicolumn{2}{|c|}{magnitudes} \\
& \multicolumn{1}{|c|}{RA (J2000)} & \multicolumn{1}{|c|}{Dec (J2000)} & \multicolumn{1}{|c|}{$B_{\rm J}$} & \multicolumn{1}{|c|}{$R$} \\
& \multicolumn{1}{|c|}{(h min s)} & \multicolumn{1}{|c|}{($\degr$ $\arcmin$ $\arcsec$)} & \multicolumn{1}{|c|}{(mag)} & \multicolumn{1}{|c|}{(mag)} \\
\hline
NVSS J001737$-$364710 & 00 17 37.66 & $-$36 47 11.5 & 21.2 & 20.6 \\
NVSS J002956$-$313701 & 00 29 56.76 & $-$31 37 02.7 & 22.6 & 19.8 \\
NVSS J003445$-$372348 & 00 34 45.59 & $-$37 23 48.9 & $> 22.5$ & 20.6 \\
NVSS J004206$-$361329 & 00 42 06.60 & $-$36 13 29.1 & $> 22.5$ & 20.9 \\
NVSS J004851$-$324623 & 00 48 51.16 & $-$32 46 23.1 & $>22.5$ & 20.7 \\
NVSS J010619$-$325713 & 01 06 20.26 & $-$32 57 32.2 & 18.3 & 17.0 \\
NVSS J011732$-$383739 & 01 17 32.77 & $-$38 37 38.5 & 22.1 & 19.7 \\
NVSS J012443$-$310820 & 01 24 43.73 & $-$31 08 22.0 & 22.2 & $> 21.5$ \\
NVSS J014353$-$331041 & 01 43 54.11 & $-$33 10 47.0 & 18.4 & 16.8 \\
NVSS J020827$-$342635 $_{_{_{+}}}$ & \multirow{2}{*}{02 08 28.55} & \multirow{2}{*}{$-$34 25 37.0} & \multirow{2}{*}{19.5} & \multirow{2}{*}{17.7} \\
NVSS J020828$-$342520 & & & & \\
NVSS J023601$-$314204 $_{_{_{+}}}$ & \multirow{2}{*}{02 36 03.99} & \multirow{2}{*}{$-$31 42 22.2} & \multirow{2}{*}{20.9} & \multirow{2}{*}{18.5} \\
NVSS J023605$-$314235 & & & & \\
NVSS J025521$-$333212 & 02 55 21.09 & $-$33 32 11.4 & 21.8 & 19.5 \\
NVSS J101215$-$394939 & 10 12 15.19 & $-$39 49 22.6 & 20.8 & 19.8 \\
NVSS J103441$-$394957 & 10 34 42.10 & $-$39 49 58.2 & 20.3 & 19.6 \\
NVSS J111539$-$375633 & 11 15 39.36 & $-$37 56 33.2 & 22.9 & 20.9 \\
NVSS J111546$-$391410 & 11 15 46.47 & $-$39 14 10.5 & $>22.5$ & 20.8 \\
NVSS J132952$-$381251 & 13 29 52.56 & $-$38 12 53.9 & 22.6 & 19.8 \\
NVSS J133317$-$383006 & 13 33 17.48 & $-$38 30 16.2 & 21.2 & 20.3 \\
NVSS J134502$-$395932 & 13 45 02.16 & $-$39 59 28.9 & 21.8 & $> 21.5$ \\
NVSS J140330$-$394130 & 14 03 30.61 & $-$39 41 35.1 & $>22.5$ & 20.0 \\
NVSS J141428$-$320637 & 14 14 28.28 & $-$32 06 38.8 & 19.4 & 18.1 \\
NVSS J144503$-$353542 & 14 45 03.24 & $-$35 35 44.6 & 22.2 & $> 21.5$ \\
NVSS J151215$-$382220 & 15 12 15.51 & $-$38 22 22.2 & 21.4 & 19.7 \\
NVSS J152921$-$362209 & 15 29 21.78 & $-$36 22 14.4 & 21.6 & 19.9 \\
NVSS J201943$-$364542 & 20 19 43.56 & $-$36 45 43.4 & 22.7 & $> 21.5$ \\
NVSS J204308$-$302527 & 20 43 08.56 & $-$30 25 28.8 & 22.2 & 19.9 \\
NVSS J214522$-$361730 & 21 45 21.90 & $-$36 17 30.7 & 21.3 & 19.1 \\
NVSS J215029$-$310457 & 21 50 29.80 & $-$31 04 59.6 & $>22.5$ & 20.1 \\
NVSS J220651$-$364752 & 22 06 51.47 & $-$36 47 51.4 & 22.6 &  $> 21.5$ \\
NVSS J221805$-$363007 & 22 18 05.61 & $-$36 30 06.0 & 22.9 & $> 21.5$ \\
NVSS J230004$-$304711 & 23 00 04.25 & $-$30 47 08.6 & $>22.5$ & 20.8 \\
NVSS J235104$-$325444 & 23 51 04.66 & $-$32 54 44.8 & $>22.5$ & 20.7 \\
\hline
\multicolumn{5}{p{80mm}}{$^{1}$Host galaxy coordinates are the SuperCOSMOS fitted positions in the UKR images, except for those cases where the counterpart is detected in the UKJ image only.}\\
\end{tabular}
\end{table}

\subsection{Flux densities}\label{408_paper1_fluxes}

The statistical properties of the flux densities of the sources in the MRCR--SUMSS sample are presented in Table~\ref{408_paper1_table:flux_properties}. Typical uncertainties are 10 per cent in the MRCR, and 3--4 per cent in SUMSS and NVSS. The ATCA 1384 and 2368 MHz flux densities were measured by fitting elliptical Gaussians to source components with the {\scriptsize MIRIAD} task {\scriptsize IMFIT}. The fitting errors were calculated following \citet{condon97} and combined in quadrature with the calibration uncertainty to obtain a total flux density uncertainty. In general, the fitting uncertainty is much smaller than the calibration uncertainty, which is $\sim$3 per cent at 1384 MHz and $\sim$5 per cent at 2368 MHz. Thus, the total flux density uncertainties are 3--4 per cent at 1384 MHz and 5--6 per cent at 2368 MHz.

The sources in the MRCR--SUMSS sample are stronger than in the SUMSS--NVSS sample. The median SUMSS and NVSS flux densities in the MRCR--SUMSS sample are 134 and 80 mJy respectively (Table~\ref{408_paper1_table:flux_properties}), compared with 47 and 22 mJy respectively in the SUMSS--NVSS sample \citep{debreuck04}. This is a consequence of the $S_{408} \geq 200$ mJy cutoff which defines the MRCR--SUMSS sample, as opposed to the $S_{1400} \geq 15$ mJy cutoff used in the SUMSS--NVSS sample; the equivalent cutoff in the SUMSS--NVSS sample at 408 MHz is 74.5 mJy for $\alpha=-1.3$.

\begin{figure}
\psfig{file=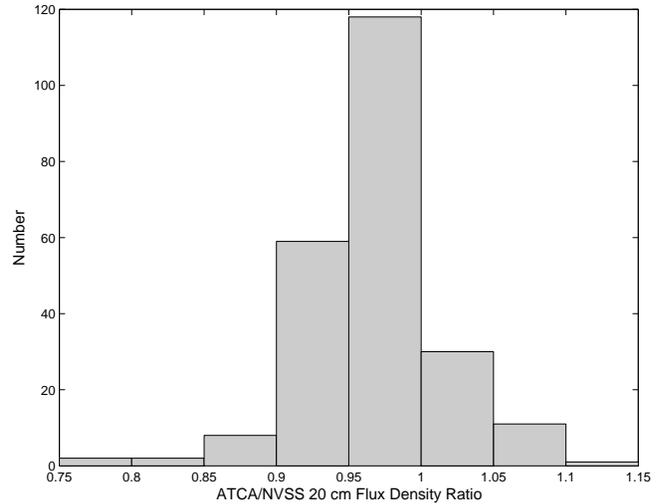,height=6.66cm}
\caption{Histogram of the ATCA/NVSS 20 cm flux density ratio for sources in the MRCR--SUMSS sample.}\label{408_paper1_fig:flux_ratio_hist}
\end{figure}

The effectiveness of our automatically constrained {\scriptsize CLEAN}ing procedure is illustrated in Fig.~\ref{408_paper1_fig:flux_ratio_hist}, which shows a histogram of the ATCA/NVSS 20 cm flux density ratio. As the sources in the MRCR--SUMSS sample have ultra-steep spectra, we would expect the ATCA 1384 MHz flux densities to be 1--2 per cent higher than the NVSS 1400 MHz flux densities, due to the slight difference in centre frequency. After automatically constrained CLEANing, the agreement between the ATCA and NVSS flux densities is excellent: $\sim$95 per cent of the sources have flux densities that agree to within $\pm 10$ per cent, and $\sim$75 per cent agree to within $\pm 5$ per cent. However, on average, the ATCA flux densities tend to be slightly lower than in NVSS by $\sim$3 per cent. This is evident in the statistical properties in Table~\ref{408_paper1_table:flux_properties}.

One possible explanation for the slight underestimation of the ATCA flux densities is that the VLA array configuration used to construct NVSS (D-array, spanning baselines 35--1030 m) is more compact than the ATCA configurations used in this study, resulting in a greater sensitivity to low-surface-brightness extended features. Indeed, we find that sources with angular sizes $> 10$ arcsec are underestimated by $\sim$4 per cent on average, compared with $\sim$2 per cent for sources with angular sizes $< 10$ arcsec. Source variability may also affect the flux densities of some sources. Even though the ATCA flux densities are still marginally underestimated after automatically constrained {\scriptsize CLEAN}ing, we emphasize that, for the vast majority of sources, these flux densities are still sufficiently accurate to permit an investigation of SEDs in the frequency range 408--2368 MHz.

\begin{table}
\centering
\caption{Flux density properties of the MRCR--SUMSS sample. $\widetilde{S_{\nu}}$ is the median flux density, $S_{\nu}$ (min.) is the minimum flux density and $S_{\nu}$ (max.) is the maximum flux density.}\label{408_paper1_table:flux_properties}
\begin{tabular}{|c|r|r|r|}
\hline
\hline
\multicolumn{1}{|c|}{Frequency} & \multicolumn{1}{|c|}{$\widetilde{S_{\nu}}$} & \multicolumn{1}{|c|}{$S_{\nu}$ (min.)} & \multicolumn{1}{|c|}{$S_{\nu}$ (max.)} \\
\multicolumn{1}{|c|}{(MHz)} & \multicolumn{1}{|c|}{(mJy)} & \multicolumn{1}{|c|}{(mJy)} & \multicolumn{1}{|c|}{(mJy)} \\
\hline
408 & 316 & 186 & 4650 \\
843 & 134 & 59 & 2029 \\
1384 & 77 & 31 & 1113 \\
1400 & 80 & 31 & 1108 \\
2368 & 44 & 14 & 573 \\
\hline
\end{tabular}
\end{table}

\subsection{Spectral indices}\label{408_paper1_spectral_indices}

\begin{figure*}
\begin{minipage}{175mm}
\psfig{file=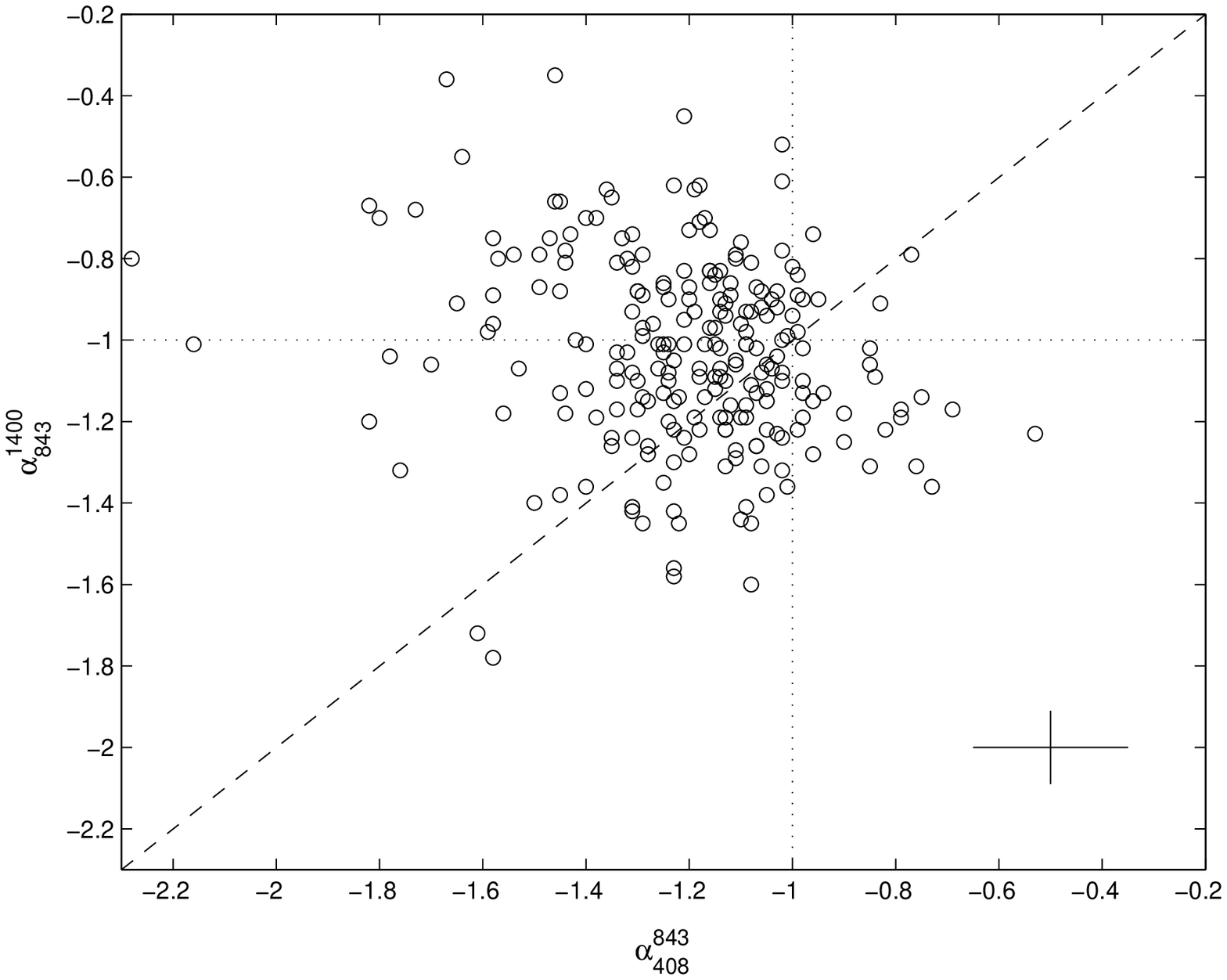,height=6.6cm}
\hspace{0.5cm}
\psfig{file=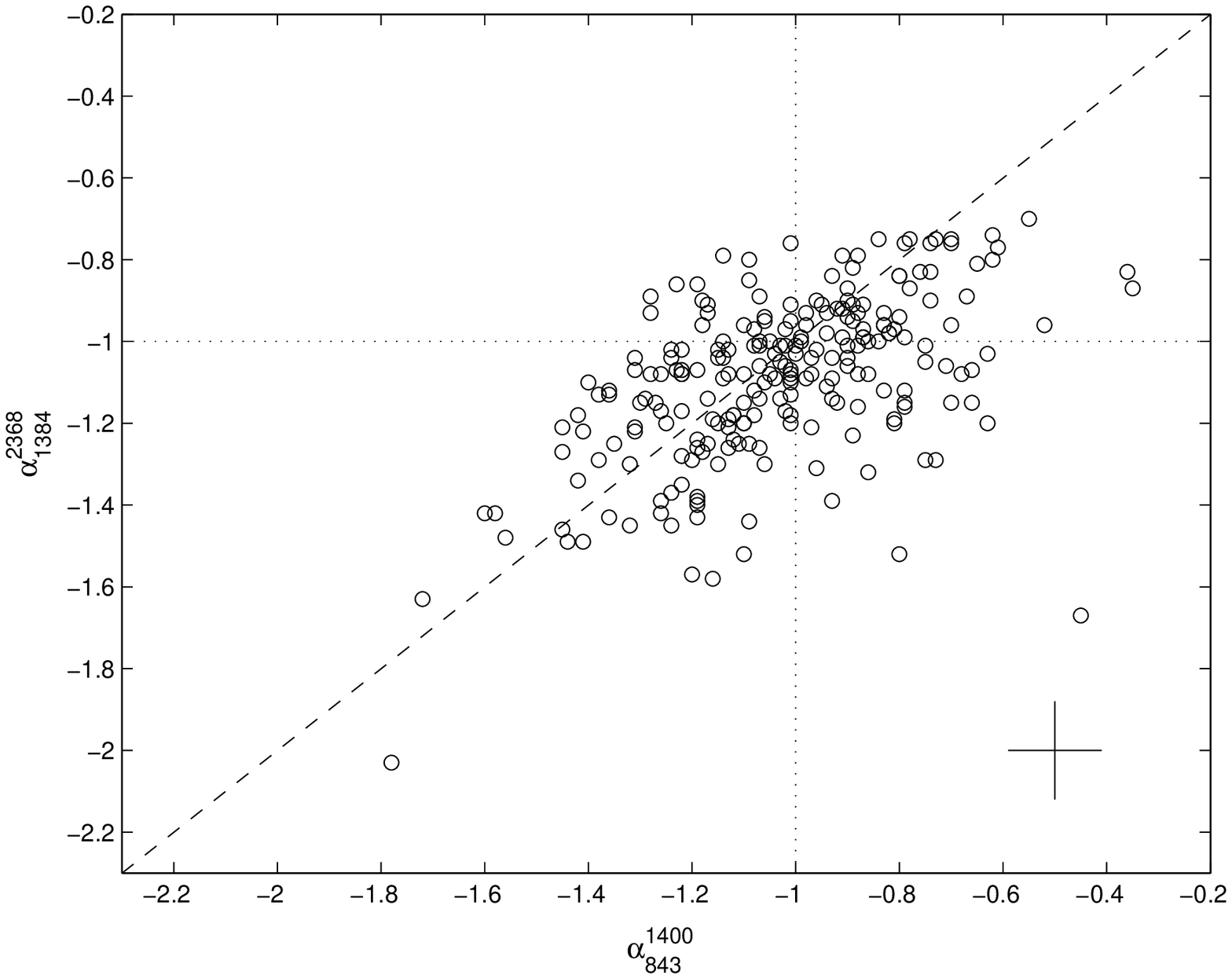,height=6.6cm}
\caption{Radio colour-colour plots for the MRCR--SUMSS sample. In the left panel, the low-frequency spectral index is $\alpha^{843}_{408}$, and the high-frequency spectral index is $\alpha^{1400}_{843}$. In the right panel, the low-frequency spectral index is $\alpha^{1400}_{843}$, and the high-frequency spectral index is $\alpha^{2368}_{1384}$. In both panels, the dashed line is a line of equality, while the dotted lines indicate where each $\alpha = -1.0$. The median individual uncertainties in $\alpha^{843}_{408}$, $\alpha^{1400}_{843}$ and $\alpha^{2368}_{1384}$ are $\pm$0.15, $\pm$0.09 and $\pm$0.12, respectively. Representative error bars featuring the median spectral index uncertainties are shown in the bottom right-hand corner of each panel.}\label{408_paper1_fig:colour_colour_plots}
\end{minipage}
\end{figure*}

The observed-frame spectral index properties of the MRCR--SUMSS sample are presented in Table~\ref{408_paper1_table:alpha_properties}. In addition, radio colour-colour plots featuring $\alpha^{843}_{408}$, $\alpha^{1400}_{843}$ and $\alpha^{2368}_{1384}$ are shown in Fig.~\ref{408_paper1_fig:colour_colour_plots}. In general, the differences among the two-point spectral indices are small: the average and median values only vary from approximately $-1.2$ to $-1.0$. The most prominent effect is visible in the left panel of Fig.~\ref{408_paper1_fig:colour_colour_plots}, where the SEDs tend to flatten over the frequency range 408--1400 MHz ($\Delta \alpha \approx 0.2$ on average). For the majority of sources, this is not a consequence of spectral curvature, and the points can be fitted to a straight spectrum (see below).

\begin{figure*}
\begin{minipage}{150mm}
\psfig{file=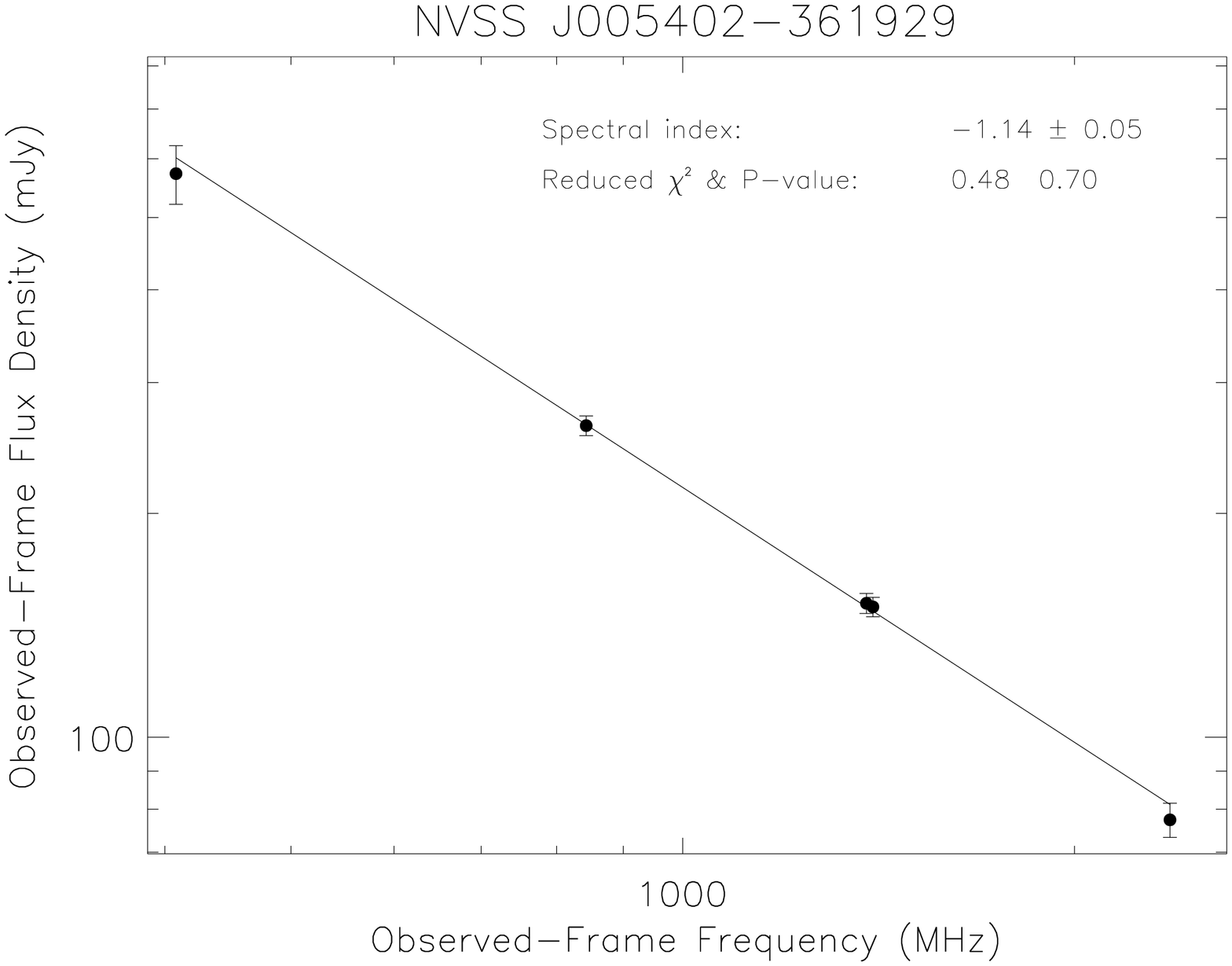,height=7.00cm,angle=90}
\hspace{0.75cm}
\psfig{file=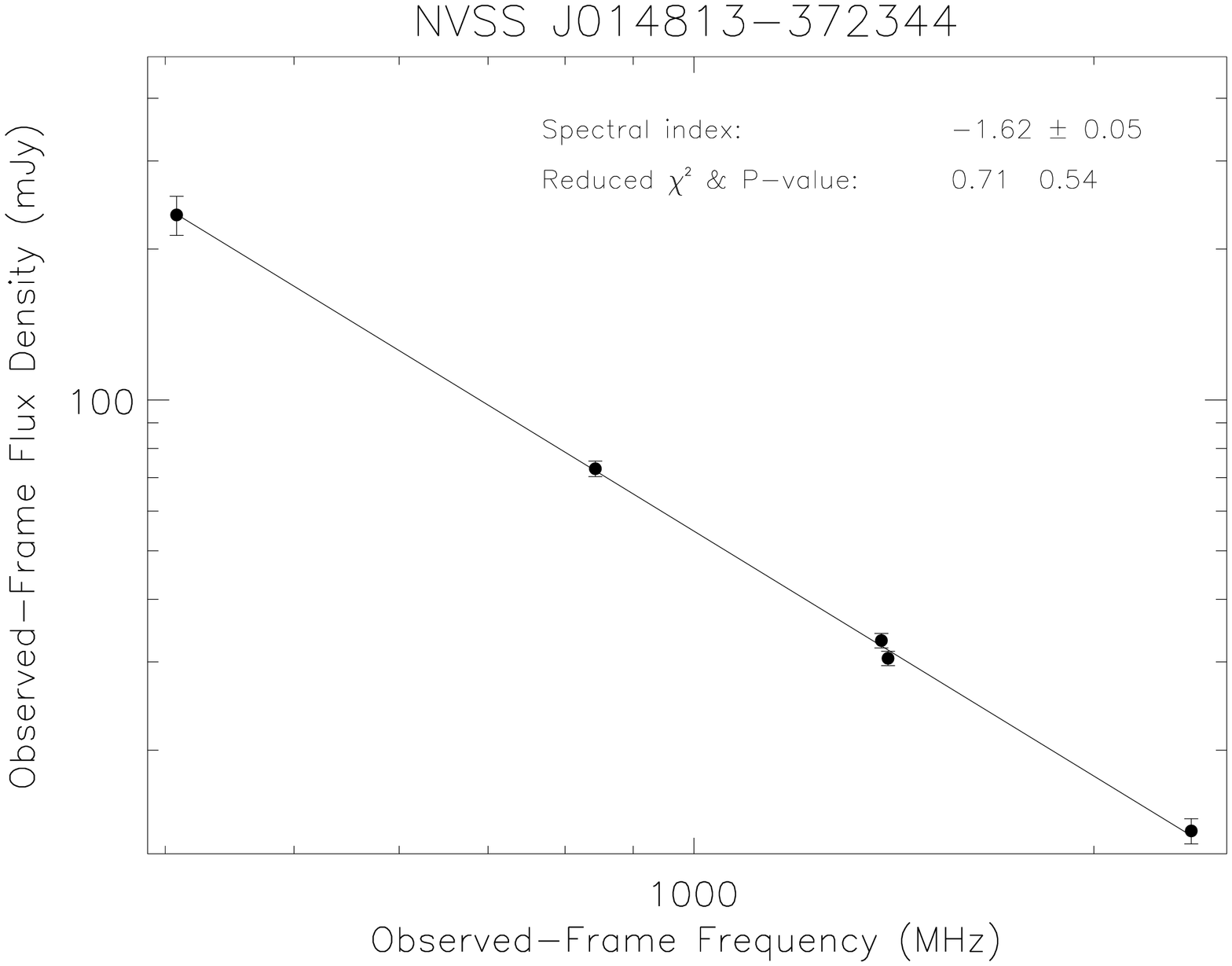,height=7.00cm,angle=90}
\newline
\newline
\newline
\psfig{file=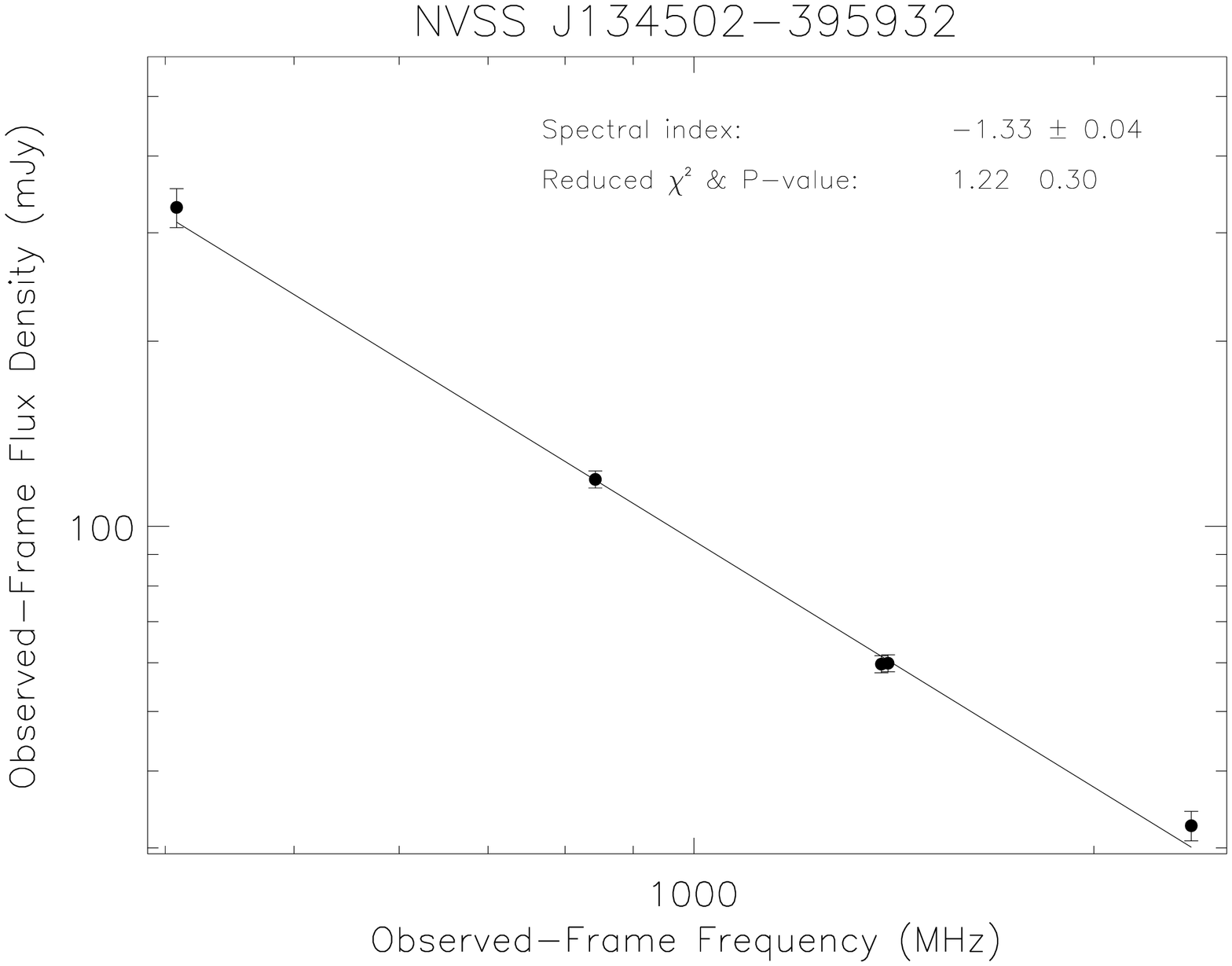,height=7.00cm,angle=90}
\hspace{0.75cm}
\psfig{file=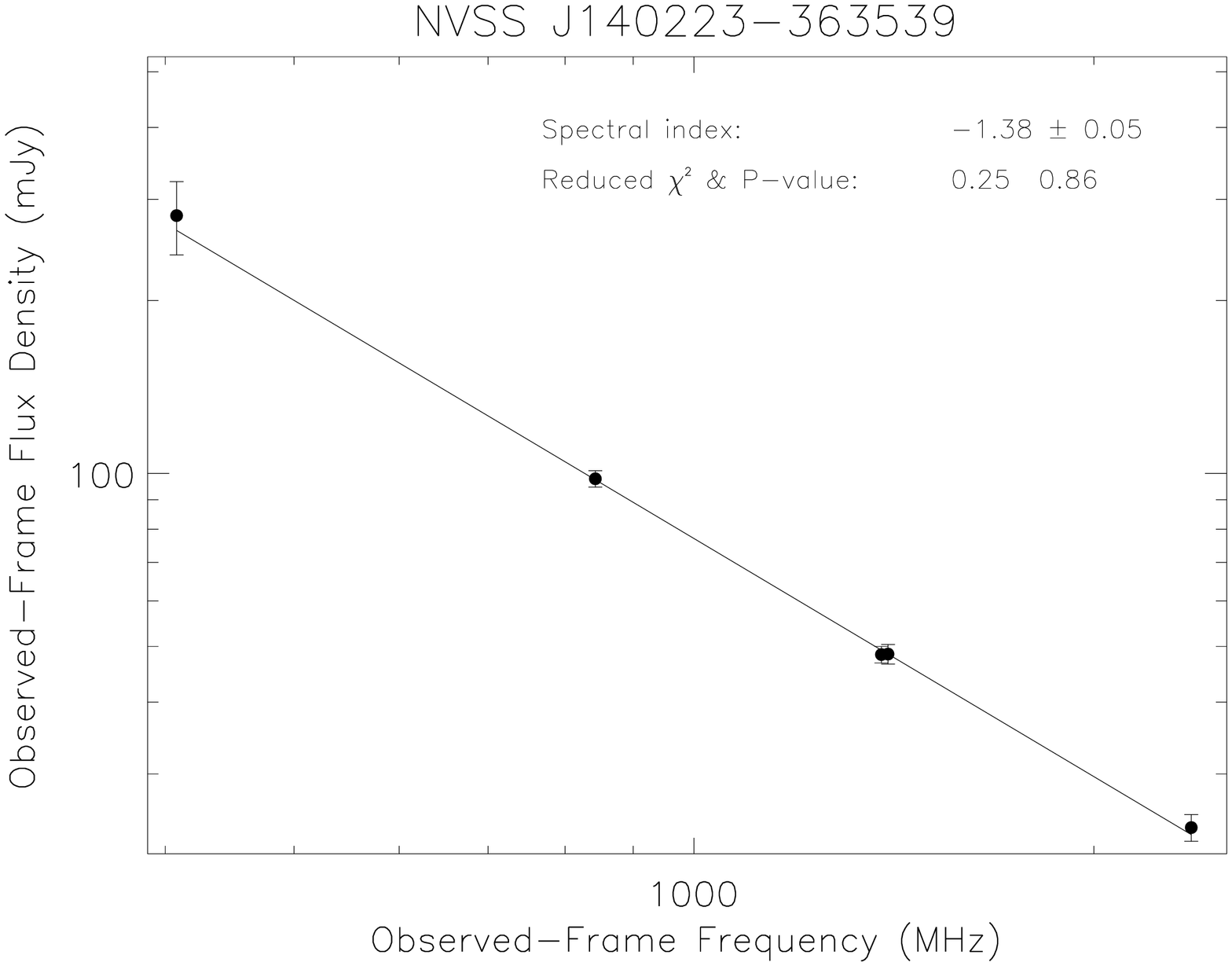,height=7.00cm,angle=90}
\caption{Observed-frame radio SEDs for a selection of sources in the MRCR--SUMSS sample. The five-point spectral index and goodness-of-fit statistics are stated for each SED.}\label{408_paper1_fig:SED_examples}
\end{minipage}
\end{figure*}

We also calculated observed-frame spectral indices based on a fit to all five data points in the $\log(\nu)$--$\log(S_{\nu})$ plane (Fig.~\ref{408_paper1_fig:SED_examples}). Fits were rejected at the 0.01 significance level. First, we attempted to fit each SED with a single power law using a least-squares linear fit with inverse-variance weighting. 198 sources (85 per cent) were found to be well described by a single power law; the statistical properties of these spectral indices are listed in Table~\ref{408_paper1_table:alpha_properties}. For the remaining sources, we then fit each SED with a second-order polynomial; four sources were found to flatten at higher frequencies, and four to steepen (see Table 3). The remaining 25 anomalous SEDs can not be fitted with a higher-order polynomial; the best fits are either linear or quadratic, as before. In these cases, the poor goodness-of-fit statistics are likely due to one or more incorrect or variable flux densities.

\begin{table}
\setlength{\tabcolsep}{4.25pt}
\caption{Observed-frame spectral index properties of the MRCR--SUMSS sample. The labelling of statistical quantities follows the same convention as in Table~\ref{408_paper1_table:flux_properties}; in addition, $\overline{\alpha}$ is the mean spectral index and SEM is the standard error of the mean.}\label{408_paper1_table:alpha_properties}
\begin{tabular}{|c|r|r|r|r|r|}
\hline
\hline
\multicolumn{1}{|c|}{$\alpha$} & \multicolumn{1}{|c|}{$\widetilde{\alpha}$} & \multicolumn{1}{|c|}{$\overline{\alpha}$} & \multicolumn{1}{|c|}{SEM} & \multicolumn{1}{|c|}{$\alpha$ (min.)} & \multicolumn{1}{|c|}{$\alpha$ (max.)} \\
\hline
$\alpha^{843}_{408}$ & $-1.18$ & $-1.21$ & 0.02 & $-2.28$ & $-0.53$ \\
\\
$\alpha^{1400}_{843}$ & $-1.03$ & $-1.03$ & 0.02 & $-1.78$ & $-0.35$ \\
\\
$\alpha^{1400}_{408}$ & $-1.12$ & $-1.14$ & 0.01 & $-1.69$ & $-0.78$ \\
\\
$\alpha^{2368}_{1384}$ & $-1.07$ & $-1.09$ & 0.01 & $-2.03$ & $-0.70$ \\
\\
$\alpha$ (5-point fit,  & $-1.11$ & $-1.12$ & 0.01 & $-1.83$ & $-0.76$ \\
single power law) & & & & & \\
\hline
\end{tabular}
\end{table}

\begin{figure}
\psfig{file=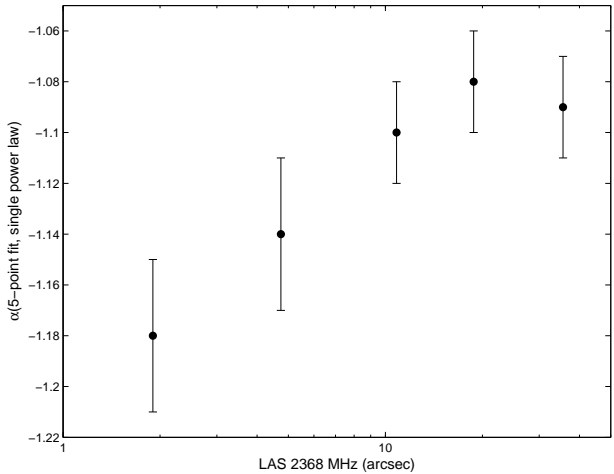,height=6.25cm}
\caption{Group means of single power law spectral indices plotted as a function of LAS. Error bars represent the standard error of the mean. Note that in the lowest LAS bin, we assume that sources with LAS upper limits have a LAS of 2 arcsec when calculating the group mean.}\label{408_paper1_fig:LAS_alpha}
\end{figure}

We now investigate the role of radio morphology in our USS selection procedure. In Fig.~\ref{408_paper1_fig:LAS_alpha}, we have divided the MRCR--SUMSS sample into five bins, and plotted the group means of the single power law spectral indices as a function of LAS. We find that, on average, $\alpha$ steepens slightly with decreasing angular size: the spectral index group mean decreases by $\sim$0.1 over the LAS range covered by our sample. A Spearman rank correlation test reveals that the correlation is significant at a confidence level $> 99.99$ per cent. This may suggest that, on average, the smallest sources are located in the densest environments; these sources have the steepest spectra because the radio lobes are pressure-confined and lose their energy slowly via synchrotron and inverse Compton mechanisms \citep[e.g.][]{klamer06}. Indeed, the sources in the smallest LAS bin in Fig.~\ref{408_paper1_fig:LAS_alpha} will consist mainly of CSS sources, in which the radio jets are confined to the interstellar medium of the host galaxy. An alternative explanation is that the smallest sources are younger and hence more luminous, resulting in a steeper electron energy distribution being injected into the lobes \citep{blundell99}. We discuss this result in more detail in Section~\ref{408_paper_1_uss_environments}.

\subsection{Linear polarization}\label{408_paper1_frac_pol}

\begin{figure*}
\begin{minipage}{175mm}
\psfig{file=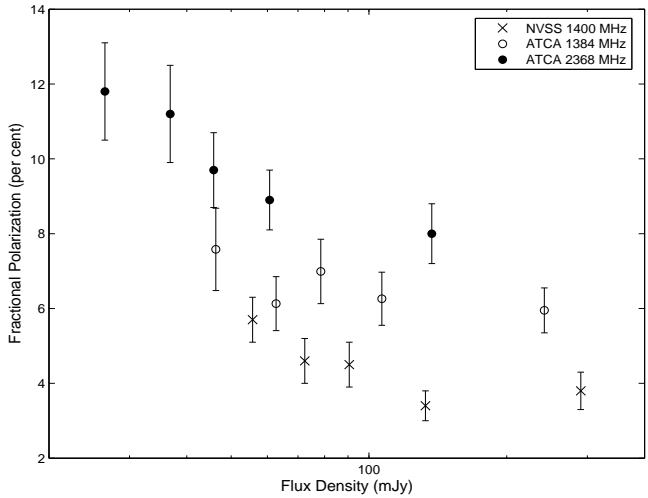,height=6.55cm,width=8.5cm}
\hspace{0.5cm}
\psfig{file=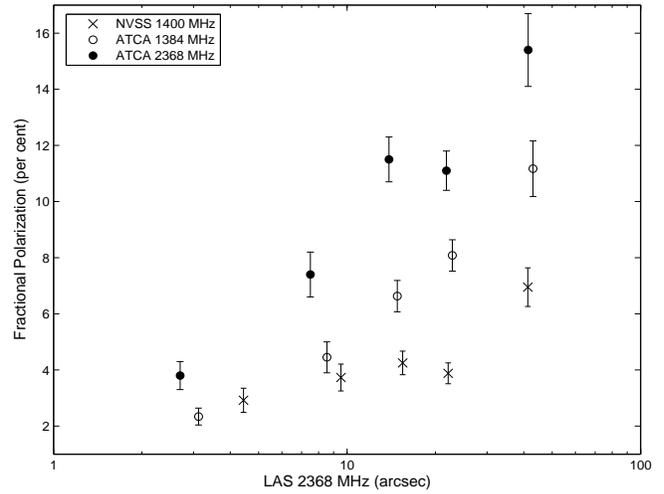,height=6.5cm,width=8.5cm}
\caption{Fractional polarization group means plotted as a function of flux density (left panel) and LAS (right panel). Error bars represent the standard error of the mean. As in Fig.~\ref{408_paper1_fig:LAS_alpha}, we assume that sources with LAS upper limits have a LAS of 2 arcsec when calculating group means. }\label{408_paper1_fig:pol_correlations}
\end{minipage}
\end{figure*}

The fractional linear polarization $m$ is given by

\begin{equation}
m = \frac{\sqrt{S_{Q}^2+S_{U}^2}}{S_{I}},
\end{equation}
and is a measure of how well ordered the magnetic field is in the emitting regions. Taking advantage of the ATCA's multi-channel continuum mode, we attempted to minimize the effects of bandwidth depolarization \citep[e.g.][]{gardner66} by dividing our effective bandwidth of 104 MHz into four approximately equally-spaced channels (centred at 1344, 1368, 1392 and 1424 MHz for the 1384 MHz bandpass, and 2332, 2360, 2384 and 2408 MHz for the 2368 MHz bandpass) and averaging the resultant images. To maximize the signal-to-noise ratio (S/N), the fractional polarization was measured at the positions of peak intensity in each source; if the S/N was too low to use this approach, we calculated a 3$\sigma$ upper limit. We were able to obtain detections above $3\sigma$ in $\sim$80 per cent of the sources at both 1384 and 2368 MHz.

The statistical properties of the ATCA and NVSS fractional polarization measurements are presented in Table~\ref{408_paper1_table:pol_properties}. Note that to increase the reliability of our statistics, we only consider NVSS fractional polarization values above $3\sigma$; these occur in approximately half of the sample. In addition, we do not include source components with upper limits at 1384 and 2368 MHz. The median relative uncertainties are 10, 17 and 14 per cent at 1384, 1400 and 2368 MHz, respectively. The differences between the 20 cm ATCA and NVSS statistical properties are due to beam depolarization in NVSS.

In their study of a flux-limited sample with $S_{1400} > 80$ mJy, \citet{mesa02} used data from NVSS to show that fractional polarization is anti-correlated with flux density for steep-spectrum ($\alpha < -0.5$) sources, possibly due to a different source population emerging at lower flux densities. \citet{tucci04} reported a similar effect in a flux-limited sample with $S_{1400} \geq 100$ mJy. In the left panel of Fig.~\ref{408_paper1_fig:pol_correlations}, we show fractional polarization group means as a function of flux density at 1384, 1400 and 2368 MHz. ATCA group means were calculated using the maximum fractional polarization measured in a source component in each of the images (excluding components with upper limits). As in the \citet{mesa02} and \citet{tucci04} studies, we find that fractional polarization and flux density are anti-correlated in NVSS; a Spearman rank correlation test shows that the relationship is significant at a confidence level $> 99.99$ per cent. In addition, a significant anti-correlation is present at 2368 MHz, though at a slightly lower confidence level (97.3 per cent). Interestingly, however, there is no evidence of an anti-correlation at 1384 MHz. Whether the anti-correlation at 2368 MHz is genuine remains unclear: as the source components become fainter, we can only measure progressively higher fractional polarization values at the 3$\sigma$ level. Higher S/N measurements are needed to clarify this issue.

\begin{table}
\centering
\setlength{\tabcolsep}{4.25pt}
\caption{Fractional polarization properties of the MRCR--SUMSS sample. The labelling of statistical quantities follows the same convention as in Tables~\ref{408_paper1_table:flux_properties} and \ref{408_paper1_table:alpha_properties}.}\label{408_paper1_table:pol_properties}
\begin{tabular}{|c|c|c|c|}
\hline
\hline
& \multicolumn{1}{|c|}{$m_{\rm ATCA, 1384\:MHz}$} & \multicolumn{1}{|c|}{$m_{\rm NVSS}$} & \multicolumn{1}{|c|}{$m_{\rm ATCA, 2368\:MHz}$} \\
& \multicolumn{1}{|c|}{(per cent)} & \multicolumn{1}{|c|}{(per cent)} & \multicolumn{1}{|c|}{(per cent)} \\
\hline
$\widetilde{m}$ & 5.2 & 3.5 & 8.1 \\
\\
$\overline{m}$ & 6.2 & 4.4 & 9.5  \\
\\
SEM & 0.3 & 0.3 & 0.4 \\
\\
$m$ (min.) & 0.2 & 0.9 & 0.5 \\
\\
$m$ (max.) & 29.2 & 12.4 & 40.1 \\
\hline
\end{tabular}
\end{table}

In the right panel of Fig.~\ref{408_paper1_fig:pol_correlations}, we show fractional polarization group means as a function of LAS at 1384, 1400 and 2368 MHz. The fractional polarization is observed to be correlated with LAS at all three frequencies, with the confidence level ranging from 98.7 per cent in NVSS to $> 99.99$ per cent with the ATCA data. This will largely be a consequence of beam depolarization for the NVSS data, and in the smallest LAS bins at both 1384 and 2368 MHz. However, the effect is still clearly present for largest angular sizes greater than the resolution at both 1384 and 2368 MHz, where we can measure the fractional polarization in each of the individual lobes. This may be a further indication that, on average, the smaller sources in the sample are embedded in denser environments, in which a dense cocoon of gas causes substantial depolarization. Another possibility is that if the smaller sources are more luminous (see Section~\ref{408_paper1_spectral_indices}), then fractional polarization and luminosity may be anti-correlated. Note that we can rule out a link between the left and right panels in Fig.~\ref{408_paper1_fig:pol_correlations}, as we do not find any correlation between flux density and LAS. To gain further insight into the polarization properties, higher-resolution radio data are necessary to accurately determine the fractional polarization of candidate HzRGs with small angular sizes.

\begin{figure*}
\begin{minipage}{150mm}
\psfig{file=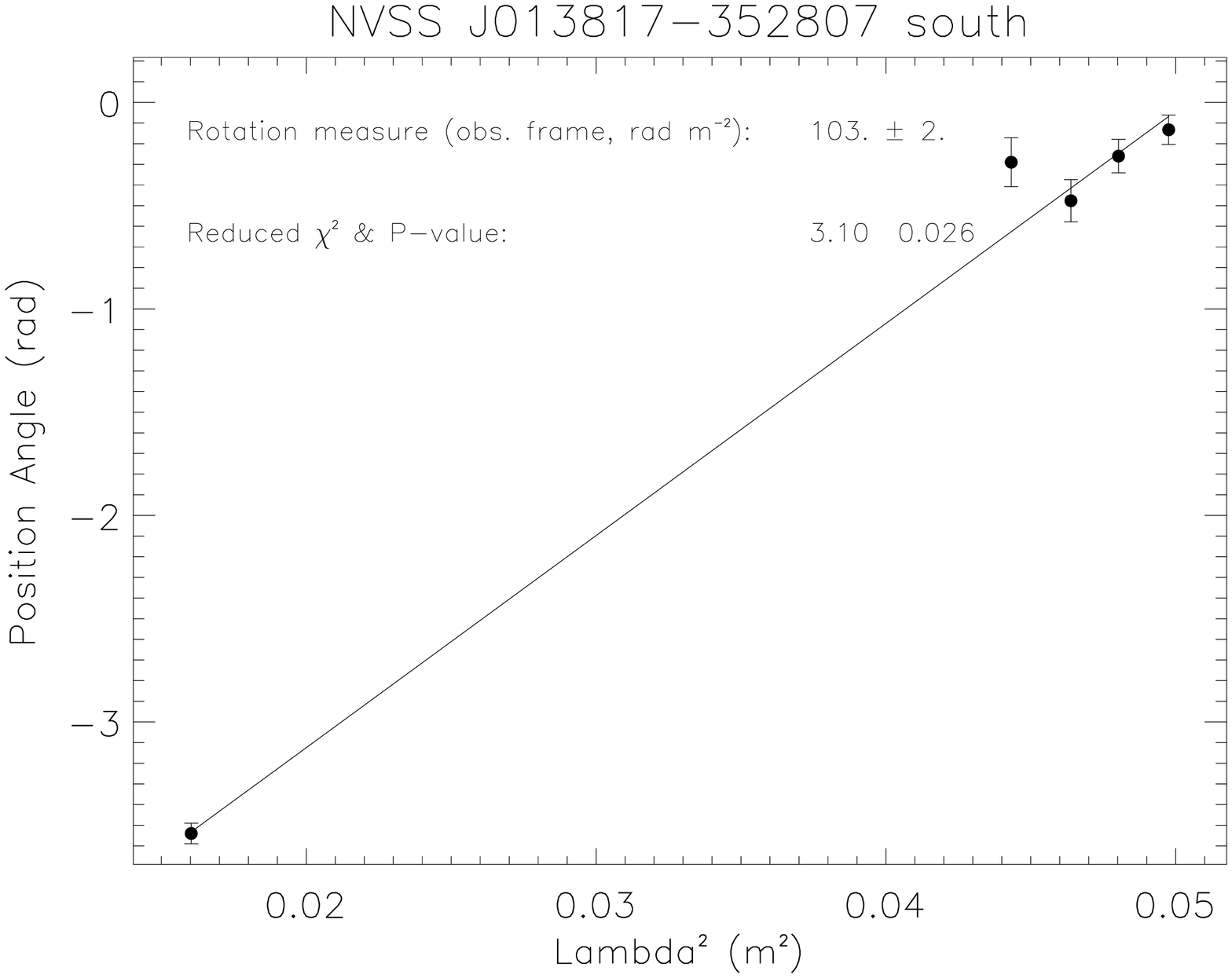,height=7.00cm,angle=90}
\hspace{0.75cm}
\psfig{file=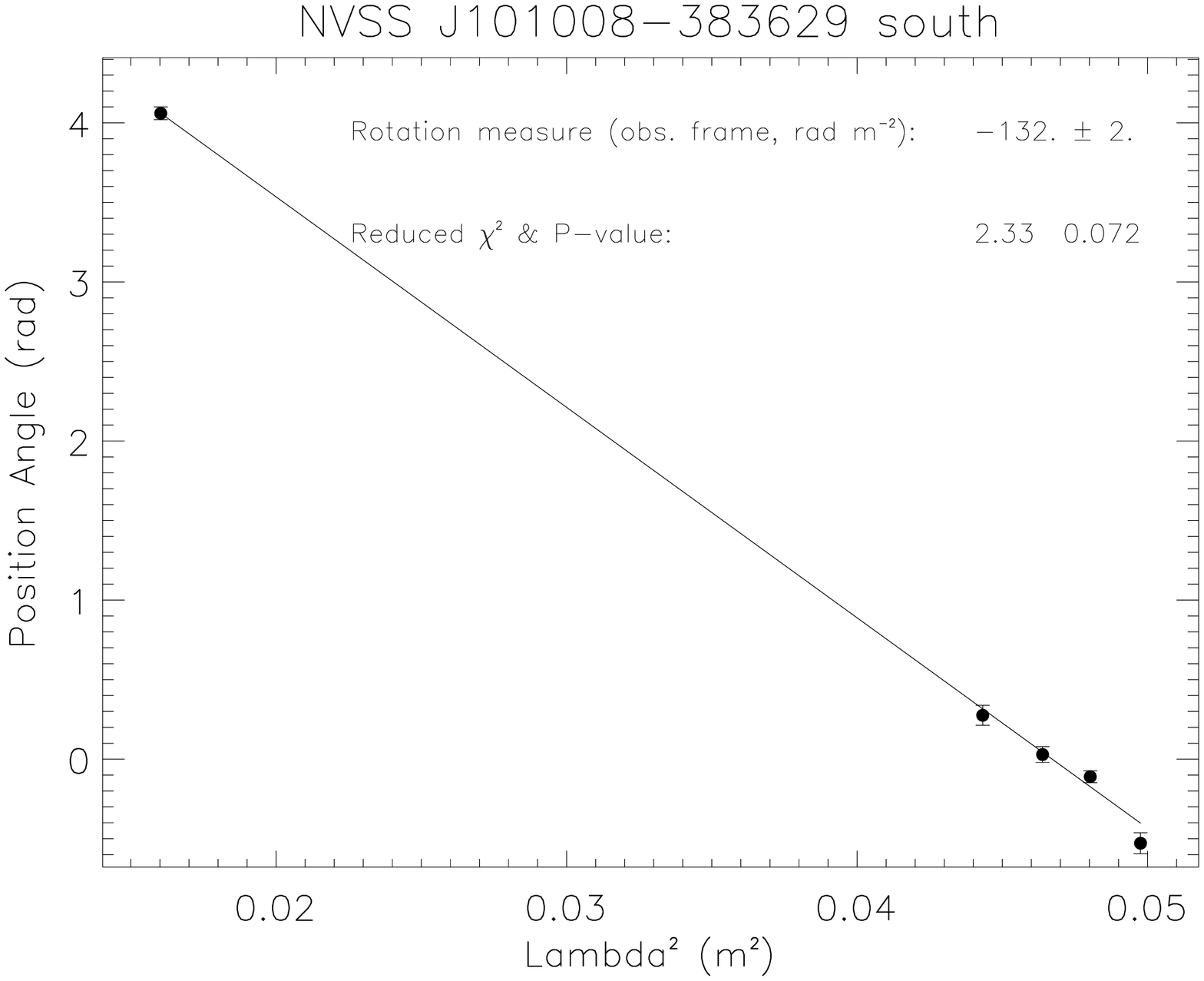,height=7.00cm,angle=90}
\newline
\newline
\newline
\psfig{file=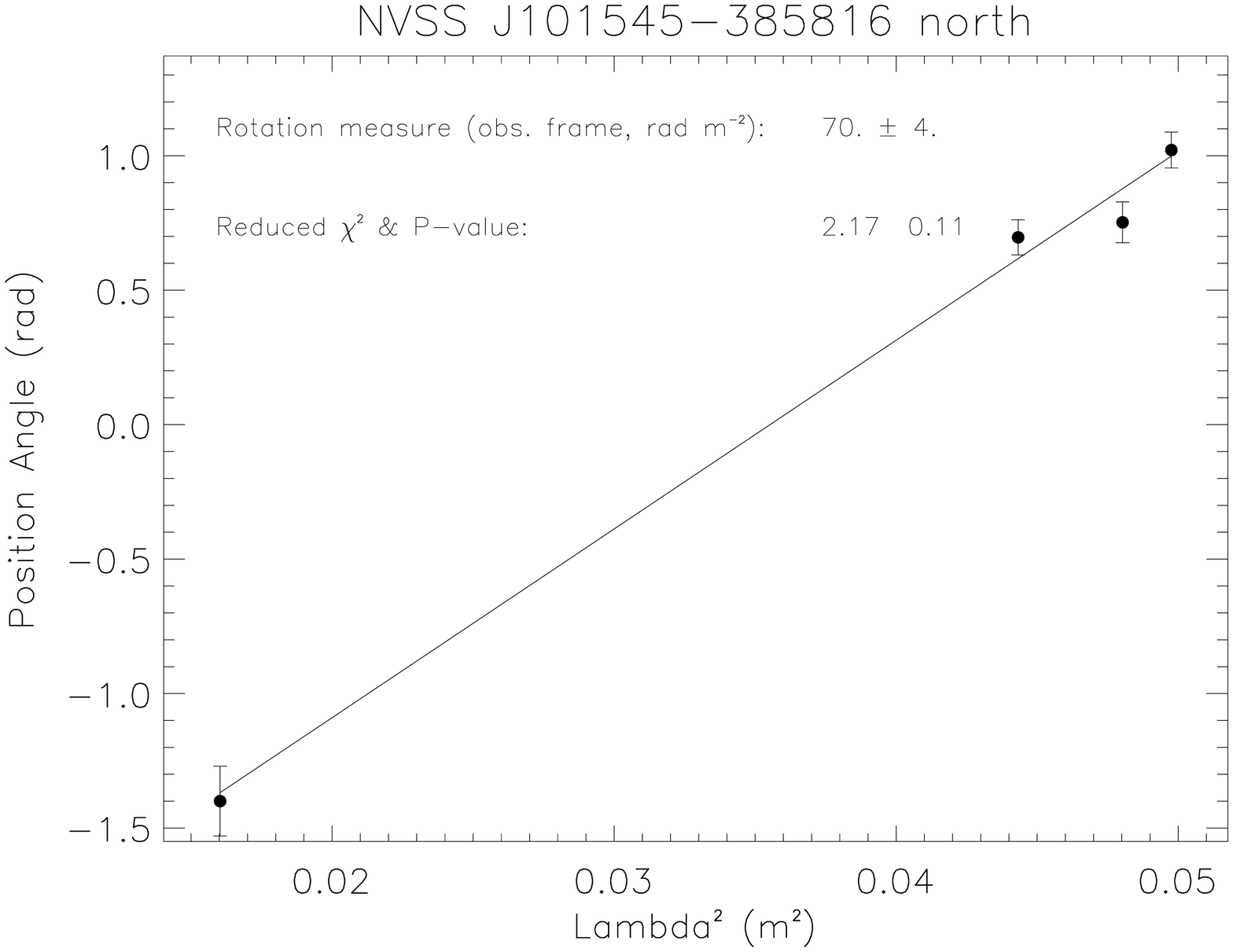,height=7.00cm,angle=90}
\hspace{0.75cm}
\psfig{file=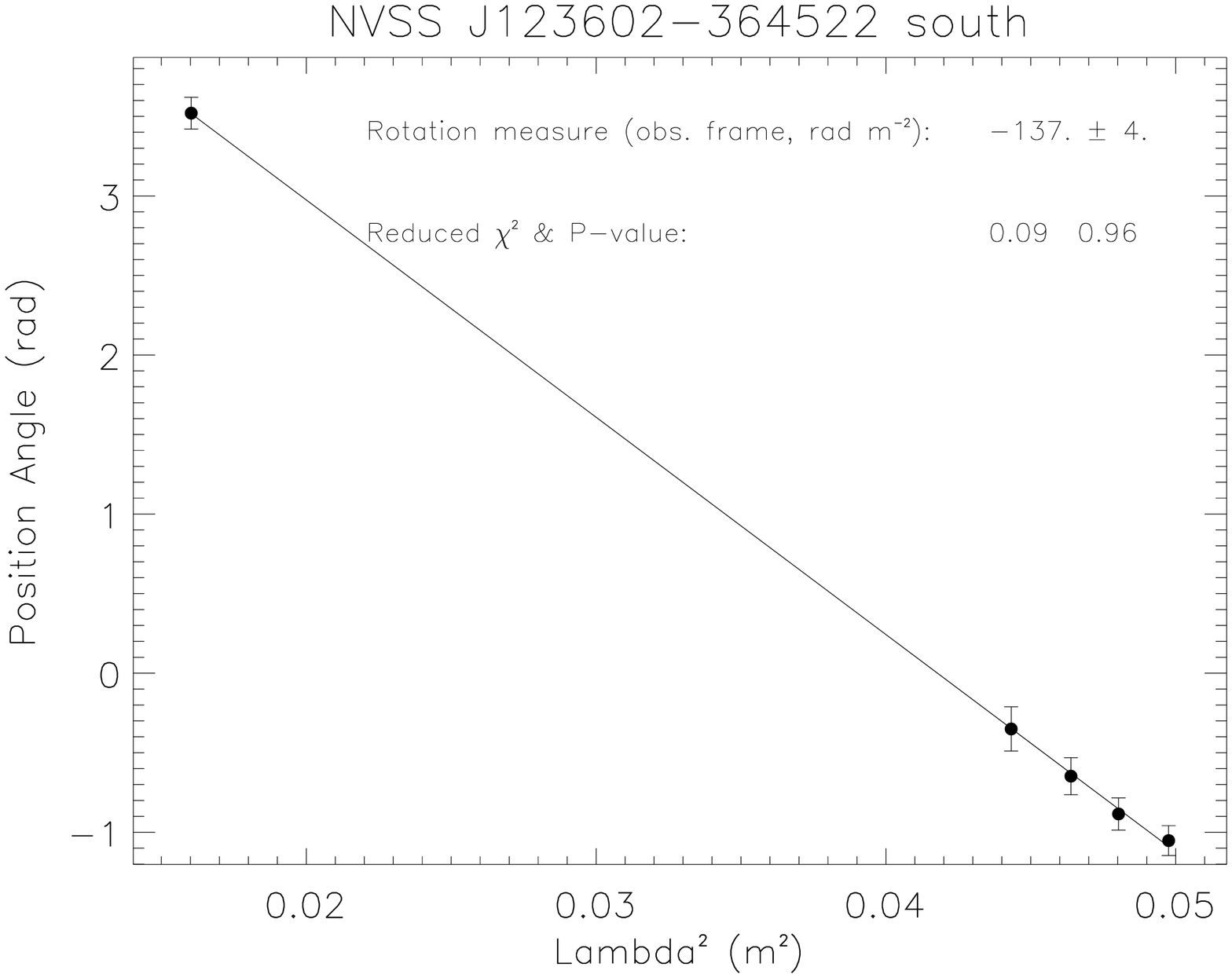,height=7.00cm,angle=90}
\caption{Examples of polarization position angle versus $\lambda^2$ for a selection of sources in the MRCR--SUMSS sample. The observed-frame RM (not corrected for the Galactic Faraday screen) and goodness-of-fit statistics are stated in each panel.}\label{408_paper1_fig:RM_examples}
\end{minipage}
\end{figure*}

\citet{mesa02} found that steep-spectrum sources have a median NVSS fractional polarization of 2.2 per cent. Using the NVSS polarization data for the MRCR--SUMSS sample, we now examine whether the fractional polarization properties of the ultra-steep spectrum population are consistent with the properties of the steep-spectrum population as a whole. As \citet{mesa02} do not explicitly state whether they have imposed a $3\sigma$ detection limit, we consider all 107 sources with $S_{1400} > 80$ mJy for which we have NVSS fractional polarization values. We derive a median NVSS fractional polarization of 2.1 per cent, which is similar to the value determined by \citet{mesa02} for the entire steep-spectrum population. This result is intriguing, as we might have expected the sources with ultra-steep spectra to be less polarized if they indeed reside in the densest environments (see Section~\ref{408_paper1_spectral_indices}). Furthermore, we find no correlation between fractional polarization and spectral index in the MRCR--SUMSS sample itself. Higher-frequency observations are needed to investigate the depolarization properties of the MRCR--SUMSS sample.

\subsection{Rotation measures}\label{408_paper_1_RMs}

As a linearly polarized wave propagates through a magnetized plasma, the plane of polarization is subject to rotation.  This phenomenon is described by the Faraday rotation measure RM, measured in rad m$^{-2}$. The RM is determined from the linear relationship between polarization position angle ($\Phi$) and observing wavelength squared ($\lambda^2$):
\begin{equation}
\Phi = \Phi_{0} + \rm RM\lambda^2,
\end{equation}
where $\Phi_{0}$ is the intrinsic polarization position angle. The polarization position angle, measured from the north to the east, is defined by
\begin{equation}
\Phi = \frac{1}{2}\tan^{-1}\left(\frac{S_{U}}{S_{Q}}\right).
\end{equation}
We used our ATCA data to calculate RMs at the positions of peak intensity. A 1.5$\sigma$ cutoff in the individual $Q$ and $U$ images was adopted when evaluating $\Phi$. We found that the n$\pi$ ambiguity in $\Phi$ was best solved by dividing the 1384 MHz bandpass into four bins (centred at 1344, 1368, 1392 and 1424 MHz), as was done when measuring the fractional polarization. To increase our $\lambda^2$ baseline, we also included a single 2368 MHz data point. Position angle values were averaged over equivalent areas at each frequency (typically a $3 \times 3$ pixel box in the 1384 MHz bandpass and a $5 \times 5$ pixel box at 2368 MHz). For the small number of sources with RMs that are also blended at 1384 MHz, we averaged the position angles over the individual source components at 2368 MHz. We only calculated an RM in a source component if there were at least three data points in the 1384 MHz bandpass with sufficient S/N that could be combined with the 2368 MHz data point. This was necessary as the S/N is too low to allow a reliable RM determination in the 1384 MHz bandpass alone. In total, we obtained RMs for 118 sources in the MRCR--SUMSS sample. Fig.~\ref{408_paper1_fig:RM_examples} shows $\Phi$ versus $\lambda^2$ and the corresponding least-squares linear fit with inverse-variance weighting for a selection of sources.

\begin{table*}
\begin{minipage}{175mm}
\scriptsize
\setlength{\tabcolsep}{3.25pt}
\centering
\caption{Selection criteria and surface densities of recent USS samples.}\label{408_paper1_table:surface_densities}
\begin{tabular}{|c|c|c|c|c|c|c|c|}
\hline
\hline
Sample & Flux density & Spectral index & Angular size & Sources & Area & Surface density & Reference \\
& cutoff (mJy) & cutoff & cutoff (arcsec) & & (sr) & (sr$^{-1}$) & \\
\hline
WENSS--NVSS & $S_{1400} \geq 10$  & $\alpha^{1400}_{325} \leq -1.3$ & $\cdots$ & 343 & 2.27 & 151 & \citet{debreuck00} \\
\\
TEXAS--NVSS & $S_{1400} \geq 10$  & $\alpha^{1400}_{365} \leq -1.3$ & $\cdots$ & 268 & 5.58 & 48 & \citet{debreuck00} \\
\\
MRC--PMN & $S_{408} \geq 700$, $S_{4850} \geq 35$  & $\alpha^{4800}_{408} \leq -1.2$ & $\cdots$ & 58 & 2.23 & 26 & \citet{debreuck00} \\
\\
WISH--NVSS & $S_{1400} \geq 10$  & $\alpha^{1400}_{352} \leq -1.3$ & $\cdots$ & 154 & 1.60 & 96 & \citet{debreuck02b} \\
\\
6C* & $960 \leq S_{151} \leq 2000$ & $\alpha^{4850}_{151} \leq -0.981$ & $<$ 15 & 29 & 0.133 & 218  & \citet{blundell98} \\
\\
6C** & $S_{151} \geq 500$ & $\alpha^{1400}_{151} \leq -1.0$ & $<$ 13 & 68 & 0.421 & 162 & \citet{cruz06} \\
\\
VLA (74 MHz)--NVSS & $S_{74} \geq 100$ & $\alpha^{1400}_{74} \leq -1.2$ & $\cdots$ & 26 & 0.05 & 520 & \citet{cohen04} \\
\\
SUMSS--NVSS & $S_{1400} \geq 15$ & $\alpha^{1400}_{843} \leq -1.3$ & $\cdots$ & 53 & 0.11 & 482 & \citet{debreuck04} \\
\\
MRCR--SUMSS & $S_{408} \geq 200$ & $\alpha^{843}_{408} \leq -1.0$ & $\cdots$ & 202 & 0.35 & 577 & This paper \\
\\
MRCR--SUMSS & $S_{408} \geq 200$ & $\alpha^{843}_{408} \leq -1.3$ & $\cdots$ & 67 & 0.35 & 191 & This paper \\
\hline
\end{tabular}
\end{minipage}
\end{table*}

If we assume that the Faraday screen producing the RM is located at the source redshift $z$, then the intrinsic RM, $\rm RM_{\rm intr}$, is related to the observed RM, $\rm RM_{\rm obs}$, by

\begin{equation}\label{408_paper1_equation:RM}
{\rm RM}_{\rm intr} = ({\rm RM}_{\rm obs} - {\rm RM}_{\rm gal}) \times (1+z)^2,
\end{equation}
where $\rm RM_{\rm gal}$ is the contribution from the Galactic Faraday screen. In Table 4, we give $\rm RM_{\rm gal}$ for each source with a measured RM using the all-sky RM map of \citet{hollitt04}; the corrections range from 0 to 63 rad m$^{-2}$ in magnitude. After correcting for RM$_{\rm gal}$, the magnitudes of the observed RMs range from 0 to 165 rad m$^{-2}$. On average, the individual uncertainties in the RMs are small ($\overline{\Delta \rm RM} = 2$ rad m$^{-2}$), due to the relatively large baseline between the 1384 and 2368 MHz bandpasses.

The intrinsic RM can also be expressed as

\begin{equation}\label{408_paper1_equation:RM2}
{\rm RM_{intr}} = 812 \int^{L}_{0} n_{\rm e} \bmath{B} \cdot d\bmath{l},
\end{equation}
where $L$ is the path length in kpc, $n_{\rm e}$ is the thermal electron density in cm$^{-3}$, and $\bmath{B}$ is the magnetic field strength in $\mu$G. The integral is taken along the line of sight. Extreme rest-frame RMs ($\ga 1000$ rad m$^{-2}$) have been observed in high-resolution, high-frequency investigations of HzRGs with $z \ga 2$ with both the VLA \citep{carilli97,athreya98,pentericci00} and the ATCA \citep{broderick07}. Such high RMs have been used as evidence to suggest that these HzRGs are situated in very dense environments, analogous to rich clusters at low redshift \citep[][and references therein]{carilli02}. A detailed analysis of the rest-frame RM distribution of the MRCR--SUMSS sample is given in Paper II.

\section{Discussion}\label{408_paper1_discussion}

\subsection{Surface density of the MRCR--SUMSS sample}\label{408_paper1_surface_density}

To assess the effectiveness of our selection criteria, we now compare the surface density of the MRCR--SUMSS sample with the surface densities of other USS samples in the literature (Table~\ref{408_paper1_table:surface_densities}). Such comparisons are affected by different selection criteria and various forms of incompleteness. While it is clear that the number of sources in a given USS sample will be strongly dependent on the adopted cutoffs in spectral index, angular size and flux density, the surface density is also influenced by the resolution of the catalogues from which the sample was drawn, and the fraction of sources that scatter in and out of the sample because of the uncertainties in the spectral indices themselves. For example, more sources will be matched to a counterpart in another catalogue if the catalogues have similar resolution, and spectral indices obtained over wide frequency baselines typically have smaller errors. Considering only the 205 sources with $\alpha^{843}_{408} \leq -1.0$ after revision of the flux densities (see Section~\ref{408_paper1_sample_definition}), and excluding the three low-redshift FR I sources discussed in Section~\ref {408_paper1_morphologies}, we derive a surface density of 577 sr$^{-1}$ for the MRCR--SUMSS sample. While this surface density is the highest in Table~\ref{408_paper1_table:surface_densities}, this is because the spectral index cutoff in the MRCR--SUMSS sample is flatter than in most of the other samples, resulting in a sharp increase in the number of sources in the steep tail of the spectral index distribution that satisfy the cutoff (Fig.~\ref{408_paper1_fig:alpha_distribution}). If we were to restrict the MRCR--SUMSS sample to $\alpha^{843}_{408} \leq -1.3$, then we only include 67 sources with a corresponding surface density of 191 sr$^{-1}$. This is $\sim$2.5 times lower than the SUMSS--NVSS surface density, but $\sim$1.3 times higher than the surface density of the WENSS--NVSS sample \citep{debreuck00}.

Assuming a spectral index cutoff of $-1.3$, there are two main reasons why the surface density of the MRCR--SUMSS sample is lower than in the SUMSS--NVSS sample. First, the sources in the SUMSS--NVSS sample are fainter (see Section~\ref{408_paper1_fluxes}). Second, the resolution of SUMSS and NVSS are much better matched than the MRCR and SUMSS, and so fewer sources are excluded due to confusion. In Paper II, we establish the true effectiveness of our selection criteria by examining the $K$-band magnitude distribution of our sample.

\subsection{Reliability of the USS cutoff}\label{408_paper_1_cutoff_reliability}

\begin{figure}
\psfig{file=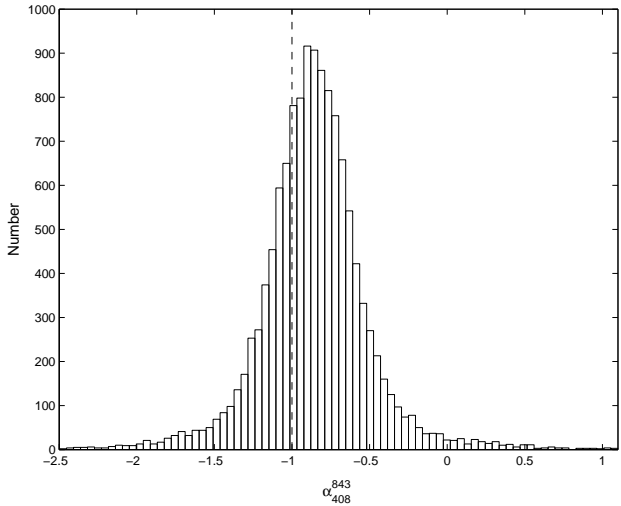,height=6.66cm}
\caption{Distribution of two-point spectral indices between the MRCR and SUMSS in the overlap region $\delta < -30\degr$. The dashed line shows our spectral index cutoff $\alpha^{843}_{408} = -1.0$.}\label{408_paper1_fig:alpha_distribution}
\end{figure}

We now estimate how the MRCR--SUMSS surface density is affected by the number of sources that scatter in and out of the sample due to spectral index uncertainties. First, the uncertainties in $\alpha^{843}_{408}$ result in a net excess of genuine $\alpha^{843}_{408} > -1.0$ sources scattering into the sample because our cutoff is on the steep tail of the spectral index distribution. To assess the magnitude of this effect, we follow a similar approach to \citet{debreuck00} and \citet{debreuck04} by compiling a random sample from the overall distribution of $\alpha^{843}_{408}$, which is then convolved with a Gaussian distribution of spectral index uncertainties with standard deviation equal to the mean uncertainty in $\alpha^{843}_{408}$ ($\overline{\Delta\alpha^{843}_{408}} = 0.15$). The distribution of spectral indices between 408 and 843 MHz in the overlap region $\delta < -30\degr$ (Fig.~\ref{408_paper1_fig:alpha_distribution}) has a mean spectral index $\overline{\alpha} = -0.86$ and standard deviation $\sigma_{\alpha} = 0.24$. Using this distribution, we expect that in a sample of 202 sources with $\alpha^{843}_{408} \leq -1.0$, there will be an excess of $\sim$20 sources ($\sim$10 per cent) that are scattered into the sample. This excess is far less than the expected surplus of sources that scatter into the SUMSS--NVSS sample \citep[35 per cent,][]{debreuck04}, despite the SUMSS--NVSS sample having a similar parent spectral index distribution and mean spectral index uncertainty \citep[$\overline{\alpha}=-0.82$, $\sigma_{\alpha} = 0.25$, $\overline{\Delta\alpha^{1400}_{843}}=0.12$;][]{debreuck04}. This behaviour occurs because there is a greater percentage increase in the area under the spectral index distribution function (i.e. the probability) for $\alpha \leq -1.3$ than for $\alpha \leq -1.0$ after the spectral index uncertainties are taken into account. Thus, while there are fewer sources in the distribution function with $\alpha \leq -1.3$, the fraction of sources that scatter into the sample is greater.

To gain further insight into the reliability of our spectral index cutoff, we have compared the values of $\alpha^{843}_{408}$ with the spectral indices derived from our five-point SEDs (see Section~\ref{408_paper1_spectral_indices}). The five-point spectral indices are ideal for this task as they give a more accurate picture of the true spectral index over a wider frequency range. Of the 198 sources with single power law SEDs, 159 have $\alpha$(5-point) $ \leq -1.0$. While it is clear that the spectral indices of the eight flattening or steepening sources are a strong function of frequency, we find that three flattening sources (NVSS J002431$-$303330, NVSS J150405$-$394733 and NVSS J233238$-$323537) are not ultra-steep over the range 843--2368 MHz. This may result from either the presence of a core component (NVSS J002431$-$303330 and NVSS J233238$-$323537 have a single-component morphology at 2368 MHz, while NVSS J150405$-$394733 is resolved into a triple at 2368 MHz), or an overestimated 408 MHz flux density. Nine of the remaining 25 sources with invalid five-point spectral indices have $\alpha^{1400}_{843}$ and $\alpha^{2368}_{1384} > -1.0$, which may indicate that they are not ultra-steep spectrum sources. Thus, in summary, we estimate that $\sim$80 per cent of the sources in the MRCR--SUMSS sample have a true spectral index steeper than $-1.0$. As there is significant scatter in the $z$--$\alpha$ correlation, we do not expect the fraction of high-redshift radio galaxies in our sample to be significantly affected.

\subsection{Environments of USS sources}\label{408_paper_1_uss_environments}

In this section, we discuss the radio spectral properties of the MRCR--SUMSS sample in relation to the environments in which the sources reside. The lack of spectral steepening seen in the vast majority of the SEDs is in close agreement with \citet{klamer06}, who found that in a subsample of 37 sources from the SUMSS--NVSS sample, 33 (89 per cent) have SEDs that are straight between 843 MHz and 18 GHz in the observed-frame, while the remainder flatten at higher frequencies. As discussed by \citet{klamer06}, the absence of spectral steepening is inconsistent with the conventional explanation that the $z$--$\alpha$ correlation results from a $k$-corrected concave radio spectrum that is further steepened at high redshift due to increased inverse Compton scattering off the cosmic microwave background \citep{krolik91}. Instead, \citet{klamer06} postulate that the $z$--$\alpha$ correlation is a consequence of source environment, driven by an increase in the fraction of radio galaxies residing in regions of high ambient density at high redshift.

If the smallest sources in the MRCR--SUMSS sample are, on average, at higher redshifts (as is suggested by Paper II), and the correlations discussed in Sections~\ref{408_paper1_spectral_indices} and \ref{408_paper1_frac_pol} are a consequence of source environment, then our results are consistent with the claim of \citet{klamer06}. However, as our sample is flux-limited, we cannot rule out an underlying luminosity dependence, since the highest-redshift sources will also be more luminous as a result of Malmquist bias. In Paper II, we use our spectral index and polarization data in conjunction with spectroscopic redshifts and redshifts estimated from the $K$--$z$ diagram to test the hypothesis of \citet{klamer06} in greater detail.

\subsection{Low-frequency spectral curvature}

\begin{table*}
\setlength{\tabcolsep}{3.25pt}
\scriptsize
\centering
\begin{minipage}{175mm}
\caption{74 MHz flux densities and spectral indices of sources detected in the VLSS. The six-point fit is derived using the flux density values in Table 3 plus the 74 MHz flux density. As in Table 3, we give the observed-frame spectral index at 1400 MHz for those sources that have flattening spectra $(*)$ at low frequency. }\label{408_paper1_table:VLSS}
\begin{tabular}{|l|r|r|r|c|c|c|l|r|r|r|}
\hline
\hline
\multicolumn{1}{|c|}{Source}  & \multicolumn{1}{|c|}{$S_{74}$} & \multicolumn{1}{|c|}{$\alpha^{408}_{74}$} & \multicolumn{1}{|c|}{$\alpha$} &  & & &  \multicolumn{1}{|c|}{Source} & \multicolumn{1}{|c|}{$S_{74}$} & \multicolumn{1}{|c|}{$\alpha^{408}_{74}$} & \multicolumn{1}{|c|}{$\alpha$} \\
 & \multicolumn{1}{|c|}{(mJy)} & & \multicolumn{1}{|c|}{(6-point fit)} & & & & & \multicolumn{1}{|c|}{(mJy)} & & \multicolumn{1}{|c|}{(6-point fit)} \\
\hline
NVSS        J000231$-$342614    & $ 1350    \pm 190 $ & $   -0.82   \pm 0.12    $ & $   -0.98   \pm 0.04    $ & & & & NVSS        J033540$-$312409    & $ 5620    \pm 680 $ & $   -0.92   \pm 0.09    $ & $   -1.03   \pm 0.03    $   \\
NVSS        J000742$-$304325    & $ 770 \pm 100 $ & $   -0.47   \pm 0.11    $ & $   -1.19 \: (*)    $ & & & & NVSS        J034027$-$331711    & $ 3800    \pm 750 $ & $   -1.02   \pm 0.13    $ & $   -0.98 \pm 0.04    $   \\
NVSS        J001210$-$342103    & $ 1280    \pm 180 $ & $   -0.94   \pm 0.11    $ & $   -1.04   \pm 0.04    $ & & & & NVSS        J120839$-$340307    & $ 22930   \pm 2340    $ & $   -0.93   \pm 0.08    $ & $   -1.10   \pm 0.03    $   \\
NVSS        J001506$-$330155    & $ 2100    \pm 240 $ & $   -0.63   \pm 0.10    $ & $   -0.89   \pm 0.03    $ & & & & NVSS        J141428$-$320637    & $ 3780    \pm 490 $ & $   -0.97   \pm 0.10    $ & $   -1.32 \: \rm (*)     $   \\
NVSS        J004147$-$303658    & $ 1190    \pm 180 $ & $   -0.78   \pm 0.11    $ & $   -0.92   \pm 0.04    $ & & & &  NVSS        J214114$-$332307    & $ 3760    \pm 460 $ & $   -0.87   \pm 0.09    $ & $   -1.25 \: \rm (*)     $   \\
NVSS        J004851$-$324623    & $ 1550    \pm 210 $ & $   -0.79   \pm 0.09    $ & $   -0.94   \pm 0.03    $ & & & & NVSS        J215009$-$341052    & $ 1710    \pm 220 $ & $   -0.76   \pm 0.10    $ & $   -1.19 \: \rm (*)     $   \\
NVSS        J005320$-$322756    & $ 810 \pm 140 $ & $   -0.65   \pm 0.12    $ & $  -1.11 \: \rm (*)     $ & & & & NVSS        J215029$-$310457    & $ 4650    \pm 490 $ & $   -0.88   \pm 0.08    $ & $   -1.53 \: \rm (*)     $   \\
NVSS        J014249$-$310348    & $ 740 \pm 130 $ & $   -0.52   \pm 0.16    $ & $   \cdots    $ & & & & NVSS        J215047$-$343616    & $ 4490    \pm 530 $ & $   -1.07   \pm 0.09    $ & $   -1.05   \pm 0.03    $   \\
NVSS        J014501$-$332648    & $ 1360    \pm 180 $ & $   -0.77   \pm 0.10    $ & $   -0.93 \pm 0.03    $ & & & & NVSS        J215226$-$341606    & $ 1670    \pm 220 $ & $   -0.99   \pm 0.10    $ & $   -0.99   \pm 0.03    $   \\
NVSS        J014918$-$301001    & $ 760 \pm 130 $ & $   -0.63   \pm 0.15    $ & $   -0.87   \pm 0.04    $ & & & & NVSS        J215547$-$344614    & $ 1780    \pm 210 $ & $   -1.00   \pm 0.09    $ & $   -1.07   \pm 0.03    $   \\
NVSS        J015745$-$310557    & $ 1700    \pm 210 $ & $   -0.77   \pm 0.09    $ & $   \cdots    $ & & & & NVSS        J215717$-$313449    & $ 620 \pm 100 $ & $   -0.62   \pm 0.12    $ & $   \cdots    $   \\
NVSS        J015938$-$302236    & $ 1860    \pm 210 $ & $   -0.87   \pm 0.10    $ & $   -1.43 \: \rm (*)    $ & & & & NVSS        J221650$-$341008    & $ 2230    \pm 290 $ & $   -0.85   \pm 0.09    $ & $   -1.00   \pm 0.03    $   \\
NVSS        J021208$-$343111    & $ 1360    \pm 220 $ & $   -0.83   \pm 0.11    $ & $   -1.00 \pm 0.04    $ & & & & NVSS        J224450$-$334326    & $ 1300    \pm 190 $ & $   -0.83   \pm 0.11    $ & $   -0.98   \pm 0.04    $   \\
NVSS        J021759$-$301512    & $ 1560    \pm 170 $ & $   -1.00   \pm 0.09    $ & $   -1.04   \pm 0.03    $ & & & & NVSS        J232007$-$302127    & $ 2110    \pm 300 $ & $   -0.76   \pm 0.10    $ & $   -1.32 \: \rm (*)      $   \\
NVSS        J022825$-$302005    & $ 1160    \pm 150 $ & $   -0.74   \pm 0.11    $ & $   -0.93   \pm 0.03    $ & & & & NVSS        J235104$-$325444    & $ 1370    \pm 160 $ & $   -0.87   \pm 0.09    $ & $   -1.03   \pm 0.03    $   \\
NVSS        J024012$-$305742  $_{_{_{+}}}$   & \multirow{2}{*}{$ 2500    \pm 490 $} & \multirow{2}{*}{$   -0.93   \pm 0.13    $} & \multirow{2}{*}{$ \cdots    $} & & & & NVSS        J235945$-$330354    & $ 3900    \pm 400 $ & $   -1.08   \pm 0.08    $ & $   -1.24   \pm 0.03    $   \\
NVSS        J024016$-$305710    &               & & & &               &               &               &               &               &                        \\
\hline
\end{tabular}
\end{minipage}
\end{table*}

\begin{figure}
\centering
\psfig{file=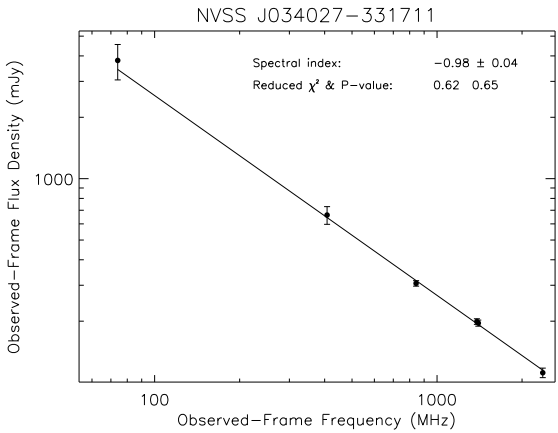,height=7.5cm,angle=90}
\caption{Observed-frame radio SED from 74--2368 MHz for the source NVSS J034027$-$331711. The six-point spectral index and goodness-of-fit statistics are shown.}\label{408_paper1_fig:SED_examples2}
\end{figure}

While a large fraction of the SEDs in the MRCR--SUMSS sample are described by a single power law above 408 MHz, other studies have found that the spectra of most USS sources tend to flatten below our selection frequencies due to the effects of synchrotron self-absorption \citep[e.g.][]{blundell98,debreuck00,bornancini07}. At these frequencies, high-redshift sources should be more easily distinguished from the low-redshift population because they are sampled at higher rest-frame frequencies and therefore exhibit less spectral curvature. Thus, it is thought that USS samples selected at very low frequencies \citep[e.g.][also see Table~\ref{408_paper1_table:surface_densities}]{cohen04} might be more efficient at finding the highest-redshift radio galaxies.

In order to investigate the amount of low-frequency spectral curvature present in the MRCR--SUMSS sample, we used data from the 74 MHz VLA Low-Frequency Sky Survey \citep[VLSS;][]{cohen07}, which has a resolution of $80 \times 80$ arcsec$^{2}$ and a typical rms noise level of $\sim$100 mJy beam$^{-1}$. However, as the southern boundary of the VLSS is at $\delta \approx -30^{\circ}$, our analysis is limited by the small amount of overlap with the MRCR--SUMSS sample. Using the 2007 June 18 VLSS catalogue, we obtained 74 MHz flux densities for 32 sources; these are listed in Table~\ref{408_paper1_table:VLSS}.

Using data from the VLSS, the 352 MHz Westerbork in the Southern Hemisphere Survey \citep[WISH;][]{debreuck02b} and NVSS, \citet{bornancini07} found significant curvature below 352 MHz in a subset of 12 USS sources with $\alpha^{1400}_{352} \leq -1.3$ (WISH--NVSS selection, see Table~\ref{408_paper1_table:surface_densities}): the median values of the two-point spectral indices are $\widetilde{\alpha^{352}_{74}} = -1.1$ and $\widetilde{\alpha^{1400}_{352}} = -1.5$. We can test whether we see the same amount of curvature in the MRCR--SUMSS sample over a similar frequency range by using our 74, 408 and 1400 MHz data. We derive median spectral indices of $\widetilde{\alpha^{408}_{74}} = -0.84$ and $\widetilde{\alpha^{1400}_{408}} = -1.06$; individual $\alpha^{408}_{74}$ values are shown in Table~\ref{408_paper1_table:VLSS}. While the flatter median spectral index between 74 and 408 MHz implies the presence of spectral curvature, the difference between the spectral index medians is about half that found by \citet{bornancini07}. This suggests that low-frequency spectral curvature is not as significant in the MRCR--SUMSS sample.

We now examine whether there is sufficient flattening below 408 MHz such that the SEDs can not be fitted by a single power law. To do this, we computed SED fits similar to those described in Section~\ref{408_paper1_spectral_indices}, but extended each fit to 74 MHz. The spectral indices obtained from these fits are also listed in Table~\ref{408_paper1_table:VLSS}. We find that 20 of the SEDs are still well described by a single power law below 408 MHz (e.g. Fig.~\ref{408_paper1_fig:SED_examples2}). Eight SEDs flatten below 408 MHz; of these, only NVSS J215029$-$310457 previously showed evidence of spectral curvature between 408 and 2368 MHz (see Table 3). The remaining four SEDs have poor goodness-of-fit statistics between 408 and 2368 MHz, and this remains so after each fit is extended to 74 MHz (see discussion in Section~\ref{408_paper1_spectral_indices}). Thus, while the relative fraction of sources with concave spectra has increased with the inclusion of VLSS data (25 per cent compared with 2 per cent in Section~\ref{408_paper1_spectral_indices}), the majority of sources have negligible spectral curvature between 74 and 2368 MHz. This may suggest that the sources with straight spectra are at higher redshift. We investigate this issue further in Paper II.

\section{Conclusions}\label{408_paper1_conclusions}

We have defined a sample of 234 USS sources with $\alpha^{843}_{408} \leq -1.0$ and $S_{408} \geq 200$ mJy, assembled to find distant radio galaxies in the region $-40\degr < \delta < -30\degr$. We have drawn the following conclusions based on our investigation of the radio properties of the sample:

\begin{enumerate}
\item 198 sources (85 per cent) have spectral energy distributions that remain straight from 408 MHz to 2368 MHz in the observed frame. Where 74 MHz data are available, we find that 20 sources (63 per cent) remain well fitted by a single power law.
\item With decreasing angular size, the average spectral index steepens slightly and the average fractional polarization decreases. Both effects are possibly due to the smallest sources residing in the densest environments.
\item The median fractional polarization varies from 3.5 per cent at 1400 MHz to 8.1 per cent at 2368 MHz. The fractional polarization is found to be anti-correlated with flux density at both 1400 and 2368 MHz, but not at 1384 MHz. Higher-resolution observations with high S/N are needed to obtain a clearer picture of the fractional polarization properties.
\item Observed-frame rotation measures have been determined for half of the sample. The maximum observed-frame RM magnitude is 165 rad m$^{-2}$.
\item The surface density of the MRCR--SUMSS sample is 577 sr$^{-1}$.
\item 14 per cent of the sources have candidate optical identifications in the SuperCOSMOS Sky Survey.
\end{enumerate}

\section*{Acknowledgments}

JWB acknowledges the receipt of both an Australian Postgraduate Award and a Denison Merit Award. RWH and JJB acknowledge support from the Australian Research Council. EMS acknowledges support from the Australian Research Council through the award of an ARC Australian Professorial Fellowship. We thank David Crawford for providing the MRCR, Bryan Gaensler for useful discussions and the anonymous referee for helpful suggestions. The Australia Telescope Compact Array is part of the Australia Telescope which is funded by the Commonwealth of Australia for operation as a National Facility managed by CSIRO. SuperCOSMOS Sky Survey material is based on photographic data originating from the UK, Palomar and ESO Schmidt telescopes and is provided by the Wide-Field Astronomy Unit, Institute for Astronomy, University of Edinburgh. This research has made use of the NASA/IPAC Extragalactic Database (NED) which is operated by the Jet Propulsion Laboratory, California Institute of Technology, under contract with the National Aeronautics and Space Administration.

{}

\bsp

\end{document}